%% file: MIT-Thesis.tex
\DocumentMetadata{ 
	pdfstandard = a-2b,
	pdfversion  = 1.7,
	lang		= en-US,
}

\documentclass[fontset=newtx-sans-text]{mitthesis} %,fontset=libertine, fontset=newtx-sans-text, fontset=heros-stix2, fontset=stix2
\usepackage{listings}%   documentation is here https://ctan.org/pkg/listings

\usepackage{lipsum}
\IfPackageAtLeastTF{lipsum}{2021/09/20}{\setlipsum{auto-lang=false}}{}

\usepackage{booktabs}% publication quality tables (https://ctan.org/pkg/booktabs)
\usepackage{graphicx} % To resize the table
\usepackage{longtable}
\usepackage[normalem]{ulem}
\useunder{\uline}{\ul}{}

\usepackage{array}% Additional options for column formats (https://ctan.org/pkg/array)

\usepackage{multirow}
\usepackage{caption}
\usepackage{subcaption}
\usepackage{amsmath}
\usepackage[listings,skins,breakable]{tcolorbox}
\usepackage{float}

\DeclareUnicodeCharacter{2060}{}
\DeclareUnicodeCharacter{05F3}{'}

\graphicspath{ {figures/} }% For details see: https://latexref.xyz/dev/latex2e.html#g_t_005cgraphicspath

\usepackage[numbers,sort&compress]{natbib}

\usepackage{setspace}%  documentation at https://ctan.org/pkg/setspace
\hypersetup{%
	pdfsubject={Template for writing MIT theses with the mitthesis class},
	pdfkeywords={Massachusetts Institute of Technology, MIT},
	pdfurl={},
	pdfcontactemail={},
	pdfauthortitle={},
}

\begin{document}

%%% edit the following commands to match your thesis %%%%%%%%%%

\title{The Power of Perception in Human-AI Interaction: Investigating Psychological Factors and Cognitive Biases that Shape User Belief and Behavior}

% \Author{Author full name}{Author department}[Author's first PREVIOUS degree][Author's second PREVIOUS degree][...
% Note that third, fourth, fifth, and sixth arguments are optional [] and may be omitted

\Author{Eunhae Lee}{Integrated Design and Management and \\ 
\hspace*{10.4em} Department of Electrical Engineering and Computer Science}[Bachelor of Arts, Bryn Mawr College (2015)]

% Use once for each degree fulfilled by thesis
% For two degrees from one department, leave the department argument blank for the second degree {}.
\Degree{Master of Science in Engineering and Management}{Integrated Design and Management Program}
\Degree{Master of Science in Electrical Engineering and Computer Science}{Department of Electrical Engineering and Computer Science}

%\Degree{Master of Science in Physics}{}
%\Degree{Bachelor of Science in Mechanical Engineering}{Department of Mechanical Engineering}

% If there is more than one supervisor, use the \Supervisor command for each.
\Supervisor{Manish Raghavan}{Assistant Professor of EECS}
\Supervisor{Pattie Maes}{Professor of the Media Lab}
% \Supervisor{Edward C. Pickering}{Professor of Physics, and \\ \> Professor of Something Else}
% \Supervisor{Secunda Castor}{Professor of Research}
% \Supervisor{Quintus Castor}{Professor of Log Dams}

% Professor who formally accepts theses for your department (e.g., the Graduate Officer, Professor Sméagol,...)
% If more than one department, use more than once
\Acceptor{Joan Rubin}{Executive Director, System Design and Management}{}
\Acceptor{Leslie A. Kolodziejski}{Professor of Electrical Engineering and Computer Science \\ \hspace*{10.4em} Chair, Department Committee on Graduate Students}{}
% \Acceptor{Tertius Castor}{Professor of Log Dams}{Graduate Officer, Department of Research}
% \Acceptor{Quarta Castor}{Professor of Lodge Building}{Graduate Officer, Department of Mechanical Engineering}
%% If you need to reduce vertical space, put the acceptor title in the second argument and leave the third blank, {}.

% Usage: \DegreeDate{Month}{year}
% Valid degree months are February, May, June, or September
\DegreeDate{September}{2024}

% Date that final thesis is submitted to department
\ThesisDate{August 16, 2024}

%%%%%%  Choose whether to have a CREATIVE COMMONS License  %%%%%%%%%%%%%%%%%%%%%%%%%%%%%%%%%%%%%%
%
% If you are using a cc license, put details of your cc license here. 
% Omit this command if you are not using a cc license.
%
\CClicense{CC BY 4.0}{https://creativecommons.org/licenses/by-nc-nd/4.0/}
%

%%%%%%%  Solutions for overflowing titlepage  %%%%%%%%%%%%%%%%%%%%%%%%%%%%%%%%%%%%%%%%%%%%%%%%%%%

% If your title page is overflowing (from too many names, degrees, etc.):
%
% (a) you can reduce the 12pt and 18pt skips between various blocks to 6pt with this command:
%
\Tighten
%
% (b)  you can scale down the Signature block at the bottom with this command:
%
% \SignatureBlockSize{\small}  %or this one 
\SignatureBlockSize{\footnotesize}
%
% (c) you can put the acceptor name and title onto two lines, rather than three like this:
%
% \Acceptor{Tertius Castor}{Professor and Graduate Officer, Department of Research}{}
%
% (d) you can change the font size of the author name[s] with
%
	% \AuthorNameSize{\normalsize}
%
% (e) and you can omit any previous degrees from the title page, instead mentioning them in the biographical sketch

% Also, if you prefer to keep the text toward the top of the page with most white space at the bottom, you
% can use this command to squash all of the vertical glue (stretchy space) with this command:
%
% \Squash 
%
% This command is useful when the text has not already reach the bottom of the page, since the glue gets squashed automatically
% when the page is too full.

%%%%%%%%%%%%%%%%%%%%%%%%%%%%%%%%%%%%%%%%%%%%%%%%%%%%%%%%%%%%%%%%%%%%%%%%%%%%%%%%%%%%%%%%%%%%%%%%%

%%% Make titlepage
\maketitle

%%%%%%%%% Contents that you need to write follows! %%%%%%%%%%%%%%%%%%%%%%%%%%%%%%%%%%%%%%%%%%%%%%

% \includeonly{acknowledgments,biography,chapter1,chapter2,...,appendixa,...} 
%   for usage of includeonly, see https://latexref.xyz/_005cinclude-_0026-_005cincludeonly.html

%%% Frontmatter (write this material in the mentioned files)  %%%%%%%%%%%%%%%%%%%%%%%%%%%%%%%%%%%

% The abstract environment creates all the required headings and footers. 
% You only need to the text of the abstract in the file abstract.tex
\begin{abstract}
	\input{abstract.tex}% use \input rather than \include because we're inside an environment
\end{abstract}

\include{acknowledgments}% acknowledgments.tex (.tex extension is presumed by \include) 

% \include{biography}% biography.tex (optional, see https://libraries.mit.edu/distinctive-collections/thesis-specs/#format)

%%% Table of contents and lists of stuff (delete unused lists, i.e., if no tables or figures) %%%%%

\tableofcontents
\listoffigures
\listoftables

%%% Chapters of thesis  %%%%%%%%%%%%%%%%%%%%%%%%%%%%%%%%%%%%%%%%%%%%%%%%%%%%%%%%%%%%%%%%%%%%%%%%%%%

%% If you want to use "double spacing", you should start here...

 \include{chapter1}% .tex extension is presumed
\include{chapter2}
\include{chapter3}
\include{chapter4}

\include{chapter5}

%%% Appendicies of thesis  %%%%%%%%%%%%%%%%%%%%%%%%%%%%%%%%%%%%%%%%%%%%%%%%%%%%%%%%%%%%%%%%%%%%%%%%

\appendix
\include{appendixa}
\include{appendixb}
\include{appendixc}

\include{appendixd}
\include{appendixe}

%%% Bibliography (biblatex)  %%%%%%%%%%%%%%%%%%%%%%%%%%%%%%%%%%%%%%%%%%%%%%%%%%%%%%%%%%%%%%%%%%%%%%

% \defbibheading{bibintoc}{\chapter*{#1}\addcontentsline{toc}{backmatter}{\refname}} 
% this sets the title of contents name for bibliography to \refname (= References)
% change "backmatter" to "chapter" if you prefer a bold face entry in the table of contents

% \printbibliography[title={\refname},heading=bibintoc]

% biblatex also supports chapter-by-chapter bibliography, https://tex.stackexchange.com/a/296502/119566
% see the biblatex manual, section 3.14.3

%%%% Option for natbib %%%%%%%%%%%%%

%%   use an appropriate style (.bst) and your own .bib file[s]

\bibliographystyle{unsrtnat}
\bibliography{references}

\end{document}

%% file: abstract.tex
% From mitthesis package
% Version: 1.01, 2023/06/19
% Documentation: https://ctan.org/pkg/mitthesis
%
% The abstract environment creates all the required headers and footnote. 
% You only need to add the text of the abstract itself.
%
% Approximately 500 words or less; try not to use formulas or special characters
% If you don't want an initial indentation, do \noindent at the start of the abstract

\singlespacing
\noindent
This thesis investigates the psychological factors that influence belief in AI predictions, comparing them to belief in astrology- and personality-based predictions, and examines the "personal validation effect" in the context of AI, particularly with Large Language Models (LLMs). Through two interconnected studies involving 238 participants, the first study explores how cognitive style, paranormal beliefs, AI attitudes, and personality traits impact perceptions of the validity, reliability, usefulness, and personalization of predictions from different sources. The study finds a positive correlation between belief in AI predictions and belief in astrology- and personality-based predictions, highlighting a "rational superstition" phenomenon where belief is more influenced by mental heuristics and intuition than by critical evaluation. Interestingly, cognitive style did not significantly affect belief in predictions, while paranormal beliefs, positive AI attitudes, and conscientiousness played significant roles. The second study reveals that positive predictions are perceived as significantly more valid, personalized, reliable, and useful than negative ones, emphasizing the strong influence of prediction valence on user perceptions. This underscores the need for AI systems to manage user expectations and foster balanced trust. The thesis concludes with a proposal for future research on how belief in AI predictions influences actual user behavior, exploring it through the lens of self-fulfilling prophecy. Overall, this thesis enhances understanding of human-AI interaction and provides insights for developing AI systems across various applications.

%% file: acknowledgments.tex
%% acknowledgments.tex

% From mitthesis package
% Version: 1.02, 2024/06/19
% Documentation: https://ctan.org/pkg/mitthesis

\chapter*{Acknowledgments}
\pdfbookmark[0]{Acknowledgments}{acknowledgments}

First, I would like to express my sincere gratitude to my advisors, Professor Pattie Maes and Manish Raghavan, for their support, guidance, and encouragement throughout my research. Their expertise and support have greatly contributed to the completion of this thesis.

I extend my sincere thanks to my collaborator, Pat Pataranutaporn, for his invaluable assistance, stimulating discussions, and boundless energy throughout this process. His support made this journey not only intellectually rewarding but also deeply enjoyable. I also want to thank Judith Amores at Microsoft Research for her valuable feedback and enthusiastic contribution. 

I am grateful to the Fluid Interfaces Group at the MIT Media Lab for providing the essential resources needed to conduct my research and for offering constructive feedback along the way. Additionally, I would like to express my appreciation to Jinjie Liu at the Institute for Quantitative Social Science at Harvard University, for sharing her insightful perspective on the analysis, which greatly enriched this work.

I also wish to acknowledge the academic advisors and administrators in the IDM/SDM program and the EECS department, whose guidance has been invaluable as I navigated the winding path of pursuing a dual Master’s degree.

My deepest thanks go to my family for their endless love and support. To my parents, Jong In Lee and Ji Young Kim, your encouragement and belief in me have been the driving force behind my success. I am also profoundly grateful to my partner, Bryan Ibarra Wong, for his patience, understanding, and unwavering support during the most challenging moments of this journey.

Finally, I would like to thank everyone who contributed to the successful completion of this thesis, including the advisors and colleagues at the MIT AgeLab who provided early guidance, my friends in the IDM program, and the external colleagues and friends who offered fresh perspectives that enriched my work along the way.

%% file: chapter1.tex
% From mitthesis package
% Version: 1.06, 2024/07/09
% Documentation: https://ctan.org/pkg/mitthesis

\chapter{Introduction}
The rapid advancement of artificial intelligence (AI) has sparked a complex dialogue that intertwines technological progress with elements of quasi-religious narratives and psychological manipulation. As AI systems, particularly generative AI and large language models (LLMs), become more integrated into daily life, understanding the psychological dynamics that influence human-AI interaction is increasingly crucial. This thesis builds on two interconnected studies that explore these dynamics through different but complementary lenses: the concept of "rational superstition" in AI and the "personal validation effect" in human-AI interactions.

The first study investigates how psychological factors influence belief in AI predictions, drawing unexpected parallels between AI and astrology. Despite AI's foundation in scientific and rational methodologies, public perception often mirrors belief systems associated with pseudoscience, such as astrology. This phenomenon, termed "rational superstition" \cite{wilson_techno-optimism_2017}, suggests that people may trust AI predictions not because of critical evaluation but due to mental heuristics and intuitive thinking. By comparing responses to AI predictions to those derived from astrology and personality psychology, this study examines how cognitive style, paranormal beliefs, AI attitudes, and personality traits influence perceptions of AI's validity, reliability, usefulness, and personalization.

Building on the insights from the first study, the second study delves into the "personal validation effect" in the context of LLM use. Here, we explore how the valence of AI-generated predictions—whether positive or negative—affects users' perceptions of these predictions' validity, reliability, usefulness, and personalization. This study uncovers the significant impact of positive AI feedback, revealing that users are more likely to perceive favorable predictions as more accurate and personalized, even when these predictions are demonstrably false. This effect highlights the potential for AI systems to manipulate user perceptions through positive reinforcement.

Together, these studies offer a comprehensive examination of how belief in AI predictions is shaped by a combination of cognitive biases, psychological factors, and the framing of AI outputs. The first study provides insights on the psychological underpinnings of trust in AI, while the second study extends this understanding by exploring specific mechanisms by which AI systems can manipulate these beliefs. By integrating these findings, this thesis aims to contribute to the broader discourse on AI ethics, transparency, and the responsible design of AI systems that foster appropriate trust and skepticism.

Finally, this thesis outlines ongoing and future work aimed at extending our understanding of how people perceive and believe in AI predictions, focusing on the underlying psychological factors. I discuss a proposed study that would explore how these perceptions influence actual behavior, drawing on the well-known phenomenon of the "self-fulfilling prophecy." Specifically, it would investigate how AI-generated predictions about individuals may shape their actions in ways that ultimately fulfill those predictions.

The flow of the research presented in this thesis reflects a logical progression from understanding the broader psychological factors influencing belief in AI predictions to exploring how AI predictions may impact actual behavior and outcomes. This integrated approach underscores the importance of considering both the cognitive and affective dimensions of human-AI interaction, ultimately informing strategies for designing AI systems that are both effective and ethically sound.

%% file: chapter2.tex
% From mitthesis package
% Version: 1.06, 2024/07/09
% Documentation: https://ctan.org/pkg/mitthesis

\chapter{Super-intelligence or Superstition? Exploring Psychological Factors Underlying Unwarranted Belief in AI Predictions}
\label{ch2_paper1}

\section{Introduction and Related Works}

Technology is often associated with scientific advancement and frequently viewed in opposition to superstition. However, the recent rapid advancements in artificial intelligence (AI) have given rise to quasi-religious perspectives among some leaders in the tech industry \cite{CultofAI}. In fact, as generative AI and large language models (LLMs) continue to evolve, a contemporary critique posits that the narrative surrounding their development bears striking resemblances to religious discourse \cite{sigal_silicon_2023}. Promoting concepts such as superintelligence and sentience, prominent figures in the tech industry have been observed making prophetic claims regarding AI's potential to either salvage or obliterate humanity \cite{danaher_techno-optimism_2022, konigs_what_2022}. These claims are often based on personal speculation and subjective views rather than rigorous scientific research. Nonetheless, empirical research has demonstrated that these popular narratives are mirrored in public perceptions of AI, frequently manifesting as exaggerated utopian or dystopian visions \cite{brauner2023public, sartori_minding_2023}. This intersection of technological advancement and quasi-religious perspective presents a complex landscape for scholars to navigate, as it blurs the boundaries between scientific progress and speculation.

While AI has been developed through predictive algorithms and statistical pattern recognition, exemplifying advancements in mathematical and rational thinking, public perception of AI may draw unexpected parallels with astrology \cite{nikolic_ecs-ecrea_2023, lazaro_pouvoir_2018, steyerl_sea_2018}. Despite astrology representing an opposing epistemology based on pseudoscientific methodologies antithetical to empirical reasoning, both AI and astrology function as prediction-generating mechanisms helping people make sense of the uncertainty in the world \cite{nikolic_ecs-ecrea_2023, lazaro_pouvoir_2018}. Anthropologist Christophe Lazaro suggests that AI, powered by data, has become a new mode of speculative thinking, with algorithms taking on the role of oracles in contemporary societies, framing AI as "artificial divination" \cite[p.~128]{lazaro_pouvoir_2018}. 

On the other hand, AI, unlike astrology, presents an aura of scientific legitimacy, backed by a vast body of scientific research in computer science, mathematics, cognitive science, and more. This foundation contributes to the belief that AI systems are rational and objective. However, studies have shown that algorithms could produce unpredictable outcomes, biased results, and inherit implicit societal biases \cite{pethig2023biased, kordzadeh2022algorithmic, johnson2020algorithmic}. Moreover, despite ongoing efforts to increase transparency, much of the inner workings of AI systems remain "black boxes" \cite{voneschenbach2021transparency, arrieta2024interpreting, linardatos2021explainable, castelvecchi2016blackbox}. Such misconceptions regarding AI may engender a form of naive technological optimism and irrational trust in the system, which Wilson refers to as "rational superstition" \cite{wilson_techno-optimism_2017}. Scholars have investigated the concept of algorithmic appreciation—the propensity to rely on algorithms for decision-making processes—which may lead to an overreliance on AI in critical decision-making contexts \cite{logg_algorithm_2019, mahmud_decoding_2024}.

Popular narratives that over-inflate the promises of AI and its capabilities is particularly alarming when considering the impact it has on mental models and perception of AI systems \cite{sartori_minding_2023}. According to the dual process theory in psychology, people engage in two types of thinking---fast and intuitive (System 1) and slow and reflective (System 2) \cite{kahneman_thinking_2011, evans_two_2009}. Incomplete or incorrect mental models shaped by speculative claims about AI rather than scientific evidence can become a mental heuristic that triggers people’s System 1 thinking rather than engaging their System 2 thinking, blurring the lines between speculation and informed reasoning \cite{dale_heuristics_2015}. This, in turn, can lead to miscalibration of trust and over-reliance on AI, which drives ineffective use of AI \cite{logg_algorithm_2019, mahmud_decoding_2024, steyvers_three_2023}. 

For instance, consider an AI system that claims to give you personalized product recommendations based on your social media usage and purchase history. A user with an incorrect mental model of the system's capabilities and limitations may hastily rely on its recommendations without considering other options. On the flip side, a user who has developed an overly skeptical mental model of AI may reject helpful recommendations and miss critical opportunities. Therefore, without ongoing effort to establish appropriate mental models, users can become more prone to manipulation and potential negative outcomes.

While belief in AI supremacy is growing despite its demonstrated limitations and biases, our study presents the first empirical investigation into the phenomenon of "rational superstition" \cite{wilson_techno-optimism_2017} in AI\footnote{This chapter is adapted from a preprint on arXiv:2408.06602, co-authored with Pat Pataranutaporn, Judith Amores, and Pattie Maes, 2024 \cite{lee2024superintelligence}.}. To do so, we examine how people's perception and trust in AI predictions may correlate with other irrational behaviors, such as belief in astrology. This research question is particularly timely and relevant given the recent surge in popularity of astrology, especially among Millennials and Gen Z \cite{page_why_2023, farrar_why_2022}, despite the lack of empirical evidence supporting its validity \cite{mcgrew_scientific_1990, nyborg_relationship_2006}. A recent consumer survey of over 2000 people in the US revealed that about a third of Millennials and a quarter of Gen Z have made financial decisions based on their horoscope \cite{safier_written_2021}. %Personality psychology, which began in the early 20th century as a scientific field studying individual differences in personality and behavior \cite{barenbaum_history_2008}, serves as another point of comparison, often consulted by those seeking self-understanding and future possibilities.

To address this research question, we conducted an experiment involving 238 participants who were presented with fictitious predictions from three sources (AI, astrology, and personality psychology) and rated their perceived validity, reliability, usefulness, and personalization. To create a reasonable basis for the predictions, participants answered questionnaires about astrology and personality and engaged in a simulated investment game that appeared to analyze their interactions and predict their future investment behavior. Participants were randomly assigned to either positive (N=119) or negative (N=119) prediction groups, receiving forecasts about their future investment behavior and outcomes accordingly.

Our study also explores how cognitive style, paranormal beliefs, gullibility, trust in AI, and personality traits may influence belief in predictions based on AI, astrology, and personality psychology. This investigation builds upon previous work examining how user characteristics such as personality traits influence the perception of AI-generated advice \cite{wester_exploring_2024} and trust in AI \cite{riedl_is_2022}. Additionally, we consider the relationship between cognitive style and paranormal beliefs, including astrology \cite{bensley_critical_2023, ballova_mikuskova_effect_2020, pennycook_everyday_2015, torres_validation_2023, pennycook_2012}, as well as how cognitive reflection affects social media usage and the sharing of (mis)information and conspiracy theories \cite{mosleh_cognitive_2021, stecula_social_2021}.

Furthermore, we explore how demographic factors such as age, gender, and education level, as well as prior experience in AI, astrology, and personality psychology, and
the level of interest in the topic influence belief in the predictions. By examining these various factors, our research aims to provide a comprehensive understanding of the "rational superstition" phenomenon in AI and its implications for decision-making in an increasingly AI-driven world.

% To summarize, our key research questions were: 

% \begin{itemize}
%     \item R1: How does the source of predictions (AI, astrology, personality psychology) influence perceived validity, reliability, usefulness, and personalization of the predictions?
%     \item R2: What are the psychological and cognitive factors that moderate the believability of these predictions?
%     % How do cognitive style, paranormal beliefs, trust in AI, gullibility, personality, prior knowledge/experience, and demographic factors such as age, gender, and education level moderate the believability of predictions? 
% \end{itemize}

Based on our research questions, we tested the following hypotheses:

\begin{itemize}
    \item H1: Individuals who are more likely to believe in predictions based on astrology and personality will be more likely to believe in AI predictions.
    \item H2: Individuals with higher cognitive reflection and need for cognition will find AI predictions more credible and reliable than astrology- or personality-based predictions.
    \item H3: Demographic factors and personal attributes (cognitive reflection, trust in AI, paranormal beliefs, gullibility, personality, level of interest, and prior experience) will moderate the perceived validity, reliability, usefulness, and personalization of predictions from different sources.
\end{itemize}

The study's findings contribute valuable insights to the ongoing discourse on AI ethics, transparency, and responsible implementation. By shedding light on the factors that influence trust and belief in AI predictions, we aim to inform policymakers, AI developers, and educators about potential pitfalls and areas for improvement in AI literacy and public engagement with AI technologies.

\section{Methods}
\label{ch2-methods}

\subsection{Participants}
\label{ch2-methods-participants}

Data were initially collected from a survey administered to 300 participants recruited through Prolific. After excluding 62 participants due to incomplete responses or failed attention checks, the final analysis was conducted with data from the remaining 238 participants. These individuals were adults aged 18 and above from diverse socioeconomic backgrounds (see Table \ref{table:demographics} for an overview of participants' demographic information).

\begin{table}[h]
\centering
\caption{Summary of participants’ demographic information}
\label{table:demographics}
\begin{tabular}{@{}llc@{}}
\toprule
\multirow{2}{*}{Number per condition} & Positive     & 119   \\  
                                      & Negative     & 119   \\ \midrule
\multirow{2}{*}{Age}                  & Mean         & 41.04 \\  
                                      & SD           & 12.38 \\ \midrule
\multirow{3}{*}{Gender}               & Female       & 53\%  \\  
                                      & Male         & 43\%  \\  
                                      & Other        & 3\%   \\ \midrule
\multirow{5}{*}{Education}            & High school  & 12\%  \\  
                                      & Some college & 21\%  \\  
                                      & Bachelor     & 43\%  \\  
                                      & Master       & 14\%  \\  
                                      & Other        & 9\%   \\ \bottomrule
\end{tabular}
\end{table}

\subsection{Experiment Protocol}
\label{ch2-methods-experiment}

\begin{figure}[h]
\includegraphics[width=15cm]{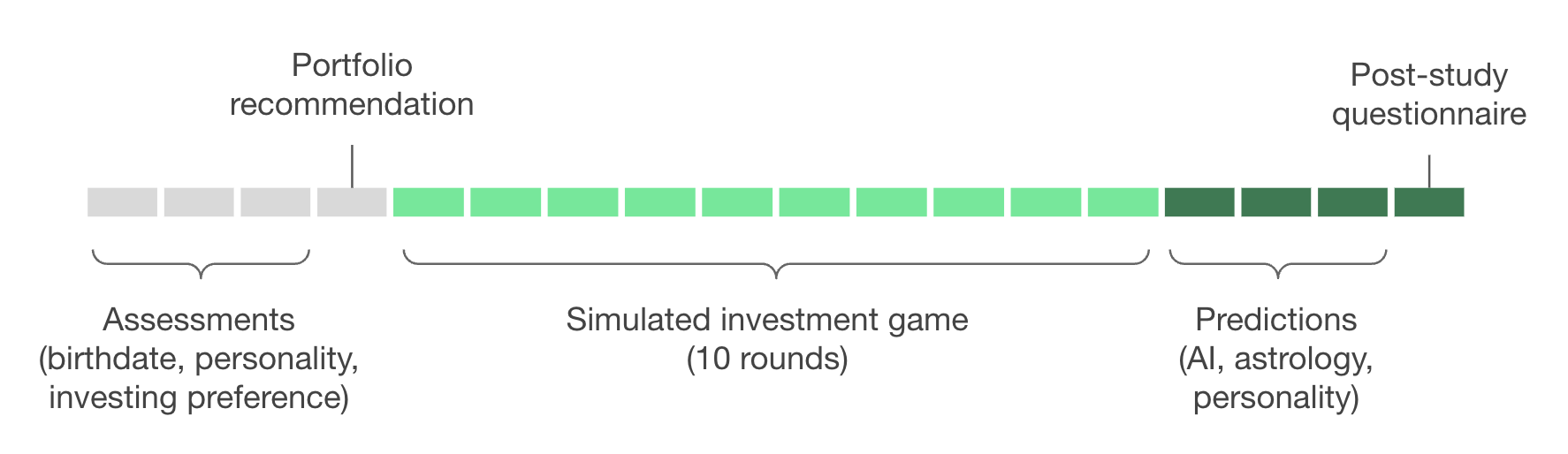}
    \centering
    \caption{Diagram of the study procedure}
\label{fig:procedure_diagram}
\end{figure}

\subsubsection{Assessments and simulated investment game}

A visualization of the study procedure is shown in Figure \ref{fig:procedure_diagram}. First, participants were asked to complete a short zodiac-related questionnaire (date, hour, and location of birth), a short personality test (adaptation of Myers-Briggs Type Indicator (MBTI) \cite{myers_myers-briggs_1962}, and a simulated investment game. In the instructions, participants were told that they will receive personalized predictions about their future investment behavior generated from three distinct sources (astrology, personality, and AI) based on their game interactions and the information they provided. What they were \emph{not} told in the beginning was that these predictions were generic, pre-determined statements that were not based on their responses, which was revealed to them after completion.

The simulated investment game was designed to elicit interactions comparable to that of modern robo-advising platforms \cite{dacunto_promises_2019}, but in a simplified way. Participants were provided with virtual currency of \$10,000 to invest across three investment categories (high risk/return, medium risk/return, low risk/return) over ten rounds (representing years). At each round, they could allocate up to 100\% of their assets across the three categories, with the goal of maximizing their total portfolio value. 

A few design choices were made to increase the sense of realism for the investing game and to increase engagement: 
\begin{itemize}
    \item \textbf{Recommended portfolio allocation:} Participants filled out a short questionnaire about their risk preferences for investing before starting the game and received a "customized" portfolio recommendation that they could view throughout the game. In reality, all participants received the same portfolio recommendation. 
    \item \textbf{Realistic market scenario:} The game was designed so that participants could experience the hypothetical ups and downs of the market, through a pre-determined series of alternating "bull market" (more gains) and "bear market" (more losses) scenarios that influenced the probability of the possible investment outcomes at each round.
    \item \textbf{Market forecasts:} To further enhance the emotional engagement, each round started off with either a positive or negative market forecast that were aligned with the behind-the-scenes market scenario. A detailed market scenario that was used in the game and examples of the market forecast can be found in Appendix \ref{appendix:game_details}. 
    \item \textbf{Financial incentives:} To motivate engagement, participants were told their investment performance would increase their total compensation for the study, inspired by previous studies that used real-life incentives to increase engagement \cite{wang_emotion_2014, shiv_investment_2005}.
\end{itemize}

\begin{figure}[h]
    \includegraphics[width=15cm]{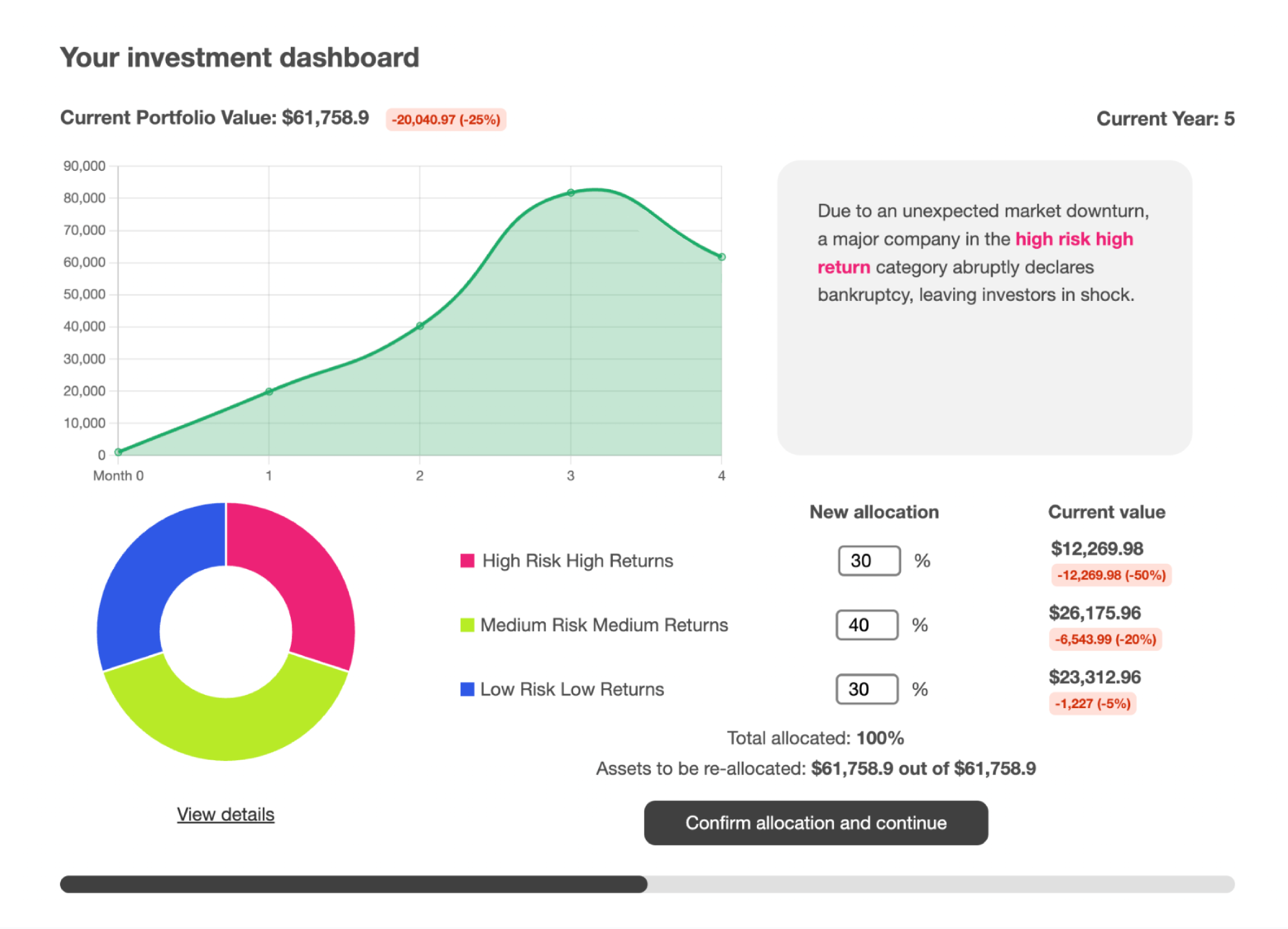}
    \centering
    \caption{Dashboard of the simulated investment game}
    \label{fig:game_dashboard}
\end{figure}

A snapshot of the simulated investment game can be seen in Figure \ref{fig:game_dashboard}. The source code for the simulated investment game and analysis is available in the GitHub repository, \url{https://github.com/mitmedialab/ai-superstition}.

\subsubsection{Predictions}

We employed a placebo approach that was proven effective in previous studies \cite{villa_placebo_2023, kosch_placebo_2022} where participants were informed that predictions were AI-generated, but were in fact pre-determined (either positive or negative), and randomly assigned. After the game, participants were presented with these pre-determined predictions from three different sources (AI, astrology, personality) in randomized order. Participants were randomly assigned to either the “Positive” prediction group (rational investor, higher returns) or “Negative” prediction group (impulsive investor, lower returns). The initial predictions were created by the authors, with variations generated using ChatGPT-4.  Examples of predictions received are shown in Table \ref{table:predictions}.  For each prediction, they were asked to respond to evaluate its perceived validity, personalization, reliability, and usefulness. The evaluation questionnaire is provided in Section \ref{sec: believability_subscales}.

\begin{table}[]
\centering
\caption{Example of predictions received by Positive and Negative prediction groups}
\label{table:predictions}
\begin{tabular}{@{}p{2.3cm}p{6.7cm}p{6.7cm}@{}}
\toprule
\textbf{Source} &
  \textbf{Positive prediction group} &
  \textbf{Negative prediction group} \\ \midrule
\textbf{Astrology} &
  As the Moon's celestial path brings it into close proximity with the stoic Saturn, this cosmic alignment sheds light on the inherent human struggle between emotion and reason when making financial decisions. Based on your astrological sign, you are more likely to make rational investment decisions based on reason rather than emotion. You have a good grasp of your emotions especially when it comes to decisions about money. Therefore, we predict that your portfolio will have higher-than-average performance in the long run. &
  As the Moon's celestial path brings it into close proximity with the stoic Saturn, this cosmic alignment sheds light on the inherent human struggle between emotion and reason when making financial decisions. Based on your astrological sign, your investment decisions are often impulsive, swayed by emotions rather than thoughtful analysis. Your emotions play a significant role in your investment decisions. As a result, it's expected that your portfolio's performance will be under the average in the long run. \\ \midrule
\textbf{Personality} &
  Whether you're an analytical thinker or led by intuition, your Myers-Briggs profile offers valuable insights into your behavior when it comes to making investment decisions. Based on your personality analysis, you are more likely to make rational investment decisions based on reason rather than emotion. You have a good grasp of your emotions especially when it comes to decisions about money. Therefore, we predict that your portfolio will have higher-than-average performance in the long run. &
  Whether you're an analytical thinker or led by intuition, your Myers-Briggs profile offers valuable insights into your behavior when it comes to making investment decisions. Based on your personality analysis, your investment decisions are often impulsive, swayed by emotions rather than thoughtful analysis. Your emotions play a significant role in your investment decisions. As a result, it's expected that your portfolio's performance will be under the average in the long run. \\ \midrule
\textbf{AI} &
  Within the digital echo of your actions, AI sifts through the noise to forecast your next move. It transforms your clicks, likes, and digital whispers into a map of future investment choices. Based on our AI model analysis, you are more likely to make rational investment decisions based on reason rather than emotion. You have a good grasp of your emotions especially when it comes to decisions about money. Therefore, we predict that your portfolio will have higher-than-average performance in the long run. &
  Within the digital echo of your actions, AI sifts through the noise to forecast your next move. It transforms your clicks, likes, and digital whispers into a map of future investment choices. Based on our AI model analysis, your investment decisions are often impulsive, swayed by emotions rather than thoughtful analysis. Your emotions play a significant role in your investment decisions. As a result, it's expected that your portfolio's performance will be under the average in the long run. \\ \bottomrule
\end{tabular}
\end{table}

\subsubsection{Post-study questionnaire}

Following the experiment, participants were asked to respond to a series of questionnaires to measure key user characteristics (cognitive style, paranormal beliefs, gullibility, trust in AI/attitudes towards AI, personality, etc.) and demographic information (age, gender, education level). A summary of the scales used in the questionnaires is provided in Table \ref{table:measures}. Detailed questionnaires can be found in Section \ref{sec:scales}.

\begin{table}[]
\centering
\caption{Summary of the measures for the user characteristics}
\label{table:measures}
\begin{tabular}{@{}p{4.5cm}p{12cm}@{}}
\toprule
\textbf{Variable name} & \textbf{Measure} \\ \midrule
Cognitive Style & A composite measure of cognitive style that combines cognitive reflection test (CRT-2) \cite{thomson_investigating_2016} and need for cognition (NCS-6) \cite{lins_de_holanda_coelho_very_2020}. Higher score indicates more analytic cognitive style. \\ \midrule
Paranormal Beliefs & A measure of participants’ belief in paranormal phenomena. A shortened version of R-PBS \cite{tobacyk_revised_2004} was used. Higher score indicates stronger belief in paranormal phenomena. \\ \midrule
AI Attitude Score & A score indicating participants’ attitudes towards AI predictions. AIAS-4 \cite{grassini_development_2023}, a 4-item scale was used. Higher score indicates more positive attitudes. \\ \midrule
Gullibility & A self-reported measure of participants' tendency to be gullible, developed by \cite{teunisse_i_2020}. Higher score indicates greater gullibility. \\ \midrule
Big Five Personality Traits & Includes extraversion, openness, agreeableness, conscientiousness, and emotional stability. Used 10-item Big Five Personality Inventory (TIPI) \cite{gosling_very_2003}. \\ \midrule
Interest in behavior & A measure of participants’ interest in the topic of prediction (personal investment behavior). "How would you rate your level of interest in understanding your future investment behavior?" (5-point Likert scale, 1=Not interested, 5=Extremely interested) \\ \midrule
Familiarity & A measure of participants' familiarity with the prediction sources (AI, Astrology, Personality). "How would you rate your level of familiarity with the following: 1) Astrology and horoscopes, 2) Personality psychology, 3) AI prediction systems" (5-point Likert scale, 1=Not at all, 5=Extremely familiar) \\ \bottomrule
\end{tabular}
\end{table}

\subsection{Measures}
\label{sec:scales}

This section includes an overview of the questionnaires that were used to measure the variables. For detailed questionnaires, see Appendix \ref{appendix_survey}. 

\subsubsection{Perceived validity, personalization, reliability, and usefulness}
\label{sec: believability_subscales}

To measure the believability of the predictions, we presented a short questionnaire that assess perceived validity, reliability, usefulness, and personalization of given predictions. These factors were chosen based on existing literature on the different qualities of a statement that contribute to persuasion, user adoption of an information technology, and positive attitudes. 

Perceived validity/accuracy of the content and reliability of the source was added according to the Elaboration Likelihood Model (ELM) in sociology, which argues that the logical validity of an argument is considered central to its believability, but people can also engage in peripheral ways of being persuaded, such as the reliability of the source \cite{petty_elaboration_1986}. Moreover, perceived usefulness was included as a metric that contributes to believability of a prediction, inspired by the Technology Acceptance Model (TAM)\cite{davis_perceived_1989}. According to this model, perceived usefulness, defined as the degree to which a person believes that using a particular system would enhance their job performance, significantly influences users' attitudes and intentions, which in turn affect actual technology usage. Lastly, perceived personalization has been found to increase attention and positive attitudes toward a statement \cite{maslowska_it_2016}.

We asked participants to rate eight statements on 7-point Likert scales (1=Strongly disagree, 7=Strongly agree), which include two statements per subscale (perceived validity, personalization, reliability, and usefulness).

\subsubsection{Cognitive Reflection (CRT-2) and Need for Cognition (NCS-6)}

Cognitive reflection, or the tendency to suppress an intuitive but incorrect response in favor of a more reflective and correct one, was chosen as a moderating factor for belief in predictions to explore how it influences the acceptance and trust in predictions about personal behavior. This approach aims to determine if higher cognitive abilities correlate with more critical evaluation and discernment of AI predictions, as well as predictions based on astrology and personality.

The Cognitive Reflection Test (CRT) was designed to predict performance in normative decision-making \cite{frederick_cognitive_2005}. In our study, we used CRT-2 \cite{thomson_investigating_2016}, an alternate form of the original 6-item CRT, in order to address issues of familiarity and learning effects that could potentially skew results. The 4-item CRT-2 maintains the same objective of measuring cognitive reflection but includes different questions to ensure a fresh and unbiased assessment, with less focus on numeric abilities \cite{thomson_investigating_2016}. 

The Need for Cognition (NFC) test is a self-report measure of an individual's tendency to engage in and enjoy effortful thinking \cite{cacioppo_need_1982}. We use the 6-item NFC assessment (NCS-6) developed by \cite{lins_de_holanda_coelho_very_2020}.

While the CRT measures objective performance, the NFC measures subjective preference, which have been found to be correlated \cite{pennycook_is_2016}. To account for the conceptual similarity of these two measures, they were combined to create a composite measure under the assumption that they have equal contribution to the final score. To do so, we summed the z-scores of the results from the two tests to account for the differences in range and variability \cite{field_discovering_2012}. 

\subsubsection{Paranormal beliefs (R-PBS)}

Paranormal beliefs refer to the conviction in phenomena beyond scientific explanation, such as supernatural events, astrology, and mystical experiences. Paranormal beliefs were chosen as a moderating factor for belief in AI, astrology, and personality predictions to investigate how these unconventional beliefs influence the acceptance of various predictive statements. To quantify the degree of paranormal belief, we adapted the Revised Paranormal Belief Scale (R-PBS) \cite{tobacyk_revised_2004} (7-point Likert scale, 1=Strongly disagree, 7=Strongly agree) to focus on the more relevant factors for our study, including traditional religious beliefs (4), spiritualism (4), precognition (4), superstition (3).

\subsubsection{AI Attitude/Trust in AI (AIAS-4)}

A 4-item, 1-factor AI attitude scale (AIAS-4) \cite{grassini_development_2023} (10-point Likert scale, 1=Not at all, 10=Completely agree) was used to measure participants' attitude towards AI. 

\subsubsection{Gullibility}

A shortened version of the originally 12-item, 2-factor scale by Teunisse et al. (2020) \cite{teunisse_i_2020} (7-point Likert scale, 1=Strongly disagree, 7=Strongly agree) was used to measure self-reported gullibility. 

\subsubsection{Big Five Personality}

Based on the Five Factor personality model, the 10-item Big Five Personality Inventory (TIPI) \cite{gosling_very_2003} (7-point Likert scale, 1=Strongly disagree, 7=Strongly agree) was used to measure participants' personality.

%The initial five-factor model was advanced by Tupes and Christal in 1961 (see Tupes & Christal, 1992).

\subsubsection{Familiarity/Level of expertise}

How would you rate your level of familiarity with the following (5-point Likert scale, 1=Not at all, 5=Extremely familiar):
\begin{itemize}
\setlength\itemsep{0em}
    \item Astrology and horoscopes
    \item Personality psychology
    \item AI prediction systems
\end{itemize}

\subsubsection{Interest in topic of prediction}

How would you rate your level of interest in understanding your future investment behavior? (5-point Likert scale, 1=Not interested, 5=Extremely interested)

\subsection{Approvals} 
This research was reviewed and approved by the MIT Committee on the Use of Humans as Experimental Subjects, protocol number E-5768. The study was pre-registered on AsPredicted.org (see 
 \url{https://aspredicted.org/QDF_8KX}).

\subsection{Analysis}
\label{sec:analysis}

\subsubsection{Variables}
\label{ch2_methods_variables}

The dependent variable (Y) was the subscale scores for perceived validity, personalization, reliability, and usefulness (7-point Likert scale, 1=Strongly disagree, 7=Strongly agree). 

The analysis included a range of predictor variables (X) to examine their effects on the subscale scores. The predictor and control variables were:

\begin{itemize}
\setlength\itemsep{0em}
    \item \textit{prophecy\_source}: A categorical variable indicating the source of the prophecy (Astrology, Personality).
    \item \textit{prophecy\_group}: A categorical variable indicating the valence of prophecy (Positive, Negative).
    \item \textit{composite\_score}: A composite measure of cognitive style.
    \item \textit{paranormal\_score}: A measure of participants' belief in paranormal phenomena. 
    \item \textit{aias\_score}:  A score indicating participants' attitudes/trust towards AI predictions.
    \item \textit{gullibility\_score}: A self-reported measure of participants' tendency to be gullible.
    \item Big Five Personality Traits: A measure including extraversion, openness, agreeableness, conscientiousness, and emotional stability.
    \item \textit{interest\_behavior}: A measure of participants' interest in the topic of prediction (personal investment behavior).
    \item \textit{familiarity}: A measure of participants' familiarity with the prediction sources (AI, Astrology, Personality).
    \item \textit{age}
    \item \textit{gender}: A categorical variable (Female, Male, Other)
    \item \textit{education}: A categorical variable (Bachelor, Master, Doctorate, Professional degree, Associate, Some college, High school, Less than high school)
\end{itemize}

\subsubsection{Multiple linear regression}

To test the hypothesis that individuals who are more likely to believe in astrology and personality-based predictions are also more likely to believe in AI predictions (H1), a multiple linear regression model was first applied to the data in wide format. The dependent variable was \textit{ai\_overall\_score}, and the main predictors were \textit{zodiac\_overall\_score} and \textit{personality\_overall\_score}, with the rest of the variables mentioned in Section \ref{ch2_methods_variables} included as control variables. The model was fitted using the lm function in R. The full results and diagnostics are in Appendix \ref{appendix:results1}.

\begin{equation}
\text{ai\_overall\_score} \sim \text{zodiac\_overall\_score} + \text{personality\_overall\_score} + \text{control variables}
\end{equation}

The following model diagnostics were performed to ensure the validity of the model:

\begin{itemize}
    \item \textbf{Linearity}: Residual plots for the linear regression model indicated no clear patterns, suggesting linearity.
    \item \textbf{Homoscedasticity}: The residual vs. fitted values plot for the linear regression model showed some heteroscedasticity, which was addressed in the mixed-effects model.
    \item \textbf{Normality of Residuals}: The Q-Q plot indicated that the residuals were approximately normally distributed.
    \item \textbf{Multicollinearity}: Variance Inflation Factors (VIFs) were calculated for the linear regression model, indicating that multicollinearity was not a concern. 
    \item \textbf{Independence of Residuals}: The Durbin-Watson test for the linear regression model showed no significant autocorrelation. 
\end{itemize}

\subsubsection{Mixed effects model}
\label{subsubsec:mixed_effects_model}
Mixed-effects models are well-suited for hierarchical data, incorporating both fixed and random effects to test multiple hypotheses simultaneously. They use partial pooling, which shifts group estimates toward the overall mean, effectively reducing Type I errors without the severe power loss of traditional multiple comparisons corrections. This approach accounts for data dependencies and correlations, minimizing the need for explicit adjustments for multiple comparisons \cite{gelman2012why}.

To examine the main effects and moderating effects of psychological and cognitive factors (H2, H3), the data was transformed to a long format, creating a three-level hierarchical structure. Each subject engaged in three types of conditions (\textit{prophecy\_source}) and evaluated across four subscales (\textit{subscale}). The outcome variable was the \textit{subscale\_score}.

The mixed effects model was specified as follows. The fixed effects were defined as:

\begin{equation}
\begin{aligned}
\text{subscale\_score} \sim & \, \text{subscale} * (\text{prophecy\_source} * \text{prophecy\_group} \\
& + \text{prophecy\_source} * \text{composite\_score} + \text{prophecy\_source} * \text{paranormal\_score} \\
& + \text{prophecy\_source} * \text{aias\_score} + \text{prophecy\_source} * \text{gullibility\_score} \\
& + \text{big5\_extraversion} + \text{big5\_openness} + \text{big5\_agreeableness} + \text{big5\_conscientiousness} \\
& + \text{big5\_emotional\_stability} + \text{interest\_behavior} + \text{familiarity} \\
& + \text{prophecy\_source} * \text{Age} + \text{education} + \text{prophecy\_source} * \text{gender})
\end{aligned}
\end{equation}

For the random effects, random intercepts were included to account for variability in baseline levels between subjects (\textit{qualtrics\_code}), and random slopes were included to account for variability in how subjects respond to different conditions (\textit{prophecy\_source}):

\begin{equation}
\text{random  } = \quad \sim \text{prophecy\_source} \mid \text{qualtrics\_code}
\end{equation}

Additionally, the model incorporated a correlation structure and variance weights to further account for the hierarchical structure and heteroscedasticity in the data:

\begin{equation}
\begin{aligned}
\text{correlation} &= \text{corSymm(form = ~1 | qualtrics\_code/prophecy\_source)}, \\
\text{weights} &= \text{varIdent(form = ~1 | subscale * prophecy\_source * prophecy\_group)}
\end{aligned}
\end{equation}

We centered the continuous predictor variables to reduce multicollinearity and improve the stability and interpretability of the coefficient estimates. Centering involved subtracting the mean of the variable from each individual value. This transforms the variable to have a mean of zero while preserving its variance and distribution. After centering the predictors, we observed that four previously non-significant main and interaction terms became significant, including the main effect of Personalization subscale, the main effect Astrology prediction source, and interaction terms between subscale (Personalization, Reliability) and Astrology prophecy source. This change is attributed to the reduction in multicollinearity, resulting in more precise coefficient estimates.

The model was implemented using the lme function from the nlme package in R. The analysis controlled for missing values through listwise deletion. Optimization settings were adjusted to ensure convergence, with maximum iterations and evaluations set to 1000. 

Several model diagnostics were performed to ensure the validity of the model:

\begin{itemize}
    \item \textbf{Linearity}: A residuals vs fitted values plot using standardized residuals indicated no clear patterns, suggesting linearity.
    \item \textbf{Homoscedasticity}: A residuals vs fitted values plot showed a random spread around the horizontal line at zero, confirming that heteroscedasticity was successfully addressed using the variance function (varIdent). One limitation to this diagnostics was the less continuous nature of the outcome variable (\textit{subscale\_score}), which was multilevel with 13 levels (value of 1 to 7 with an increment of 0.5), which caused the residual plot to exhibit a slight diagonal pattern.
    \item \textbf{Normality of Residuals}: Normality of residuals was assessed using a Q-Q plot, using standardized residuals to account for both fixed and random effects. While there were slight deviations at the extremes, they were not severe and thus considered as normal deviations from real-world data. 
    \item \textbf{Multicollinearity}: 
    To address potential structural multicollinearity, we centered the continuous predictor variables. This process reduced the Variance Inflation Factor (VIF) values for all fixed effects to below 5, indicating that multicollinearity was reduced to an acceptable level. The model diagnostics, including standard errors and confidence intervals, confirmed the stability and reliability of the coefficient estimates.
    
   \item \textbf{Independence of Residuals}: 
    We applied the Durbin-Watson test at the subscale level to check for autocorrelation, finding significant positive autocorrelation in each group. Initially, we used an autoregressive correlation structure (corAR1), which resolved the issue for the "Validity" subscale and slightly reduced autocorrelation in others. However, model fit statistics suggested that an unstructured symmetric correlation structure (corSymm) was a better fit, though it did not fully address the positive autocorrelation. Comparison of the p-values, standard errors, and estimates between models with corAR1 and corSymm structures revealed no notable differences. We also attempted to model the order effect of the prophecies and subscales to mitigate this autocorrelation, but this approach did not lead to notable improvements. This unresolved autocorrelation may limit the validity of our results, and future studies should explore alternative approaches or additional data collection to address this issue.
    
    \item \textbf{Normality of Random Effects}: Q-Q plots of random effects were used to assess normality. The points closely followed the reference line, indicating overall normality, with slight deviations at the tails suggesting minor departures. However, these deviations were not significant enough to challenge the model assumptions.

    \item \textbf{Random Effects Evaluation:} The variance components and Intraclass Correlation Coefficient (ICC) were analyzed to evaluate the random effects structure. The ICC values (Adjusted ICC: 0.939, Unadjusted ICC: 0.598) indicated that a substantial portion of the variance was due to random effects. To further assess their contribution to overall model variability, we examined the variance components in the mixed effects model, which included random intercepts and slopes for \textit{prophecy\_source} at the \textit{qualtrics\_code} level. The estimated variance components are presented in Table \ref{table:random_effects_correlations}. 
\end{itemize}

\begin{table}[]
\centering
\caption{Random Effects Variance, Standard Deviation, and Correlations}
\label{table:random_effects_correlations}
\begin{tabular}{@{}p{5.5cm}>{\centering\arraybackslash}p{2.2cm}>{\centering\arraybackslash}p{2.2cm}>{\centering\arraybackslash}p{2.2cm}>{\centering\arraybackslash}p{2.2cm}@{}}
\toprule
\textbf{Random Effect} & \textbf{Variance} & \textbf{SD} & \textbf{Corr w Intercept} & \textbf{Corr w Astrology} \\ \midrule
Intercept                   & 1.560 & 1.249 & -      & -     \\ \midrule
prophecy\_sourceAstrology   & 1.211 & 1.100 & -0.260 & -     \\ \midrule
prophecy\_sourcePersonality & 1.002 & 1.001 & -0.221 & 0.293 \\ \midrule
Residual                    & 0.121 & 0.348 & -      & -     \\ \bottomrule
\end{tabular}
\end{table}

The results of the model diagnostics and the full results table of the mixed effects model are provided in Appendix \ref{appendix:results1}. The code for analysis is in Appendix \ref{appendix:code}, and the full code for implementation is available on GitHub (\url{https://github.com/mitmedialab/ai-superstition}).

\section{Results}

\subsubsection{People who are more likely to believe in astrology and personality-based predictions are more likely to believe in AI predictions.}

To test the first hypothesis (H1), a multiple linear regression model was applied to the data in wide format, with believability score for AI predictions as the outcome variable and believability scores for astrology- and personality-based predictions as predictor variables, along with control variables (cognitive style, paranormal beliefs, gullibility, AI attitude/trust in AI, big five personality, familiarity with sources, interest in topic, age, gender, education level). The scores across the four subscales (perceived validity, reliability, usefulness, and personalization) were averaged as overall "believability" scores to provide a straightforward test of the hypothesis. 

\begin{figure*}[h]
    \centering
    \begin{minipage}{0.5\textwidth}
        \centering
        \includegraphics[width=0.96\textwidth]{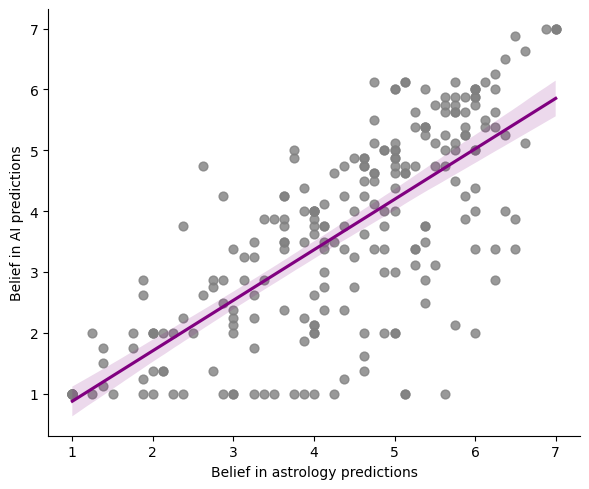} 
    \end{minipage}\hfill
    \begin{minipage}{0.5\textwidth}
        \centering
        \includegraphics[width=0.96\textwidth]{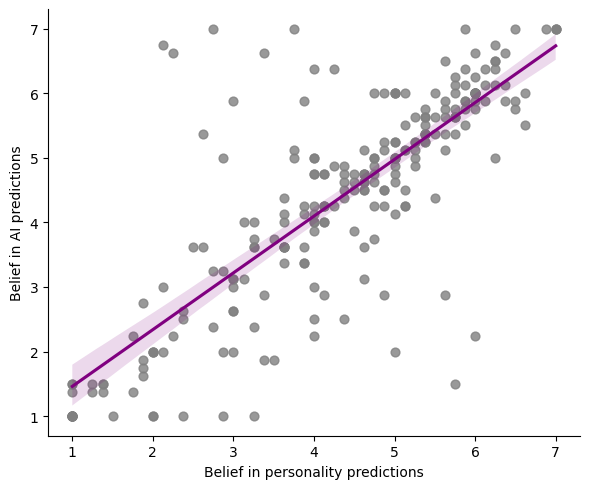} 
    \end{minipage}
    
    \caption{Scatterplots showing the relationship between belief in AI predictions and astrology predictions (left) and belief in AI predictions and personality predictions (right)}
    \begin{minipage}[c]{\textwidth}
        \subcaption*{The scatterplots show individual scores, with a linear regression line indicating a positive correlation. The shaded area represents the 95\% confidence interval for the regression line.}
    \end{minipage}
    
\label{fig:scatterplots}
\end{figure*}

The multiple linear regression analysis explained a significant proportion of the variance in the AI overall score (\(R^2 = 0.7606\), Adjusted \(R^2 = 0.7337\), \(F(24, 213) = 28.2\), \(p < 0.001\)). Significant predictors included the zodiac overall score (Estimate = 0.3119, \(p < 0.001\)) and the personality overall score (Estimate = 0.4585, \(p < 0.001\)), supporting the hypothesis that belief in astrology and personality-based predictions is positively associated with belief in AI predictions. This association can be visually observed in the scatterplots in Figure \ref{fig:scatterplots}.

% Additional significant predictors for belief in AI predictions included AI Attitude/Trust in AI (Estimate = 0.0168, \(p = 0.022\)), education level (Master's degree; Estimate = -0.5940, \(p = 0.029\)), and being in the rational prophecy group (Estimate = 0.2404, \(p = 0.032\)). The significance of these additional factors were further explored in the subsequent mixed-effects model analysis.

\subsubsection{People generally find fictitious AI predictions about their personal behavior convincing.}
% \subsection{Main effects of mixed effects model.}

To examine the influence of moderating factors including cognitive style (H2), paranormal beliefs, trust in AI, personality traits, demographic factors, etc. (H3), a mixed-effects model was fitted to the data in a nested long format, with prediction source (AI, astrology, personality) and believability subscale (perceived validity, reliability, usefulness, and personalization) as categorical predictors, and subscale score as the outcome variable. This approach allowed examination of the main effects and moderating effects of psychological and cognitive factors on belief in AI predictions. 

The model included fixed effects for subscale, prophecy source, prophecy group, cognitive style, paranormal beliefs, gullibility, AI attitude/trust in AI, big five personality, familiarity with sources, interest in topic, age, gender, education level. Interaction terms were included to assess the combined effect of prophecy source and select moderating factors on subscale scores. The model also included random effects to account for individual variability, with the random intercepts capturing the baseline differences among subjects and the random slopes capturing the variability within each participant across different conditions. The inclusion of random intercepts and slopes improved the model fit. For more on our methodology, see Section \ref{sec:methods}.

% Random intercepts for each participant and random slopes for prophecy source to capture between-subject and within-group variability, respectively. 

Based on the mixed effects model analysis, the baseline subscale score, associated with "AI" prophecy source, "Validity" subscale, "Positive" prophecy group, "Female" gender, and "Bachelor" education level, was 5.10 (p < 0.001) on a 7-point Likert scale. The main effects of prophecy source on believability of AI predictions showed that on average, perceived validity of astrology-based predictions was 0.66 points lower compared to AI predictions (95\% CI [-0.93, -0.39], p < 0.001). The difference was not statistically significant for personality-based predictions (0.07, 95\% CI [-0.18, 0.32], p = 0.593). These results suggest that people generally found fictitious predictions from AI, astrology, and personality convincing.

\begin{figure*}[h]
    \centering
    \includegraphics[width=0.96\textwidth]{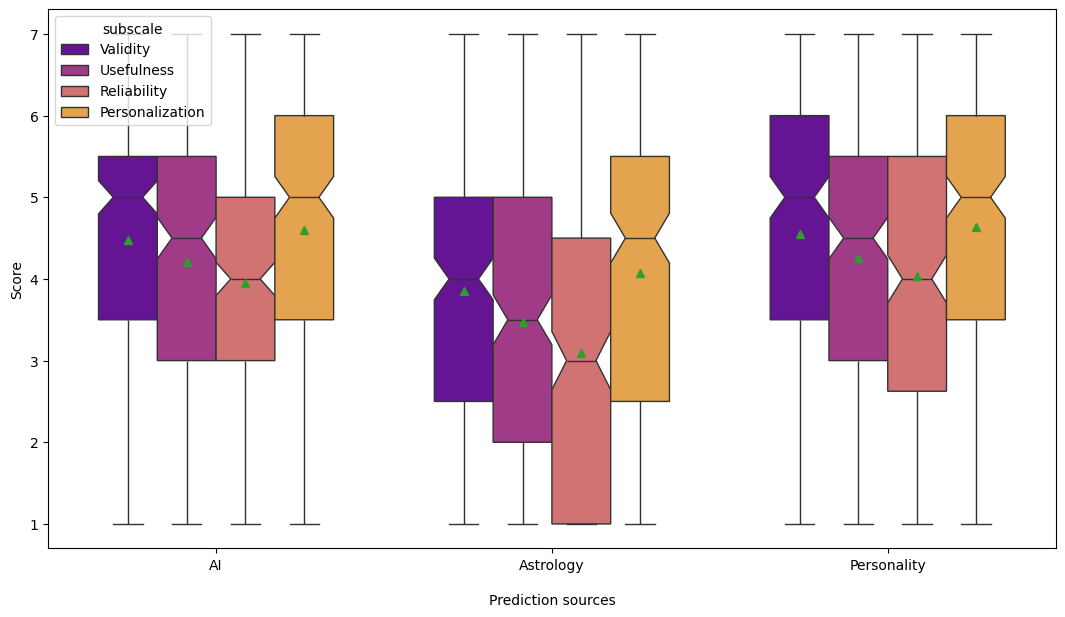} 
    \caption{Boxplot showing the distribution of subscale scores by prophecy source and subscale}
    \begin{minipage}[c]{\textwidth}
        \subcaption*{The boxes represent the interquartile range, with the means indicated in green triangles and notches representing the 95\% confidence intervals for the medians.}
    \end{minipage}
    \label{fig:boxplot}
\end{figure*}

There were significant main effects of subscales on the perception of AI predictions. On average, perceived personalization was rated 0.23 point higher than perceived validity (95\% CI [0.08, 0.38], p = 0.002). On the other hand, perceived reliability was rated 0.91 points lower than perceived validity (95\% CI [-1.13, -0.68], p < 0.001), and perceived usefulness was rated 0.63 points lower than perceived validity (95\% CI [-0.83, -0.44], p < 0.001). This relationship between the subscales by prediction source can be visualized in Figure \ref{fig:boxplot}. The interaction between subscales and prophecy source showed that the main effects of subscales were largely consistent across AI, astrology, and personality, with some subscales being more pronounced for astrology. Perceived personalization was further increased by 0.23 points (95\% CI [0.02, 0.44], p = 0.035) and perceived reliability further decreased by 0.31 (95\% CI [-0.62, -0.00], p = 0.048) for astrology-based predictions, compared to the AI baseline. Other interaction terms were not significant, suggesting that the effects are not significantly different from the main effects.

Furthermore, we found that compared to the baseline “Positive” prediction group, the “Negative” prediction group was associated with a decrease in perceived validity by 1.19 points (95\% CI [-1.53, -0.85], p < 0.001). We found no significant interactions between prophecy group and prophecy source, suggesting that this effect was consistent across AI, astrology, and personality. This suggests that while people generally found fictitious predictions believable, it was not the case for those who received negative predictions. 

\subsubsection{There is no evidence of correlation between belief in predictions and cognitive style.}

Based on our hypothesis (H2), we expected to see either a significant positive association between cognitive style and belief in AI predictions, significant negative interactions between cognitive style and other prediction sources (astrology, personality), or both. However, our results showed that the composite cognitive score, as measured as the composite of the performance-based Cognitive Reflection Test (CRT-2) \cite{thomson_investigating_2016} and the preference-based Need for Cognition (NFC-6) \cite{lins_de_holanda_coelho_very_2020}, did not significantly increase the perceived validity of AI predictions. The estimated increase in perceived validity was 0.13 points with one point increase in the composite cognitive score, but this effect was not statistically significant (95\% CI [-0.01, 0.26], p = 0.065). The composite cognitive score ranged from -4.29 to 2.61 (Mean = 0.0, SD = 1.44). 

However, we found some significant negative interactions between cognitive score and the subscales. Compared to perceived validity, higher composite cognitive score was associated with a decrease in perceived reliability (-0.11, 95\% CI [-0.20, -0.02], p = 0.021) and perceived usefulness (-0.12, 95\% CI [-0.20, -0.04], p = 0.004), making the main effect less positive. The interaction with perceived personalization was not statistically significant (-0.05, 95\% CI [-0.11, 0.01], p = 0.092).

The interactions between cognitive style and prediction sources showed that while there was some negative interaction for astrology (-0.10, 95\% CI [-0.21, 0.01], p = 0.079) and personality (-0.07, 95\% CI [-0.18, 0.03], p = 0.170), the effects were not statistically significant. This suggests a lack of evidence that cognitive style is an influential factor in how people perceive predictions based on AI, astrology, and personality. Three-way interactions between subscales, prophecy source, and composite cognitive score did not lead to any significant results.

\subsubsection{Higher paranormal beliefs increases perceived validity, reliability, usefulness, and personalization of AI predictions.}

One of the most interesting findings was that having paranormal beliefs was positively associated with belief in AI predictions. The results from the mixed effects model showed that the paranormal beliefs score, as measured by a shortened version of the Revised Paranormal Belief Scale (R-PBS) \cite{tobacyk_revised_2004}, significantly increased perceived validity of AI predictions. With each point increase in the paranormal beliefs scale, perceived validity of AI predictions increased by an average of 0.02 points (95\% CI [0.01, 0.03], p = 0.001) on a 7-point scale. The paranormal score, which was centered for the analysis, originally ranged from 15 to 95 (Mean = 45.6, SD = 20.08).

The interaction between paranormal beliefs and subscales showed that the effect was stronger for perceived reliability (0.01, 95\% CI [0.00, 0.01], p = 0.020) and perceived usefulness (0.01, 95\% CI [0.01, 0.02], p < 0.001), both being statistically significant. There was no significant interaction effect for perceived personalization (0.00, 95\% CI [-0.00, 0.01], p = 0.427), suggesting that the effect did not differ significantly from perceived validity.

Paranormal beliefs score was an even stronger predictor for belief in astrology-based predictions, perhaps less surprisingly given that the scale includes questions around astrology (see Section \ref{sec:scales}). Each standard deviation increase in paranormal beliefs led to an additional 0.01 point increase in perceived validity (95\% CI [0.01, 0.02], p = 0.001) compared to the AI baseline. Differences for personality-based predictions were not significant compared to the AI prediction baseline (-0.00, 95\% CI [-0.01, 0.00], p = 0.274). No significant interactions were observed between the subscales, prophecy source, and paranormal score.

\subsubsection{Positive AI attitudes increases belief in AI predictions, especially perceived reliability.}

Individuals with more positive attitudes towards AI found predictions based on AI more believable. One point increase in AIAS score \cite{grassini_development_2023} led to an increase of perceived validity of AI predictions by 0.04 points (95\% CI [0.01, 0.06], p = 0.001) on a 7-point scale. The AIAS score ranged from 4 to 40 (Mean = 25.79, SD = 8.68).

The main effect was augmented by an additional 0.03 points for perceived reliability (95\% CI [0.02, 0.05], p < 0.001), while the interactions were not significant for perceived personalization (-0.00, 95\% CI [-0.01, 0.01], p = 0.941) and usefulness (0.01, 95\% CI [-0.00, 0.03], p = 0.071). This suggests that positive attitudes toward AI/trust in AI is more closely related to perceived reliability among all subscales, which is aligned with the findings of studies that explore the relationship between trust/attitude and reliance in AI \cite{kahr_understanding_2024, klingbeil_trust_2024, chiou_trusting_2023, kuper_psychological_2023, lee_trust_2004}.

The interaction between AI attitude (AIAS) and prediction source showed that the positive impact of positive AI attitudes on perceived validity was mostly reversed for astrology-based predictions (-0.24, 95\% CI [-0.41, -0.08], p = 0.004), while the effect did not differ significantly for personality-based predictions (0.04, 95\% CI [-0.11, 0.19], p = 0.590). The interaction effects between subscales, prophecy source, and AI attitude score were not statistically significant.

\subsubsection{People with high conscientiousness are less likely to believe in predictions about personal behavior.}

Out of the big five personality traits (Extraversion, Openness, Agreeableness, Conscientiousness, Emotional stability), conscientiousness was negatively associated with perceived validity of predictions across all sources. With each point increase in the conscientiousness score (Mean = 5.29, SD = 1.32, Range = [1.0, 7.0], perceived validity was estimated to decrease by an average of 0.15 points (95\% CI [-0.30, -0.01], p = 0.032) on a 7-point scale. There were no significant interaction effects with subscales observed, suggesting the main effect was consistent across the four subscales.

While the other domains of the five-factor model were not found to have significant influence, there were some variation in interaction terms. Extraversion was associated with a 0.06 point increase in perceived usefulness (95\% CI [0.02, 0.10], p = 0.004), while Openness was associated with a 0.06 decrease in perceived personalization (95\% CI [-0.10, -0.01], p = 0.012).

\subsubsection{Level of interest in the topic of prediction increases perceived validity, reliability, usefulness, and personalization.}

While often overlooked in the context of studying trust in AI predictions, individuals' interest in the topic of prediction (in our case, personal investing behavior) was an influential factor in perceived validity, personalization, reliabiliity, and usefulness of predictions. With each point increase in the level of interest in the topic, perceived validity significantly increased on average by 0.27 points (95\% CI [0.11, 0.43], p = 0.001) on a 7-point scale. The interest in behavior scale ranged from 1 to 5 (Mean = 3.18, SD = 1.09).

The positive association was observed across other subscales. Compared to perceived validity, perceived personalization increased slightly less, reduced by 0.06 points (95\% CI [-0.11, -0.01], p = 0.018), while interaction was not statistically significant for perceived reliabilty (-0.07, 95\% CI [-0.14, 0.01], p = 0.068) and perceived usefulness (0.02, 95\% CI [-0.04, 0.09], p = 0.475), suggesting that the effects were comparable to perceived validity. The results suggest that when all else is held constant, the more people are interested in the topic, the more likely they will perceive fictitious predictions to be valid, reliable, useful, and personalized.

\subsubsection{There is no evidence that the level of familiarity with the prediction sources influences belief in predictions.}

On the other hand, familiarity with the prediction sources (AI, astrology, personality psychology), which measures the level of self-reported familiarity/prior knowledge in each source, did not have a significant effect on perceived validity (-0.02, 95\% CI [-0.11, 0.07], p = 0.650) across all sources. The familiarity scale ranged from 1 to 5 (Mean = 2.96, SD = 1.05).

The interaction between familiarity and subscales shows that perceived personalization increases slightly but significantly by 0.05 points (95\% CI [0.00, 0.11], p = 0.035), while differences were not statistically significant for perceived reliability (0.02, 95\% CI [-0.06, 0.10], p = 0.685) and perceived usefulness (0.01, 95\% CI [-0.06, 0.08], p = 0.781).

Unlike what the literature suggests about the familiarity effect, a cognitive phenomenon where people tend to prefer things they are familiar with \cite{hansen2009liking}, our results showed inconclusive evidence that familiarity in the prediction sources (AI, astrology, and personality) neither increased nor decreased belief in the respective predictions.

\subsubsection{Other results}

Gullibility was not found to be a significant predictor of perceived validity of AI predictions (-0.02, 95\% CI [-0.04, 0.01], p = 0.128). While the interaction  between gullibility and Personalization subscale was significant (0.01, 95\% CI [0.00, 0.02], p = 0.021), the results were inconclusive when compared to the main effects. Interactions between gullibility and prophecy source were not significant (Astrology: -0.00, 95\% CI [-0.02, 0.02], p = 0.977, Personality: -0.02, 95\% CI [-0.04, 0.00], p = 0.096), suggesting the effects were similarly insignificant across different prophecy sources. Gullibility scale ranged from 6 to 39 (Mean = 14.68, SD = 7.69). %Similarly, three-way interaction between gullibility, subscale, and prophecy source showed that Personalization subscale, and Personality prophecy source was significant (-0.13, 95\% CI [-0.24, -0.01], p = 0.029), but it did not lead to conclusive findings when compared with the main effect.

We found that older age was associated with a decrease in belief in predictions across all sources. One year increase in age (Mean = 41.04, SD = 12.38, Range = [19, 75]) was associated with a decrease of perceived validity of AI predictions by 0.02 points (95\% CI [-0.03, -0.00], p = 0.014). Interactions between age and subscales were not significant, nor were interactions between age and prophecy source. The non-significant interaction terms suggested that the effect was consistent across subscale and prophecy source.

For gender, we did not find significant main effects relative to the Female reference level. While male participants were associated with lower perceived validity scores for AI predictions than female participants, it was not statistically significant (-0.31, 95\% CI [-0.69, 0.06], p = 0.103). However, they tend to perceive them as more reliable than female participants (0.41, 95\% CI [0.14, 0.68], p = 0.003). Interactions with other subscales were not significant, suggesting they were similar to the perceived validity baseline. Interactions between gender and prophecy source were not significant.

There were no significant main effects for level of education compared to the Bachelor reference level. Interactions between education level and subscale showed some significant effects between High school and Personalization (-0.23, 95\% CI [-0.40, -0.06], p = 0.009), and Less than high school and Reliability (-0.89, 95\% CI [-1.56, -0.22], p = 0.009), but these results did not lead to conclusive results. The detailed results table can be found in Appendix \ref{appendix:results1}.

\section{Discussion}

Belief is subjective and context-dependent; it is also socially constructed through external stimuli such as media narratives, contextual factors, and internal disposition \cite{berger1966social, oliver2019media}. When looking at belief, trust, and reliance in AI predictions, it is important to consider not just the features of the AI system, but also the characteristics of the users and the broader context. The results from our study show that people who are more likely to believe in astrology- and personality-based predictions were more likely to believe in AI predictions. Moreover, we found that belief in AI predictions on personal behavior can be influenced by various psychological, cognitive, and contextual factors that also drive belief in astrology- or personality-based predictions. By breaking down the concept of believability into four subscales (perceived validity, reliability, usefulness, and personalization), we were able to observe interesting patterns and interactions with different subscales.

\subsubsection*{Cognitive style} While prior literature seemed to suggest that an analytic cognitive style would lead to more skepticism in predictions based on astrology and possibly personality \cite{bensley_critical_2023, pennycook_everyday_2015, pennycook_2012}, our results did not show significant evidence to support these claims. Moreover, cognitive style was not a significant predictor in belief in AI predictions. Thus, our hypothesis (H2) that people with more analytic cognitive style would be more likely to believe in AI predictions than predictions based on astrology and personality was not supported by our findings. This suggests that analytic cognitive style may not necessarily lead to rational skepticism of fictitious predictions, and perhaps there may be a missing link or mechanism between cognitive style and belief/trust in AI predictions, which could be a topic of future study.

% Pennycook et al. finds that “those with a more analytic thinking style are more skeptical about religious, paranormal, and conspiratorial concepts” (Pennycook et al., p. 425). Similarly, Bensley (2023) claims that critical thinking is linked to how people "reject false conspiracy theories, paranormal and pseudoscientific claims, psychological misconceptions, and other unsubstantiated claims" and tend to be more "scientifically skeptical and possess a more rational–analytic cognitive style" \cite{bensley_critical_2023}. 

\subsubsection*{Paranormal beliefs} Paranormal beliefs were found to be an influential predictor of belief in predictions across all sources, including AI predictions. While the association may seem unlikely at first, it can be interpreted within the context of the influence of popular AI narratives. The popular portrayal of AI as rational and objective creates the impression that people would also treat AI in a rational way, and thus more scientifically-inclined people would find AI predictions more believable. However, our results show that belief in AI predictions is more closely associated with paranormal beliefs than one's cognitive style. This result points to the existence of the phenomenon of "rational superstition" in AI that was described in the introduction. Moreover, while the effect of paranormal beliefs on belief in AI predictions was significant across all subscales, the effect was larger for perceived usefulness and reliability.  Social science research suggests that the rise in belief in astrology is an indicator of rising uncertainties and anxieties as a result of disintegration of community in the modern era \cite{bauer_belief_1997}. As such, people turn to pseudo-scientific approaches such as astrology for insights to help them navigate their lives. Our findings suggest that people may perceive AI predictions in a way that is similar to the way they perceive astrology-based predictions, which could help explain the results.

\subsubsection*{AI Attitude/Trust in AI}
Our results showed that people with positive attitudes towards AI found AI predictions more valid, reliable, useful, and personalized. Among the subscales, perceived reliability was found to be most closely related to attitude towards AI. This supports findings from prior literature that positive attitudes toward AI lead to higher reliance in AI predictions \cite{solberg_conceptual_2022, lee_trust_2004}.

\subsubsection*{Personality}
Our findings using the Big Five personality traits \cite{mccrae_validation_1987} show that conscientiousness had a negative influence on perception of validity, personalization, reliability, and usefulness, while other traits did not have significant effects. This contrasts previous findings by Riedl \cite{riedl_is_2022}, who found that agreeableness, openness, and extraversion positively influenced trust and high neuroticism negatively impacted trust, while results on conscientious were mixed. 

Conscientiousness is defined as “socially prescribed impulse control that facilitates task- and goal-directed behavior, such as thinking before acting, delaying gratification, following norms and rules and planning, organizing, and prioritizing tasks” \cite[p.~120]{john_paradigm_2008}. Prior studies have found a positive relationship between conscientiousness and cognitive ability \cite{corbeanu_conscientiousness_2023, meyer_conscientiousness_2024}. Contrary to our expectations that these two traits may mirror each other in direction, it was not the case---cognitive style did not have significant effect on believability in our study, while conscientiousness had a significant negative influence on perception of validity, personalization, reliability, and usefulness.

\subsubsection*{Interest in the topic of prediction} 
Our findings show that people's interest in the topic of prediction is a strong predictor when it comes to their belief in predictions, regardless of the source. While our results do not confirm any causal relationship, it could suggest that the more interested one is in a topic, the more exposure they may have to information about it, and the more likely they may be susceptible to cognitive biases that could influence their belief. Some biases that could explain this tendency are confirmation bias and belief bias. Confirmation bias is when individuals seek out and remember information that confirm their existing beliefs \cite{pohl_cognitive_2012}. Belief bias occurs when individuals judge the strength of an argument based on the believability of its conclusion rather than the logical structure of the argument itself \cite{evans_conflict_1983, klauer_belief_2000}. 

% When situated within the dual process theory, belief bias is known to be largely attributed to System 1 thinking (especially under time pressure \cite{evans_rapid_2005}), but it can also be intervened by System 2 thinking to construct mental models around existing beliefs in a way that can be biased, leading to overriding of logic \cite{de_neys_conflict_2017, evans_two_2009}. Known as the selective processing model (cite), this explains why people may engage logic when a statement contradicts their beliefs, but fail to engage logical thinking when a statement aligns with what they believe. This bias could explain why explanations of AI systems could lead to overreliance \cite{gonzalez_interaction_2021}.

This has implications in the context of AI-assisted decision-making in different fields; e.g. in marketing, companies may take advantage of people's interest in a certain topic and expose them to more targeted information (e.g. advertisements) that may create an impression of hyper-personalization and lead to more positive attitudes toward the brand \cite{chandra2022personalization}. Moreover, in the medical field, an individual may be more prone to believing AI-based predictions about their health if it confirms their prior beliefs. 

\subsubsection*{Familiarity/Prior knowledge}
Familiarity with the prediction source was not found to be a significant predictor of belief in predictions. Earlier, we discussed how people's positive AI attitudes and interest in the topic of the prediction were positively correlated with belief in the predictions, which may be explained by its likelihood to fuel certain cognitive biases such as belief bias. On the other hand, an individual's level of familiarity in AI, astrology, and personality psychology may not be necessarily indicative of their attitude towards the field. For instance, someone who is not acquainted with the details of latest AI developments may have either a utopian or dystopian view of AI's impact on society, as mentioned in the introduction. Similarly, just because an individual is very familiar with astrology does not inherently mean they are more likely to believe or adopt the predictions, as that may depend on other personal and contextual factors. This suggests two things: 1) simply knowing more about the prediction source and underlying technology does not predict whether one finds a prediction more or less believable, and 2) it does not make one better at calibrating one's trust, providing implications for the design of trustworthy AI systems.

% While familiarity/level of expertise in the prediction sources did not significantly influence believability, there could be a different angle of familiarity that could be worth further exploration--namely, familiarity of the predictions themselves. Does frequent exposure to an incorrect prediction increase belief? \cite{swire2017role} suggests that it might be the case; role of familiarity in influencing the perception of incorrect information 

\subsubsection*{Gullibility}
While prior literature suggested a positive relationship between gullibility and paranormal beliefs \cite{bensley_critical_2023, torres_validation_2023}, our findings showed inconclusive results on whether self-reported gullibility had a positive or negative effect on the belief of fictitious predictions based on AI, astrology, and personality. There may be a few potential explanations for this. One is that the order of the study potentially biased participants' answers. The survey was completed after participants had seen and evaluated the predictions, at which point their perception of the believability of the predictions may shadow their self-perception of gullibility---in other words, they did not want to contradict their decisions so quickly by admitting that they are more prone to being fooled. Another possible explanation is the limitation of a self-report nature of the gullibility scale. Some people may be hesitant to admit that they are gullible for different reasons. As such, future studies can further explore the relationship between gullibility and belief in AI predictions.

\subsubsection*{Demographic factors}
Our results found that older people are more skeptical of predictions about personal behavior, across AI, astrology, and personality sources. This is aligned with prior literature that seems to suggest that older age is associated with lower trust in AI \cite{chu_age-related_2023}, due to many reasons including bias, lack of learning avenues, and concerns about privacy \cite{shandilya_understanding_2024}. For gender, while the main effects were insignificant, interactions revealed that male participants were more likely to perceive AI predictions as reliable than female participants. This is supported by a prior study that found that being male with higher education and with a Western background were predictors of trust in AI among the general population \cite{yakar_people_2022}. Moreover, while prior work suggested a positive association between level of education and belief in AI \cite{chu_age-related_2023}, we did not find significant main effects for level of education, and the interaction terms did not lead to conclusive results. 

\subsection{General Discussion}

Our results highlight the irrational side of how humans perceive and believe in AI predictions by comparing the perception of fictitious predictions based on AI, astrology, and personality. Our analysis showed that belief in AI prediction about personal behavior may be more related to paranormal beliefs than cognitive style, and that positive attitudes toward AI and interest in the prediction topic enhance this belief, regardless of familiarity with the sources. 

These findings suggest that highly accurate and reliable performance is not a prerequisite for people to put high trust in predictions; models do not need to be completely trustworthy for users to trust them. This disconnect between perception and actual performance raises important discussions about how people perceive and engage with AI systems. 

\subsubsection*{1) Incorrect mental models of AI can override perception of validity.}

Belief in AI predictions is often rooted in mental models rather than actual performance. Hoff and Bashir's three-layer framework of trust in automation---dispositional, situational, and learned---illustrates that system performance contributes to learned trust through active engagement with the system \cite{hoff_trust_2015}. Our study shows that even without validating system performance, people rated fictitious AI predictions as highly valid, reliable, useful, and personalized. This placebo effect of AI has been observed by other researchers, where mere priming with AI descriptions alters perception and behavior \cite{villa_evaluating_2024, villa_placebo_2023, kloft_ai_2024, pataranutaporn2023influencing}.

This underscores the significant role mental models play in shaping trust in AI \cite{steyvers_three_2023, larghi_mentalistic_2024, nourani_anchoring_2021, bansal_beyond_2019, kulesza_too_2013, rouse_looking_1986}. Mental models, comprising beliefs and expectations about AI, are influenced by prior beliefs and ongoing interactions \cite{steyvers_three_2023, larghi_mentalistic_2024}. By framing the scenario in certain ways, including the task domain and the information provided about the AI system, we observed that people may perceive fictitious AI predictions to be accurate, personalized, and trustworthy \cite{steyvers_three_2023}.

Moreover, our results suggest that people might view AI predictions similarly to astrology, supporting the idea of "rational superstition" in AI and automation \cite{wilson_techno-optimism_2017}. We argue that this phenomenon potentially stems from the idealized portrayal of AI as enchanted yet rational and deterministic \cite{crawford_atlas_2021}, which can bias interactions and reinforce existing beliefs about AI.

The dual-process theory in psychology provides insight into how larger narratives can bias perceptions of AI. This theory describes two types of thinking: fast and intuitive (System 1) and slow and reflective (System 2) \cite{kahneman_thinking_2011, evans_two_2009}. Mental models, shaped by public narratives, may act as heuristics that trigger System 1 thinking, leading to speculation rather than informed reasoning \cite{dale_heuristics_2015}. Incomplete or incorrect mental models, shaped by speculative claims rather than evidence, can lead to overreliance or miscalibrated trust, affecting effective AI use \cite{logg_algorithm_2019, mahmud_decoding_2024, steyvers_three_2023, klingbeil_trust_2024}.

This highlights the need to demystify AI in popular narratives and promote awareness of both its capabilities and limitations to foster accurate mental models. This leads to the importance of AI literacy efforts and how narratives are shaped by providers, media, etc. As technology advances, users' mental models should evolve with a balanced level of skepticism to prevent overreliance or undue skepticism.

\subsubsection*{2) Mental models can influence and be influenced by cognitive biases.}

Cognitive biases help explain the irrational belief in AI predictions. Belief bias, where individuals evaluate statements based on prior beliefs rather than logical validity \cite{evans_conflict_1983, klauer_belief_2000}, is largely driven by System 1 thinking (especially under time pressure \cite{evans_rapid_2005}) but can also involve System 2, reinforcing mental models aligned with existing beliefs \cite{de_neys_conflict_2017, evans_two_2009}. This selective processing model explains why people may fail to engage logical thinking when a statement aligns with their beliefs, especially concerning self-related beliefs \cite{evans2000thinking}.

The Barnum effect, where individuals believe vague, general statements to be highly accurate for them personally, also plays a role \cite{forer_fallacy_1949}. This effect, studied in contexts like astrology and personality tests \cite{glick_fault_1989, johnson_barnum_1985}, can explain how people view generic AI-generated predictions as personalized and valid. Cognitive biases like belief bias and the Barnum effect shape mental models and influence AI reliance \cite{nourani_anchoring_2021}, while incorrect mental models can, in turn, induce these biases.

\subsubsection*{3) Design of AI systems must consider mental models and cognitive biases to support the development of an appropriate level of trust.}

AI systems should be designed to help users form accurate mental models. In contexts like mental health chatbots, incorrect mental models can lead to unrealistic expectations and greater disappointment among users. Thus, these systems should include mechanisms that promote an appropriate level of trust to avoid over or under-reliance, which should be accompanied by efforts to increase AI literacy \cite{lee_trust_2004}. Fortunately, mental models are dynamic and can evolve with more interaction \cite{steyvers_three_2023, gero_mental_2020}.

Awareness of the potential impact of cognitive biases on human-AI interaction is crucial to reduce the risk of negatively exploiting these human tendencies. AI systems that merely validate users' pre-existing beliefs may be favored but may not support better decision-making, particularly in critical areas like personal finance or health. Thus, we could think about how to design systems that can present information in a way that helps people think critically.

There are also important considerations for designing explainable AI systems. While explanations are intended to build trust, they can inadvertently increase over-trust, especially if users have inaccurate mental models \cite{danry2022deceptive}. Studies have shown that belief bias can lead to overreliance on AI explanations when they align with prior beliefs \cite{gonzalez_interaction_2021}. To mitigate this, explanations should be designed to engage users in critical thinking, perhaps by questioning the user or presenting examples that clarify the AI model’s reasoning \cite{danry2023_ask_me, nguyen2023visual}.

% Moreover, a study has found that narcissism was a strong predictor of belief in astrology \cite{andersson_even_2022}.  That can be interpreted as those with narcissistic tendencies are more likely to fall into cognitive biases such as belief bias and confirmation bias that confirm their existing beliefs about themselves.

\subsubsection*{4) Our understanding of human-AI interaction needs to include psychological and contextual factors.}

How people believe, trust, and rely on AI predictions for their decision-making should be seen in a larger context \cite{lee_trust_2004}. Contextual factors, including individual differences, task domain, information provided, media discourse, organizational and social environment, and more are all factors that could influence user perceptions and behavior in AI-assisted decision-making \cite{kordzadeh2022algorithmic, steyvers_three_2023}. 

Our findings emphasize that beyond system performance, user-related factors are critical in evaluating human-AI interaction. Psychological factors, including prior attitudes toward AI, paranormal beliefs, personality traits, and interest in the topic, significantly influence how people perceive AI predictions. These findings must be considered within the broader societal narratives, mental models, and cognitive biases that shape human-AI interactions.

\section{Limitations and Future Work}

While our study has presented valuable insights into the phenomenon of "irrational superstition" in AI predictions and cognitive biases, there are several important limitations that can be addressed in future work. First, our experiment was specifically designed to provide predictions on personal behavior. There are some limitations to the extent to which our results can be generalized to the general use of AI. Therefore, future studies could look at how people perceive AI predictions in different contexts and prediction topics, while also exploring the relationship with underlying psychological and cognitive factors at play. For instance, the effect can be tested within the context of AI-assisted decision-making in personal health or well-being, or in educational settings using an AI tutor. We believe that our study opens up more questions about understanding the intricate ways in which humans perceive, trust, and interact with AI systems. Second, like most survey designs, our survey-based approach to measuring psychological and cognitive traits may be affected by social desirability bias (the tendency to respond in a way that seems socially favorable) \cite{nederhof_methods_1985}, as well as common method biases \cite{podsakoff2003common}, and reliance on self-reported data. Therefore, we hope future studies can validate our findings using different methods and approaches. Lastly, while our study mainly focused on user-related factors, future studies could also look at other contextual factors such as social and organizational environments and their impact on user perception and trust.

\section{Conclusion}

Our study empirically investigated the phenomenon of “rational superstition” in AI by providing a side-by-side comparison of people's perceptions of fictitious predictions based on AI, astrology, and personality. Even without any validation of system
performance, people perceived fictitious AI predictions as highly believable, and their perceptions were positively associated with their perception of astrology- and personality-based predictions. Our findings showed that people's belief in AI predictions was positively associated with paranormal beliefs, positive attitude towards AI, interest in the topic of prediction, and negatively associated with conscientiousness and age. We also explored interactions with other prediction sources and the subscales including perceived validity, reliability, usefulness, and personalization. In our discussion, we highlighted the role of mental models and cognitive biases in shaping people's perceptions of AI predictions, and the importance of including psychological and contextual factors in studying human-AI interaction. 

%% file: chapter3.tex
% From mitthesis package
% Version: 1.06, 2024/07/09
% Documentation: https://ctan.org/pkg/mitthesis

\chapter{Personal Validation Effect in LLM Use: Positive AI Responses Bias Perceptions of Validity, Personalization, Reliability, and Usefulness of Fictitious Predictions}
\label{ch3_paper2}

\section{Introduction}
Danish author Hans Christian Andersen published "The Emperor's New Clothes" in 1837. This cautionary tale highlights human vulnerability to suggestion and flattery. In the story, two swindlers persuade an emperor they can weave fabric invisible to the foolish. Through false praise, and always telling the emperor what he wants to hear, thereby gaining his trust, the swindlers convince the emperor to parade around naked, believing he's wearing magnificent attire. This psychological manipulation bears a striking resemblance to the phenomenon of "sycophancy" observed in Large Language Models (LLMs), where the use of human feedback to fine-tune AI assistants inadvertently encourages responses that align with users' preferences rather than truthful information, creating generative echo-chambers \cite{sharma2023understanding, sharma_generative_2024}.

The deliberate fine-tuning of AI models to consistently provide positive affirmation, while ostensibly aimed at building user trust and retention, may inadvertently serve as a subtle form of manipulation. This concern is underscored by studies revealing that people's evaluations of AI models are often irrational and subjective \cite{pataranutaporn2023influencing, kloft_ai_2024, kosch_placebo_2022, villa_placebo_2023}. Just as the emperor's subjects perceived nonexistent clothing because of suggestion, users may interpret AI-generated content as uniquely insightful, trustworthy, or personalized, even when it lacks substance. 

In psychology, the "personal validation effect," also known as the Barnum or Forer effect, describes the tendency of individuals to accept vague and general personality descriptions as uniquely applicable to themselves \cite{forer1949fallacy}. Since its initial discovery in 1949, this phenomenon has been extensively studied and replicated, showing that people often rate supposedly personalized descriptions more favorably than general ones, even when the content is identical \cite{snyder1972further}. This effect has been explored in various contexts, including personality tests and astrology \cite{dickson1985barnum, johnson_barnum_1985, glick_fault_1989}.

One of the working principles of the personal validation effect is the Pollyanna principle, or more generally known as the positivity bias, which refers to the tendency to focus on the positive and to recall positive memories more accurately than negative ones \cite{matlin_pollyanna_2016}. This effect is particularly concerning in the context of AI, especially in critical fields such as healthcare, where an AI assistant's praise of a physician's choices could unduly influence medical decision-making. In educational settings, AI systems offering unwarranted positive feedback to students may hinder genuine learning and critical self-assessment, potentially fostering a false sense of achievement reminiscent of "toxic positivity" \cite{lecompte2022you}.

As sophisticated AI systems become increasingly integrated into our daily lives, understanding the psychological dynamics at play in human-AI interactions is crucial. This paper investigates the phenomenon of personal validation in the context of LLM use, highlighting how positive AI responses can significantly influence users' perceptions of the validity, personalization, reliability, and usefulness of predictions—even when those predictions are demonstrably false. While the personal validation effect is a general phenomenon, its implications are especially dangerous in the context of AI, where users may unknowingly place undue trust in AI-generated content, leading to misguided decisions and beliefs.

The impetus for this research stemmed from our previous study \cite{lee2024superintelligence} (see Chapter \ref{ch2_paper1}), which was inspired by the tech industry's quasi-religious narratives around AI development. This study explored how participants' trust in and responses to AI bear similarities to beliefs in astrology and pseudo-scientific personality predictions. The primary objective was to elucidate how people's reactions to AI predictions could be less rational and analytical than commonly assumed.

In the course of that research, a novel and intriguing research question was uncovered that extended beyond the initial investigation: \textit{How does the valence (positive or negative) of AI-generated predictions influence perceptions of validity, personalization, reliability, and usefulness?} While this was initially pre-registered as a secondary hypothesis under our broader investigation, we realized that the implications and scope of this inquiry marked a significant shift from our prior focus on AI and superstition. It became clear that the distinctive nature of this inquiry, coupled with its potential for wide-ranging implications for real-world contexts, necessitated a separate, dedicated study. 

As a result, this study aims to uncover the nuanced ways in which the valence, or positive or negative framing, of AI communications affects user perceptions and decision-making in real-world scenarios\footnote{This work is co-authored with Pat Pataranutaporn, Judith Amores, and Pattie Maes, 2024.}. Based on prior literature, we formulated the following hypothesis: Positive responses will significantly increase the perceived validity, personalization, reliability, and usefulness of fictitious predictions, compared to negative ones.

In addition to the pre-registered main hypothesis, we also explored the following research questions for exploratory analysis: To what extent do cognitive style, paranormal beliefs, gullibility, AI attitude/trust in AI, and demographic factors such as age and gender moderate the effect of the \textit{valence of predictions} on people's belief in those predictions?

To explore this hypothesis and research questions, we built upon the approach from the previous study \cite{lee2024superintelligence} to draw additional insights related to the Pollyanna principle or positivity bias in the context of human-LLM interaction and the psychological factors that influence belief in AI predictions. We employed an experimental design where participants are exposed to fictitious AI-generated predictions with either positive and negative valences to systematically examine these effects. This study contributes to the growing body of knowledge on human-AI interaction in three ways: 1) examining the extent to which the Barnum effect manifests in human-AI interactions, 2) quantifying the impact of positive framing on users' perceptions of AI-generated information, and 3) proposing strategies to mitigate the negative effects of the personal validation effect in AI design and implementation.

The results from our study (N=238) revealed that positive AI-generated predictions are perceived as significantly more valid (36\% increase), personalized (42\% increase), reliable (27\% increase), and useful (22\% increase) compared to negative predictions. These findings underscore the substantial impact of prediction valence on user perceptions, with implications for the design and deployment of AI systems across various applications.

The implications of this research reach areas of AI ethics, interaction design, and digital literacy. By understanding how personal validation and other cognitive biases operate in the context of human-LLM interaction, we can develop more responsible AI systems that balance the benefits of personalization with the need for accuracy and objectivity. Moreover, this work emphasizes the importance of educating users about the limitations and potential biases of AI systems, fostering a more critical and informed approach to human-AI interaction.  Furthermore, as we continue to navigate the evolving landscape of AI technology, this work highlights that it is crucial to consider not only the technical capabilities of these systems but also the psychological factors that shape our interactions with them.

\section{Related Works}
\subsection{Psychology of Human-AI Interaction}

The evolution of human-AI interaction has been shaped by the complex interplay between human psychology and our interactions with technology, beginning with the foundational work in Human-Computer Interaction (HCI) and advancing through the emergence of Human-Centered AI (HCAI), which emphasizes balancing human control with automation to enhance capabilities and promote ethical, responsible outcomes \cite{shneiderman2020human}. The rapid development and widespread adoption of technologies, particularly Large Language Models (LLMs), have dramatically transformed human-AI interaction, presenting new opportunities and challenges in designing conversational AI interfaces \cite{yang2024human, marri_conceptual_2023} and requiring more dynamic evaluation methods beyond static assessments \cite{ibrahim2024beyond}. In this context, studying the psychology of human-AI interaction is more important than ever, as it influences how people engage with AI systems, impacting user trust, decision-making, and the responsible use of AI systems in daily life. Understanding these psychological dynamics is essential for designing AI that is intuitive, ethical, and aligned with human needs.

A fundamental concept in this field is the Media Equation theory, proposed in 1996 \cite{lee2008media, reeves1996media}. This research revealed that people often treat computers and other media as if they were real social actors, unconsciously assigning them human characteristics. This phenomenon highlights the human tendency to form social relationships and anthropomorphize non-human entities, a concept also known as the "Eliza Effect," named after the ELIZA chatbot created in the 1960s \cite{weizenbaum1966eliza, natale2019if}. Anthropomorphism significantly influences human-AI interaction by shaping how users perceive and engage with AI systems, as these human-like characteristics can lead to stronger emotional connections and altered expectations.

Researchers have identified various observable factors that influence human-AI interaction, including appearance (especially personified appearance) \cite{knijnenburg2016inferring, zlotowski2016appearance, li2010cross, komatsu2008effect, pi2022influences, paetzel2016influence, koda1996agents}, voice characteristics like tone, pitch, and style \cite{pias_impact_2024, goodman_pitch_2021, seaborn2021voice, seaborn2021measuring, ehret2021prosody, kim2021designers, aylett2019siri, lewis2015investigating, hwang2022ai, cohn_expressiveness_2019, niculescu_making_2013}, dialogue \cite{volkel2021eliciting, kraus2020effects, feine2019taxonomy}, movement and behavior \cite{castro2016effects, van2014robot, feine2019taxonomy}, and emotional and social expression \cite{song2017expressing, paradeda2016facial, feine2019taxonomy, mallick_what_2024, moridis_affective_2012, breazeal2004designing}. Fine-tuning these factors allows AI system designers to enhance the impact of virtual agents or physical robots and strengthen their relationship with users. However, observable factors are only part of the equation. The internal processes of human psychology play a crucial role, as people construct "mental models" of the world and the entities they interact with \cite{johnson1983mental, norman2014some, bansal2019beyond, rutjes2019considerations, gero2020mental, kieras1984role, kulesza2012tell}. These models shape perceptions and expectations, influenced by cultural context, individual beliefs, and specific use cases. 

Recent studies have shown that altering a person's mental model of an AI system can affect their interaction with it. For example, one study found that participants primed to believe an AI had caring motives perceived it as more trustworthy, empathetic, and better-performing \cite{pataranutaporn2023influencing}. This research highlighted a feedback loop where the AI and the user reinforce the user's mental model, emphasizing the importance of how AI systems are introduced.

The role of expectations in human-AI interaction has also been explored. One study demonstrated that heightened AI expectations can improve performance through placebo effects \cite{kloft_ai_2024}. Participants with high expectations performed better when a sham AI system was present. Other studies have shown similar placebo effects in adaptive interfaces, where the belief in receiving adaptive AI support boosts expectations about task performance, positively correlating with actual performance and behavior \cite{kosch_placebo_2022, villa_placebo_2023}. These findings suggest that system descriptions can elicit placebo effects, potentially biasing the outcomes of user studies and influencing the future assessment of AI-based interfaces and technologies.

In addition, previous work \cite{sharma2023understanding} has investigated sycophantic model behavior in LLMs that try to win the users' favor, exerting greater influence on user perception and behavior. A recent paper \cite{usman2024persuasive} has studied the persuasive influence of AI ingratiation on consumer behavior, and found that such behavior leads to increased acceptance of AI-based product recommendations. The authors propose that the effect is caused by the potential bias of consumers viewing ingratiating comments as more accurate and objective.

As conversational agents powered by large language models become more human-like, users are starting to view them as companions rather than mere assistants. This shift in perception underscores the importance of understanding the psychological factors influencing human-AI interaction, as they play a crucial role in shaping user experiences, expectations, and the overall effectiveness of AI systems.

\subsection{Personal Validation Effect and Related Cognitive Biases}

The personal validation effect is a psychological phenomenon where individuals highly rate vague personality descriptions as personalized and accurate, despite these descriptions being general enough to apply to many people. Forer’s seminal study in 1949 provided early evidence of this effect, demonstrating that participants rated generic personality statements as highly accurate when told they were tailored specifically for them \cite{forer1949fallacy}. Subsequent research in the early 1970s further corroborated these findings, showing that students often accept generalized personality interpretations as accurate when they believe the feedback is personally tailored to them \cite{snyder1972further}. The role of favorability, referring to whether personality descriptions are positive or negative, has been studied, revealing that participants perceived favorable descriptions as more personally true and attributed greater skill to the source of these descriptions \cite{snyder1976effects}. The study also found that the valence of the descriptions was more influential on acceptance than the modality (whether presented orally or in writing).

Underpinning this preference toward positive descriptions is the Pollyanna principle, or the positivity bias, which has been studied in the context of self-report measures, such as quality of life \cite{cummins_maintaining_2002}, future expectations \cite{lench_automatic_2012}, and when describing other people \cite{bergsieker_stereotyping_2012}, where people generally gravitate towards positively-valenced responses. A study on social networks found that observers were more accurate in perceiving positive interpersonal ties than negative ones, with negative ties being more frequently overlooked or imagined, which aligns with the Pollyanna principle \cite{marineau2021positive}.

More recently, the influence of the personal validation effect have extended to various domains, including recommendation systems. A recent crowd-sourced experiment examined the perceived quality of personalized versus non-personalized movie recommendations \cite{suwanaposee2023specially}. Contrary to expectations, results showed numerically lower mean quality scores for personalized recommendations, albeit without statistical significance. This suggests that Barnum-like effects of personalization may have limited influence on perceived quality in recommender systems. 

In the context of large language models (LLMs), research has explored people's perceptions of LLM-generated advice and the role of advice style and user characteristics in shaping these perceptions. A recent study found that different LLM advice styles influenced user perceptions, with "skeptical" style rated as most unpredictable and "whimsical" style as least malicious \cite{wester2024exploring}. Moreover, the study also revealed that individuals who are more agreeable and with higher technological insecurity tend to like the advice more and find it more useful. While this study touches upon the style of LLM outputs, the effect of different valence (positive or negative) of the AI predictions on user perception have yet to be directly studied, which leads to the goal of this study. 

Furthermore, cognitive biases including belief bias and confirmation bias could potentially explain the mechanisms that power the effect of personal validation. Belief bias occurs when people evaluate an argument based on how believable its conclusion is, rather than on its logical validity \cite{evans_conflict_1983, klauer_belief_2000}. According to dual process theory in psychology \cite{kahneman_thinking_2011, evans_two_2009}, this bias is mainly associated with System 1 thinking, especially when people are under time pressure \cite{evans_rapid_2005}. However, System 2 thinking can also play a role by creating biased mental models based on pre-existing beliefs, sometimes overriding logical reasoning \cite{de_neys_conflict_2017, evans_two_2009}. The selective processing model \cite{evans2000thinking} suggests that people are more likely to apply logic to arguments that contradict their beliefs, but may neglect logical thinking for arguments that confirm their beliefs, particularly regarding self-related beliefs.

Moreover, confirmation bias is the tendency for individuals to seek out and recall information that supports their existing beliefs \cite{pohl_cognitive_2012}. Confirmation bias has been studied in the context of AI-assisted decision-making \cite{bashkirova_confirmation_2024, nazaretsky_confirmation_2021}. In human-AI interaction, this bias can cause people to perceive statements that align with their existing beliefs as more trustworthy and accept them \cite{bashkirova_confirmation_2024}. This selective attention and preference can intensify the personal validation effect, where people give undue credibility to AI outputs that match their expectations %, reinforcing their existing mental model of AI. This tendency can lead to a skewed understanding of AI capabilities and limitations.

These studies collectively highlight the complex interplay between personal validation effect, perception of personalized information, and the influence of positive versus negative information in various contexts, from personality assessments to social network perceptions and LLM-generated advice. Building on the rich literature in social and cognitive psychology, this study uniquely contributes by empirically testing how the valence of AI-generated predictions affects user perception and trust, specifically in human-LLM interactions. It offers novel insights into the potential biases at play in human-LLM interaction and suggests strategies to mitigate
the negative effects of the personal validation effect in designing AI systems.

\section{Methods}
\label{ch3-methods}

To address our research question, we utilize the study design, data collection, and analysis framework established in the previous study \cite{lee2024superintelligence} (See Section \ref{ch2-methods}). This section provides a brief overview of the overall approach and the modifications specifically made for this study to address the distinct aspects of the current research question.

\subsection{Participants}

Participants and data collection procedures were consistent with our previous study (See Section \ref{ch2-methods-participants}). Briefly, data were collected from 238 participants aged 18 and over from various socioeconomic backgrounds, recruited on Prolific. Participants were randomly assigned to either the "Positive" prediction group (N=119) or the "Negative" prediction group (N=119). See Table \ref{table:demographics} for an overview of participants' demographic information.

\subsection{Experiment Protocol}

The experimental protocol, including the initial astrology and personality assessments and simulated investment game, followed the procedures outlined in our previous study \cite{lee2024superintelligence} (see Section \ref{ch2-methods-experiment}). 

The study used a simulated investment game that was designed to elicit interactions comparable to those of modern robo-advising platforms \cite{dacunto_promises_2019}, but in a simplified way. Participants were given \$10,000 in virtual currency to allocate across three investment categories (high risk/return, medium risk/return, and low risk/return) over ten rounds (representing years). They could invest up to 100\% of their assets across the three categories at each round, with the goal of maximizing the total returns. The game simulated market fluctuations through a series of pre-determined "bull" (gains) and "bear" (losses) market scenarios, affecting potential investment outcomes. 

After the final round, participants were exposed to predictions about their future investment behavior. We employed a placebo approach that was proven effective in previous studies \cite{villa_placebo_2023, kosch_placebo_2022} where participants were informed that predictions were AI-generated, but were in fact pre-determined (either positive or negative), and randomly assigned. The predictions were either of positive valence (e.g., rational investor, higher returns) or negative valence (e.g., impulsive investor, lower returns), with participants randomly assigned to each condition. Each person received predictions from three sources (AI, astrology, personality-based) which were presented in a random order. Examples of these predictions are listed in Table \ref{table:predictions}.  Participants then evaluated each prediction for its perceived validity, personalization, reliability, and usefulness (see Section \ref{sec: believability_subscales}).

Following the experiment, participants were asked to respond to a series of questionnaires on participant characteristics (cognitive style, paranormal beliefs, gullibility, trust in AI/attitudes towards AI, personality, etc.) and demographic information (age, gender, education level). A summary of the scales used in the questionnaires is provided in Table \ref{table:measures}. Detailed questionnaires can be found in Appendix \ref{appendix_survey}.

\subsection{Analysis}

We employed an approach similar to the analysis outlined in our previous study, with some major changes. The variables included in this analysis were the same as those used in the previous study, with subscale scores as the dependent variable and predictors such as prophecy source, prophecy group, cognitive style (composite score), paranormal beliefs, AI attitude/trust in AI, gullibility, Big Five personality traits, interest in the prediction topic, familiarity with prediction sources, and demographic variables (age, gender, and education level). See Section \ref{ch2_methods_variables} for the description of key variables.

A mixed-effects model was fitted to the data in a nested long format with a three-level hierarchical structure. Participants were divided into two prediction groups (positive, negative), and each subject engaged in three types of predictions (AI, astrology, personality) and evaluated across four subscales (perceived validity, reliability, usefulness, and personalization). The outcome variable was the \textit{subscale\_score}, which represented the scores for each subscale on a 7-point Likert scale. The mixed effects model included fixed effects for subscale, prophecy source, prophecy group, cognitive style, paranormal beliefs, gullibility, AI attitude/trust in AI, big five personality, familiarity with the prediction sources, interest in the prediction topic, age, gender, education level. Interaction terms were included to assess the combined effect of prophecy group and select moderating factors on subscale scores. For the random effects, random intercepts and slopes were included to account for the variability between subjects across different conditions.

A few changes were made to the model specification from the previous study to focus on the impact of prophecy group based on the new hypotheses. Specifically, the focus shifted from \textit{prophecy\_source} to \textit{prophecy\_group} for interactions with the moderator variables, allowing for an examination of how different prediction groups influence believability, moderated by these variables. To ensure consistency and enable valid comparisons across models, the random effects structure, correlation, and weighting schemes were retained from the previous model, appropriately accounting for the data structure.

The mixed effects model was specified as follows (changes in the model specification are in bold). The fixed effects were defined as:

\begin{equation}
\begin{aligned}
\text{subscale\_score} \sim & \, \text{subscale} * (\text{prophecy\_group} * \text{prophecy\_source} \\ 
& + \textbf{prophecy\_group} * \text{composite\_score} + \textbf{prophecy\_group} * \text{paranormal\_score} \\
& + \textbf{prophecy\_group} * \text{aias\_score} + \textbf{prophecy\_group} * \text{gullibility\_score} \\
& + \text{big5\_extraversion} + \text{big5\_openness} + \text{big5\_agreeableness} + \text{big5\_conscientiousness} \\
& + \text{big5\_emotional\_stability} + \text{interest\_behavior} + \text{familiarity} \\
& + \textbf{prophecy\_group} * \text{Age} + \text{education} + \textbf{prophecy\_group} * \text{gender})
\end{aligned}
\end{equation}

For the random effects, we included random intercepts for each participant (\textit{qualtrics\_code}) to account for between-subject variability and random slopes for prophecy source to capture the variability in responses across participants to different conditions (\textit{prophecy\_source}):

\begin{equation}
\begin{aligned}
\text{random  } &= \quad \sim \text{prophecy\_source} \mid \text{qualtrics\_code}
\end{aligned}
\end{equation}

Additionally, the model incorporated a correlation structure and variance weights to further account for the hierarchical structure and heteroscedasticity in the data:

\begin{equation}
\begin{aligned}
\text{correlation} &= \text{corSymm(form = ~1 | qualtrics\_code/prophecy\_source)}, \\
\text{weights} &= \text{varIdent(form = ~1 | subscale * prophecy\_source * prophecy\_group)}
\end{aligned}
\end{equation}

Continuous predictors were centered to reduce multicollinearity and to stabilize coefficient estimates. The model was fit using the \textit{lme} function from the \textit{nlme} package in R, with settings to exclude missing values and optimize convergence. 

Several diagnostic tests were conducted to ensure the validity of the model. Linearity was confirmed through a residuals vs. fitted values plot, which showed no clear patterns, indicating that the linearity assumption was met. Homoscedasticity was supported by a random spread around zero in the residuals plot, although the multilevel nature of the outcome variable caused some diagonal patterns. The normality of residuals was assessed with a Q-Q plot, showing slight deviations at the extremes but overall consistency with normal distribution. Multicollinearity was addressed by standardizing the continuous predictor variables, reducing Variance Inflation Factor (VIF) values to acceptable levels (below 5). Independence of residuals was evaluated using the Durbin-Watson test, which revealed significant positive autocorrelation at the subscale level. An autoregressive correlation structure (corAR1) initially helped with the "Validity" subscale but wasn't fully effective for others. Switching to an unstructured symmetric correlation (corSymm) improved the model fit but didn't completely resolve the autocorrelation. This unresolved autocorrelation may pose a limitation in our analysis, suggesting the need for alternative methods or additional data in future studies. Normality of random effects was confirmed through a Q-Q plot, with minor deviations noted. The evaluation of random effects, including variance components and Intraclass Correlation Coefficient (ICC), demonstrated significant variability attributable to these effects. 

The results of the model diagnostics and the full results table of the analysis are provided in Appendix \ref{appendix:results2}. The code for analysis is in Appendix \ref{appendix:code}, and the full code for implementation is available on GitHub (\url{https://github.com/mitmedialab/ai-superstition}).

\section{Results}
\label{ch3_results}

The results from our experiment (N=238) showed that positive predictions are perceived as significantly more valid, personalized, reliable, and useful. 

\subsection{Descriptive statistics}
Table \ref{table:descriptive_stats} provides the mean, standard deviation, and the 95\% confidence interval for perceived validity, personalization, reliability, and usefulness across Positive and Negative prophecy groups. The Positive prophecy group had higher mean scores across all subscales compared to the Negative group. For perceived validity, the Negative prophecy group had a mean score of 3.70 (SD = 1.79, 95\% CI [3.51, 3.88]), compared to 4.89 (SD = 1.32, 95\% CI [4.75, 5.03]) in the Positive prophecy group. For perceived personalization, Negative prophecy group had a mean Personalization score of 3.70 (SD = 1.90, 95\% CI [3.50, 3.89]), while the Positive prophecy group had a higher mean score of 5.18 (SD = 1.32, 95\% CI [5.04, 5.32]). The mean score for perceived reliability in the Negative prophecy group was 3.32 (SD = 1.84, 95\% CI [3.13, 3.51]), compared to 4.07 (SD = 1.62, 95\% CI [3.90, 4.23]) in the Positive prophecy group. For perceived usefulness, the Negative prophecy group had a mean score of 3.61 (SD = 1.92, 95\% CI [3.41, 3.81]), whereas the Positive prophecy group had a mean score of 4.34 (SD = 1.56, 95\% CI [4.18, 4.51]). 

\begin{table}[]
\centering
\caption{Descriptive statistics by prediction valence (prophecy group) and subscale}
\label{table:descriptive_stats}
\begin{tabular}{@{}ccccc@{}}
\toprule
\textbf{prophecy\_group} & \textbf{subscale} & \textbf{Mean} & \textbf{SD} & \textbf{95\% CI} \\ \midrule
\multirow{4}{*}{Positive (N=119)} & Validity        & 4.89 & 1.32 & {[}4.75, 5.03{]} \\  
                                  & Personalization & 5.18 & 1.32 & {[}5.04, 5.32{]} \\ 
                                  & Reliability     & 4.07 & 1.62 & {[}3.90, 4.23{]} \\ 
                                  & Usefulness      & 4.34 & 1.56 & {[}4.18, 4.51{]} \\ \midrule
\multirow{4}{*}{Negative (N=119)} & Validity        & 3.70 & 1.79 & {[}3.51, 3.88{]} \\  
                                  & Personalization & 3.70 & 1.90 & {[}3.50, 3.89{]} \\  
                                  & Reliability     & 3.32 & 1.84 & {[}3.13, 3.51{]} \\ 
                                  & Usefulness      & 3.61 & 1.92 & {[}3.41, 3.81{]} \\ \bottomrule
\end{tabular}
\end{table}

\subsection{Positive predictions are perceived as significantly more valid, personalized, reliable, and useful than negative predictions.}
\label{subsec:results_main}

The mixed effects model analysis showed that positive AI predictions generally elicited significantly more favorable responses compared to negative ones, consistent with the Pollyanna principle. The intercept was 5.20 (95\% CI [4.81, 5.59], p < 0.001) on a 7-point Likert scale, and this was associated with "AI" prophecy source, "Validity" subscale, "Positive" prediction group, "Female" gender, and "Bachelor" education level, which served as the reference level for subsequent interpretation. The main effects of the prediction valence on perceived validity score showed that compared to the baseline Positive prediction group, the Negative prediction group was associated with a decrease in perceived validity by 1.37 points (95\% CI [-1.83, -0.91], p < 0.001).

Interactions between the prediction groups and subscales revealed significant results that suggested differential impact of prediction valence across subscales. Compared to perceived validity baseline, perceived personalization further decreased by 0.25, which was significant (95\% CI [-0.43, -0.06], p = 0.008). On the other hand, perceived reliability significantly increased by 0.46, partially reversing the negative main effect (95\% CI [0.17, 0.75], p = 0.002). Similarly, perceived usefulness significantly increased by 0.56 (95\% CI [0.31, 0.81], p < 0.001). This suggested that while all subscales were impacted by the valence of predictions, the influence was stronger for perceived personalization and less strong for perceived reliability and usefulness compared to the baseline.

There were significant main effects of subscales on the perception of positive AI predictions. On average, perceived personalization was rated 0.30 point higher than perceived validity (95\% CI [0.15, 0.44], p < 0.001). On the other hand, perceived reliability was rated 0.90 points lower than perceived validity (95\% CI [-1.12, -0.68], p < 0.001), and perceived usefulness was rated 0.67 points lower than perceived validity (95\% CI [-0.86, -0.48], p < 0.001). The detailed results are provided in the supplementary materials.

\begin{figure}[h]
    \centering
    \caption{Comparison of estimates by prediction valence for each subscale}
    \begin{minipage}[c]{\textwidth}
        \subcaption*{Standard errors are represented as error bars, and subscale score range is from 1 to 7. All differences were significant (p < 0.01**).}
    \end{minipage}
    \includegraphics[width=13cm]{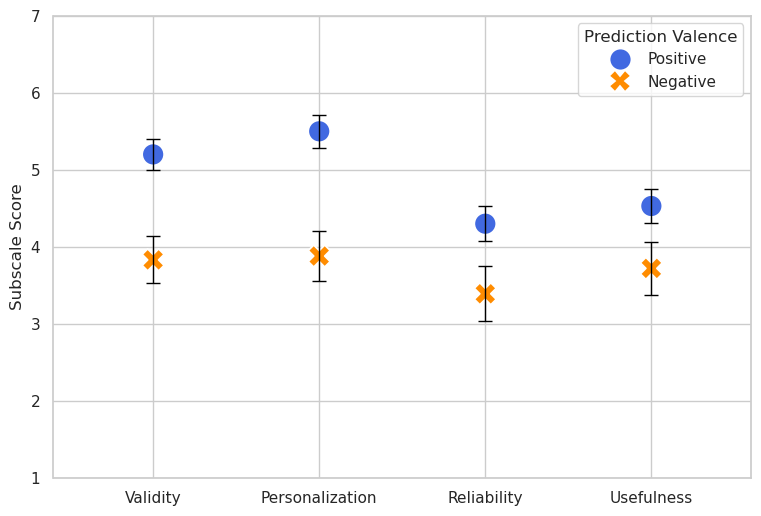}
    \label{fig:pointplot}
\end{figure}

To get a sense of the magnitude of the impact of prediction valence across subscales, the percentage differences in the estimated values of subscale scores were calculated based on the coefficients from the mixed effects model results. Positive predictions increased perceived validity by 36\% and perceived personalization by 42\%. Perceived reliability increased by 27\% and perceived usefulness was increased by 22\% compared to negative predictions. This comparison is visualized in Figure \ref{fig:pointplot} and Table \ref{table:comparison}.

\begin{table}[]
\centering
\caption{Percentage differences of estimates for each subscale by prediction valence}
\label{table:comparison}
\begin{tabular}{@{}lccc@{}}
\toprule
\textbf{}                & \textbf{Negative} & \textbf{Positive} & \textbf{$\Delta$ Percentage} \\ \midrule
\textbf{Validity}        & 3.83              & 5.20              & 36\%                       \\ 
\textbf{Personalization} & 3.88              & 5.50              & 42\%                       \\ 
\textbf{Reliability}     & 3.39              & 4.30              & 27\%                       \\ 
\textbf{Usefulness}      & 3.72              & 4.53              & 22\%                       \\ \bottomrule
\end{tabular}
\end{table}

Moreover, the main effects of prophecy source on believability of predictions showed that on average, perceived validity of astrology-based predictions was 0.66 points lower compared to AI predictions (95\% CI [-0.89	-0.43], p < 0.001). The difference was not statistically significant for personality-based predictions (0.06, 95\% CI [-0.14, 0.27], p = 0.549). The interaction between Astrology prediction source and Negative prediction group was not significant (0.09, 95\% CI [-0.23, 0.41], p = 0.598) as well as the interaction between Personality prediction source and Negative prediction group (0.02, 95\% CI [-0.27, 0.31], p = 0.881), which suggested that the negative effect of receiving the negative predictions was comparable across different prediction sources.

\subsection{Analysis of moderating factors}
Although not initially pre-registered, the inclusion of psychological factors such as paranormal belief provided valuable insights. We observed that these factors moderated the effect of AI prediction valence on believability, highlighting a complex interplay between individual predispositions and the nature of AI communications.

\subsubsection*{Cognitive style} 

The composite cognitive score was measured with a combined score of the cognitive reflection test (CRT-2) \cite{thomson_investigating_2016} and need for cognition (NCS-6) \cite{lins_de_holanda_coelho_very_2020}, and this continuous variable ranged from -4.29 to 2.61 (Mean = 0.0, SD = 1.44). Cognitive style was not found to be a significant moderator of the impact of AI prediction valence on perceived validity (Estimate = -0.16, 95\% CI [-0.39, 0.08], p = 0.196). 

However, interactions across subscales revealed that cognitive style significantly moderated the impact of negative prediction valence on perceived reliability. The previous section discussed that perceived reliability was significantly reduced with negative predictions, but this was less compared to the perceived validity baseline, as seen in the slight increase in the estimate (Estimate = 0.46, 95\% CI [0.17, 0.75], p = 0.002). However, this effect was slightly lessened with higher composite scores, which represents a more analytic cognitive style (Estimate = 0.11, 95\% CI [0.00, 0.22], p = 0.046). This suggests that those with more analytic cognitive style are less influenced by the difference in prediction valence, thus less prone to the positivity bias.

\subsubsection*{Paranormal beliefs}
We explored whether paranormal beliefs moderated the relationship between prediction valence and believability. Paranormal beliefs, as measured by a revised version of R-PBS \cite{tobacyk_revised_2004}, were considered as a continuous moderating variable, with scores ranging from 15 to 95 (Mean = 45.6, SD = 20.08). 

The interaction between prediction valence and paranormal beliefs was not significant (Estimate = 0.01, 95\% CI [-0.01, 0.02], p = 0.539), indicating that the effect of prediction valence on perceived validity does not depend on the level of paranormal beliefs. 

Looking at interactions across subscales, paranormal beliefs significantly moderated the effect of prediction valence on perceived personalization. While negative predictions generally led to a larger decrease in perceived personalization compared to perceived validity baseline (Estimate = -0.25, 95\% CI [-0.43, -0.06], p = 0.008), this effect was slightly attenuated by higher paranormal beliefs score (Estimate = 0.01, 95\% CI [0.00, 0.01], p = 0.022). On the other hand, the effect of negative valence on perceived reliability (Estimate = 0.46, 95\% CI [0.17, 0.75], p = 0.002) was slightly exacerbated by higher paranormal beliefs score (Estimate = -0.01, 95\% CI [-0.02, -0.00], p = 0.028). This indicates that the level of paranormal beliefs slightly moderates the effect of prediction valence on perceived personalization and reliability.

\subsubsection*{Gullibility}
The moderating effect of gullibility on the impact of prediction valence on believability was assessed. Gullibility, as measured by a self-report gullibility scale \cite{teunisse_i_2020}, was a continuous moderating variable, with scores ranging from 6 to 39 (Mean = 14.68, SD = 7.69). 

The significant main effect of negative predictions on perceived validity (Estimate = -1.37, 95\% CI [-1.83, -0.91], p < 0.001) as described in Section \ref{subsec:results_main} was slightly reduced for individuals with higher gullibility score, as seen in the significant interaction between Negative prophecy group and gullibility score (Estimate = 0.06, 95\% CI [0.01, 0.10], p = 0.010), suggesting a moderating role of gullibility. However, this effect did not vary significantly across subscales. This implies that the more gullible an individual is, the less likely that the valence of predictions will influence their belief in those predictions, which may be counter-intuitive. 

Additionally, contrary to the findings from the previous study \cite{lee2024superintelligence}, we found the negative relationship between gullibility and perceived validity to be significant (Estimate = -0.05, 95\% CI [-0.09, -0.02], p = 0.001). However, this effect was reversed for those who received negative predictions (Estimate = 0.06, 95\% CI [0.01, 0.10], p = 0.010). These seemingly contradictory results for gullibility are discussed in Section \ref{subsec:moderating_factors}.

\subsubsection*{AI attitude/Trust in AI}
We investigated whether AI attitude/trust in AI influenced the relationship between prediction valence and believability. Attitude toward AI was measured using AIAS-4 \cite{grassini_development_2023}, and this continuous moderating variable ranged from 4 to 40 (Mean = 25.79, SD = 8.68), where higher scores indicated more positive attitudes toward AI.

While the interaction between AIAS score and Negative prophecy group approached significance (Estimate = 0.04, 95\% CI [-0.00, 0.07], p = 0.058), it was not statistically significant. This indicates that the effect of prediction valence on perceived validity does not depends on the level of AI attitude/trust. 

Interactions with other subscales revealed significant moderating effects of AI attitude on the effect of negative predictions on perceived reliability and usefulness. Higher AIAS scores, indicating more positive attitudes toward AI, led to a slight decrease in the impact of negative predictions on perceived reliability (Estimate = -0.02, 95\% CI [-0.04, -0.00], p = 0.021) as well as perceived usefulness (Estimate = -0.02, 95\% CI [-0.03, -0.00], p = 0.034). This significant interaction suggests that for perceived reliability and usefulness, higher AI attitude actually amplifies the negative effect of negative predictions.

\subsubsection*{Age}
Interaction between age and the negative prophecy group was not significant (Estimate = 0.01, 95\% CI [-0.02, 0.04], p = 0.519), suggesting that age is not a significant moderator on the impact of prediction valence on perceived validity. Age is a continuous variable and ranged from 19 to 75 (Mean = 41.04, SD = 12.38). 

While the impact of negative predictions on perceived reliability was significant but less pronounced than the impact on perceived validity or personalization (Estimate = 0.46, 95\% CI [0.17, 0.75], p = 0.002), older age further reduced the impact of negative predictions on perceived reliability (Estimate = 0.02, 95\% CI [0.00, 0.03], p = 0.010). This suggests that older age may lead to decreased influence of the positivity bias.

\subsubsection*{Gender}
Gender was not a significant moderator of the impact of prediction valence on believability of AI predictions. Interactions between the Negative prophecy group and gender was not significant (Male: Estimate = 0.35, 95\% CI [-0.35, 1.05], p = 0.322; Other: Estimate = 0.83, 95\% CI [-0.97, 2.64], p = 0.365) compared to the female baseline, suggesting that the impact of prediction valence on perceived validity is consistent across gender. Similarly, there were no significant interaction effects across other subscales. The detailed results table can be found in Appendix \ref{appendix:results2}

\section{Discussion}
Our results demonstrate that positive AI-generated predictions are perceived as significantly more valid, personalized, reliable, and useful compared to negative predictions. This finding aligns with the personal validation effect, also known as the Barnum effect, where individuals tend to accept vague and general personality descriptions as uniquely applicable to themselves. Moreover, the findings confirm the Pollyanna principle, or positivity bias, that people find positive predictions more believable than negative ones. The significant main effect of prediction group and the non-significant interaction with astrology and personality prediction sources together support the hypothesis that people generally tend to believe in positive predictions more, irrespective of the source of the predictions.

\subsubsection*{Perceived validity}
The substantial difference in perceived validity between positive and negative predictions suggests that participants were more likely to believe in the accuracy of favorable forecasts. This tendency may be attributed to a combination of cognitive biases, including confirmation bias and belief bias, as described in \cite{lee2024superintelligence}. Belief bias refers to the tendency to evaluate statements based on their believability and alignment with their existing beliefs rather than their logical validity \cite{evans_conflict_1983, klauer_belief_2000}. This can lead to people to accept AI predictions as believable even when they are fictitious or misleading. Moreover, in terms of predictions about their personal behavior, people generally prefer information that confirms their existing beliefs or desires about themselves, and positive predictions likely align more closely with individuals' hopes and self-perceptions.

\subsubsection*{Perceived personalization}
While the positivity bias can be observed across all subscales, the marked increase in perceived personalization for positive predictions is particularly noteworthy. The perception of personalization is stronger than other subscales when predictions are positive, but similar to others when predictions are negative, as seen in Figure \ref{fig:pointplot}. This finding validates that the positivity bias is an influential working principle of the personal validation effect. 

This effect may also be explained by the self-serving bias \cite{shepperd2008exploring, mezulis_is_2004}, the tendency to attribute our successes to internal, personal factors, and our failures to external, situational factors. When presented with favorable predictions, participants may have been more inclined to view them as tailored to their unique qualities or circumstances.

\subsubsection*{Perceived reliability and usefulness}
In addition, the higher ratings for reliability and usefulness of positive predictions further underscore the influence of personal validation. Participants may have judged favorable predictions as more dependable and actionable, possibly due to the illusion of control \cite{langer1975illusion}. Positive forecasts might be seen as more reliable because they align with desired outcomes, and more useful because they support optimistic future planning.

\subsubsection{Factors that moderate the personal validation effect}
\label{subsec:moderating_factors}

Our analysis revealed several factors that moderate the impact of the personal validation effect on the perception of AI predictions. These findings provide nuanced insights into how individual differences and psychological factors influence the susceptibility to positive bias in AI-generated predictions.

Interestingly, we found that cognitive style significantly moderated the effect of prediction valence on perceived reliability, but not on other subscales. This indicates that individuals with a more analytical cognitive style are less likely to have changed perceptions of reliability when the predictions are negative versus positive, possibly making them less susceptible to the personal validation effect. While cognitive style was not found to a significant predictor of believability of predictions based on the previous study \cite{lee2024superintelligence}, our results reveal that it could moderate the level of influence of the positivity bias and the personal validation effect.

While paranormal beliefs did not significantly moderate the overall effect of prediction valence on believability, they differentially moderated the impact of negative predictions across subscales, particularly for perceived personalization and reliability. Paranormal beliefs were found to mitigate the negative impact of negative predictions for perceived personalization, suggesting that those who hold stronger paranormal beliefs were less likely to perceive personalized predictions negatively when faced with negative predictions. On the other hand, paranormal beliefs amplified the negative impact of negative predictions, meaning that individuals with higher paranormal beliefs saw negative predictions as more damaging to the reliability of the information. This highlights the role of individual differences on the effect of personal validation.

Gullibility emerged as a significant moderator of the effect of prediction valence on believability. Those who reported to be more gullible were less influenced by the different valence of predictions, suggesting they are more resistant to positivity bias. While the results seem counter-intuitive, this contradiction may stem from the self-report nature of the gullibility scale as well as the fact that it was administered after participants had rated their belief in the predictions, making them biased in their self-perception of gullibility. Moreover, the negative impact of gullibility on perceived believability was reversed when predictions were negative. This could suggest that people were more likely to be biased in their responses to the self-reported gullibility scale when predictions were positive, but not when predictions were negative, as they were less likely to admit that their trust in the positive predictions were unfounded.

Attitudes toward AI played a role in moderating the effect of prediction valence on perceived reliability and usefulness. Participants with more positive attitudes toward AI showed a stronger effect of prediction valence on these dimensions. This suggests that individuals who are more favorable towards AI technology may be more susceptible to the positivity bias of the personal validation effect, potentially due to higher trust or expectations in AI capabilities. 

Age was found to moderate the impact of prediction valence on perceived reliability, with older participants showing a reduced effect of negative predictions. This finding implies that older individuals might be less swayed by the positivity bias in AI interactions, possibly due to greater life experience or a more critical perspective developed over time.

Notably, gender did not significantly moderate the impact of prediction valence on the believability of AI predictions. This suggests that the personal validation effect and its influence on perceptions of AI-generated content may be relatively consistent across gender identities.

The exploratory findings on the role of moderating factors highlight the complex interplay between individual characteristics and the personal validation effect in AI interactions. Understanding these nuances is crucial for developing more effective and ethical AI systems that can adapt to individual differences and mitigate potential biases in user perceptions. Future research should further explore these moderating factors and their implications for designing AI interfaces and interaction paradigms that can accommodate diverse user characteristics and cognitive styles.

\subsection{General Discussion}

Our study contributes to and expands upon the existing body of knowledge in human-AI interaction.  While previous studies have focused on how users' perceptions of AI motives affect their evaluations of AI \cite{pataranutaporn2023influencing}, our research shows that the content itself---specifically its positive or negative framing---can have a profound impact on user judgments, even in the absence of perceived AI intentions.

Furthermore, our study contributes to the growing body of research on placebo effects in AI interactions \cite{kloft_ai_2024, kosch_placebo_2022, villa_placebo_2023}. By demonstrating that users' biases extend beyond performance expectations to their evaluation of AI-generated content, we provide a more comprehensive understanding of how psychological factors shape human-AI interactions. The observed preference for positive predictions in our study adds a new dimension to recent investigations into sycophantic behavior in language models \cite{sharma2023understanding, usman2024persuasive}. Our work empirically demonstrates that positively-biased model behavior affects user perception beyond simple misinformation, potentially reinforcing users' positivity bias and creating a feedback loop. Models that prioritize user favor over accuracy can be more persuasive and influential, affecting users beyond the accuracy of their outputs.

These findings have important implications for AI system design and deployment, particularly in generating predictions or personalized recommendations. Users' tendency to favor positive AI-generated content could lead to over-reliance on AI, potentially resulting in poor decisions if the predictions are inaccurate or biased. This bias could also be exploited by malicious actors to manipulate user behavior or beliefs through strategically positive messaging.

In the context of human-AI interaction, these results highlight the need for increased awareness and education about cognitive biases. Users should be encouraged to critically evaluate AI-generated predictions, regardless of their valence. Additionally, designers of AI systems should consider implementing mechanisms to mitigate the effects of personal validation, such as using neutral tones in predictions, providing balanced perspectives, or explicitly highlighting the potential for bias in user perceptions.

\subsection{Limitations and Future Directions}
While our study provides valuable insights into the influence of personal validation on human-AI interaction, it has several limitations that point to important avenues for future research. 

First, our experiment focused on single-turn interactions with the predictions. In real-world scenarios, human-AI interactions often involve multiple turns of dialogue, which could potentially amplify the effects we observed. Future studies should investigate how the personal validation effect evolves over extended interactions, similar \cite{pataranutaporn2023influencing} examining whether the effect strengthens or diminishes with repeated exposure to positive or negative predictions. This longitudinal approach could reveal important dynamics in the formation and persistence of biases in human-AI relationships.

Second, the predictions used in our experiment were presented in a uniform format. However, AI systems in practice often employ various features that could influence user perceptions. Future studies should investigate how AI characteristics such as personalization, the use of virtual avatars, or speech-based interactions could potentially augment or alter the observed personal validation effect. For instance, would an embodied AI agent delivering positive predictions have a stronger influence on user perceptions compared to text-based interactions? Understanding these nuances could provide valuable insights for designing more effective and responsible AI interfaces.

Third, our study was limited to a specific domain of predictions, which emphasizes the need to extend this research into high-impact fields where the consequences of biased perception could be more severe. For instance, in the medical field, an AI system that consistently reinforces a doctor’s decisions might inadvertently foster a personal bias, potentially leading to skewed medical judgments and compromising patient outcomes. Similarly, in the realm of education, an AI tutor that provides continuous positive reinforcement without accurate assessment of a student's progress could lead to overconfidence, undermining genuine skill development and long-term learning. These scenarios represent critical areas for future investigation, as they could have significant real-world implications for patient care and student development.

Furthermore, while this study focused on the perception of AI predictions, it raises important questions about how these perceptions influence actual user behavior. Expanding the experimental framework to include measures of behavioral changes and decision-making outcomes could provide valuable insights into the broader impact of cognitive and psychological factors in human-AI interaction. Understanding this could help in designing AI systems that not only foster appropriate trust but also encourage sound decision-making in users.

Additionally, future research should explore potential mitigation strategies for the personal validation effect in AI interactions. This could involve developing and testing different approaches to presenting AI-generated predictions. Investigating the effectiveness of user education programs in reducing susceptibility to this bias could also yield valuable insights for improving AI literacy. 

% Also, similar to the previous study [Retract for anonymity], we were unable to fully resolve the positive autocorrelation issue in our mixed effects model, which may pose a slight limitation to the interpretation of our results. Future studies could explore alternative approaches or additional data collection to better address this issue. 

Lastly, cross-cultural studies could provide a more comprehensive understanding of personal validation effect in AI interactions. Cultural differences in communication styles, attitudes towards technology, and responses to positive feedback might influence how this bias manifests across different populations. Such research could inform the development of culturally sensitive AI systems and interaction guidelines.

By addressing these limitations and pursuing these future directions, researchers can build a more comprehensive understanding of the personal validation effect in human-AI interaction, ultimately contributing to the development of more effective, ethical, and user-centered AI systems.

\section{Conclusion}

In conclusion, our study demonstrates that the personal validation effect and positivity bias significantly influence how users perceive AI-generated predictions. Positive predictions are consistently rated as more valid, personalized, reliable, and useful compared to negative predictions. Moreover, the study presents an exploratory analysis of how cognitive style, paranormal beliefs, gullibility, AI attitude/trust, age, and gender moderate the negative impact of negative predictions on perceived validity, personalization, reliability, and usefulness, producing some nuanced findings that open up opportunities for future work. The findings of this study underscore the complex psychological dynamics at play in human-AI interaction and highlight the need for careful consideration of cognitive biases in the design and deployment of AI systems. As AI continues to play an increasingly prominent role in decision-making and personal assistance, understanding and mitigating the effects of personal validation will be crucial for fostering more balanced and effective human-AI collaboration.

%% file: chapter4.tex
% From mitthesis package
% Version: 1.06, 2024/07/09
% Documentation: https://ctan.org/pkg/mitthesis

\chapter{Future Directions: Investigating Self-Fulfilling Prophecies in Human-AI Interaction}
\label{ch4_future_work}

With the rapid advancement and adoption of AI in society, individuals are increasingly exposed to algorithms that make personalized predictions and recommendations based on their data. These applications range from user modeling based on web interactions for tailored experiences \cite{purificato_user_2024} to sophisticated AI assistants that continuously adapt user representations through ongoing interactions \cite{zhong_memorybank_2023}. Such technologies are widely applied in e-commerce, adaptive learning systems, targeted advertising, personalized content delivery, and healthcare. Consequently, users are constantly subjected to the “expectations” set by AI systems regarding their preferences, behaviors, and future actions. 

The concept of the self-fulfilling prophecy refers to situations where prior expectations influence subsequent behavior in a way that ultimately confirms those expectations \cite{merton1948self, jussim_self-fulfilling_1986}. Originally conceptualized by sociologist Robert K. Merton in 1948, this phenomenon has been widely studied across disciplines such as psychology, sociology, business, medicine, education, economics, and law \cite{almadi_meta-narrative_2022, henshel_boundary_1982}. A related concept, the Pygmalion effect, occurs on an interpersonal level when one person’s expectations influence another's behavior, leading to the fulfillment of the initial expectations \cite{rosenthal_pygmalion_1968, rosenthal2002pygmalion, jussim_self-fulfilling_1986}. This effect has been notably explored in education, where teachers' expectations significantly impact students' academic performance \cite{rosenthal2002pygmalion, friedrich2015pygmalion, jussim_self-fulfilling_1986}. Despite the broad exploration of these concepts, their implications within human-AI interaction remain underexplored, offering a valuable opportunity to investigate how AI's expectations might influence user behavior. This raises a critical question: Can AI's expectations of individuals shape their actual behavior in a way that fulfills those predictions?

This exploration is particularly relevant in light of concerns about advanced AI models, such as large language models (LLMs). These models have been critiqued for displaying sycophantic behavior, where they tend to agree with users' inputs regardless of accuracy \cite{sharma2023understanding}, and for creating echo chambers that reinforce existing beliefs rather than challenging them \cite{sharma_generative_2024}. Such tendencies can exacerbate the self-fulfilling prophecy effect, as users might increasingly rely on AI-generated feedback that mirrors their existing views, further entrenching those views. For instance, if an LLM consistently affirms a user's overconfidence in investment decisions, it could lead the user to make riskier choices, thereby validating the AI's predictions and perpetuating a cycle of reinforcement. These concerns highlight the importance of critically examining how AI systems influence user behavior and the potential for these technologies to inadvertently shape, rather than simply reflect, human actions. Understanding these dynamics is crucial for designing AI systems that avoid the pitfalls of manipulation and instead foster more balanced and beneficial interactions.

This chapter outlines the direction of a future work in studying the phenomenon of self-fulfilling prophecy in the context of human-AI interaction. Through an experimental design that appears to provide personalized predictions on their investment behavior and future outcome, we investigate the extent to which fictitious AI predictions can serve as self-fulfilling prophecies. More specifically, we aim to investigate whether AI predictions on one’s emotional discipline in an simulated investment game context can impact their actual emotional discipline and investment behavior in later stages. The study also explores the role of moderating factors such as affective state and trust in the realization of AI predictions. By studying the phenomenon of self-fulfilling prophecy and key moderating factors in the context of human-AI interaction, we hope to draw generalizable insights on the design of AI systems in a way that minimizes the risk of manipulation and increases potential benefits.

% User modeling is the process of creating representations of individual users’ preferences, behaviors, and characteristics through the analysis of their interaction data, enabling personalized experiences and predictions in AI systems. User modeling has been applied in various domains, including personalized recommendations in e-commerce, adaptive learning systems in education, targeted advertising, personalized content delivery in media and entertainment, healthcare for tailored treatment plans, and human-computer interaction to enhance user experience.

\section{Related Works}

\subsection{Self-fulfilling prophecy in human-AI interaction}

% Self-fulfilling prophecy has been defined by Merton as a scenario where a false definition of a situation evokes new behaviors that make the original false conception become true \cite{merton1948self}. 

The concept of self-fulfilling prophecies in human-AI interaction is an emerging area of study, with significant implications for how AI systems influence user behavior. Lawrence Switzky discusses the "ELIZA Effect" in his work Pygmalion and the Early Development of Artificial Intelligence, highlighting the human tendency to attribute greater intelligence and intention to responsive computers than is warranted \cite{switzky2023eliza}. This phenomenon underscores the potential for AI systems to shape user perceptions and actions in ways that align with the AI's predictions, even when those predictions are based on flawed or simplistic models.

De-Arteaga and Elmer (2023) have identified several ways in which machine learning systems can perpetuate or create self-fulfilling prophecies, particularly in healthcare. These include encoding flawed human beliefs into algorithms, reinforcing biases through human-machine interaction cycles, making inaccurate predictions that users act upon, and relying on outdated clinical decisions despite advancements in medical knowledge \cite{de-arteaga_self-fulfilling_2023}. Such processes highlight the potential for AI systems to not only reflect but also amplify existing biases, resulting in outcomes that may seem accurate simply because they have shaped user behavior to match the predictions.

Feedback loops play a critical role in the self-fulfilling prophecy effect within AI systems. King and Mertens (2023) argue that misunderstandings between predictions and their subjects can lead to accountability issues and feedback loops that reinforce these prophecies \cite{king_self-fulfilling_2023}. Similarly, Bauer (2023) discusses how algorithmic score disclosures can influence individual behavior, creating self-fulfilling prophecies where people align their actions with erroneous assessments \cite{bauer_mirror_2023}. This phenomenon is particularly concerning in settings where biased machine learning models are continuously retrained on new data that the models themselves help generate. In such cases, the AI's predictions may appear increasingly accurate, not because they reflect reality, but because users adjust their behavior to fit the AI's expectations, reinforcing the model's biases.

Similarly, Bauer and colleagues (2023) found that disclosing algorithmic scores can lead to anchoring bias, where individuals adjust their behavior to align with the scores, thus reinforcing the predictions and creating a feedback loop that perpetuates bias \cite{bauer_mirror_2023}. This is particularly problematic in scenarios where machine learning models are continuously retrained on data they have influenced, as biased outputs can lead to behaviors that further entrench the biases in the model.

The implications of these feedback loops are evident in areas like predictive policing, where biased models can lead to runaway cycles of reinforcement, sending police to the same neighborhoods repeatedly regardless of the true crime rate \cite{ensign_runaway_2017}. Similarly, in economic decision-making, speculative forecasts can influence behavior before any actual changes occur, as demonstrated by Petropoulos Petalas et al. \cite{petropoulos_petalas_forecasted_2017}. This study, though not directly related to AI, highlights how predictions can alter decision-making and risk-taking, further illustrating the broader impact of self-fulfilling prophecies in contexts where accurate predictions are critical.

Self-fulfilling prophecies have also been observed in the context of placebo AI treatments. Recent studies have revealed that individuals' beliefs about an AI's capabilities can influence their responses to the system, even when the AI is not functioning as claimed \cite{villa_placebo_2023, kloft_ai_2024, kosch_placebo_2022}. This highlights the power of perception in shaping user behavior and the potential for AI systems to unintentionally manipulate or reinforce certain actions.

Given this backdrop, it is crucial to explore the phenomenon of self-fulfilling prophecies in human-AI interaction, particularly in contexts like financial decision-making, where risk and outcomes are heavily influenced by user behavior. Understanding how AI expectations influence attitudes toward risk can shed light on the complex interplay between human psychology and artificial intelligence, offering insights into how to design AI systems that minimize the risk of manipulation and maximize their benefits.

\subsection{Emotions and risk-taking behavior}

Emotions play a significant role in influencing risk-taking behavior, often serving as a critical determinant in decision-making processes. One key theory is the "mood maintenance hypothesis," which posits that individuals in a positive emotional state are less likely to engage in risky behavior, as they wish to preserve their current affective state. This theory is supported by findings that suggest individuals experiencing positive affect exhibit more conservative decision-making behavior compared to those in neutral or negative states \cite{juergensen_more_2018, isen_influence_1988, nygren_influence_1996, hermalin_effect_2000}. Conversely, those in a negative emotional state may engage in riskier behavior in an attempt to improve their mood, often focusing on the potential positive outcomes rather than the more probable negative ones \cite{lerner_beyond_2000, raghunathan_all_1999}.

Integral emotions, which arise directly from the decision at hand, play a crucial role in shaping risk-taking behavior. Unlike incidental emotions, which are unrelated to the decision but still influence it, integral emotions are considered a normatively defensible input to judgment and decision-making \cite{lerner_emotion_2015}. Immediate emotions, a subset of integral emotions, refer to the feelings experienced in response to the options being considered, such as fear or excitement, which can motivate individuals to either approach or avoid certain risks. These immediate emotions differ from anticipated emotions, which involve predictions about how one will feel after the outcomes of a decision \cite{loewenstein_risk_2001, schlosser_what_2013, lerner_emotion_2015}. The framing of risky choices also affects immediate emotional responses, which can influence decision-making. When a choice is framed as a potential loss, individuals are more likely to experience negative emotions, leading to a higher likelihood of risk-taking as they seek to avoid the perceived loss \cite{young_sure_2019}. Together, these emotional processes highlight the complexity of how emotions influence risk-taking behavior, often dictating the choices individuals make under uncertainty.

\subsection{Emotional regulation in investment decision-making}

Emotion regulation involves efforts to alter the course of an ongoing emotional response \cite{gross_emotion_2015}. Individuals typically employ various strategies to manage their emotions, which significantly affect their emotional well-being, including cognitive reappraisal, where individuals modify their thoughts about a situation to change its emotional impact, and expressive suppression, which involves restraining the external display of emotions \cite{gross_individual_2003}. 

Emotional regulation plays a critical role in decision-making, particularly in high-stakes environments like investing. Emotional regulation involves a sequence of stages, including the identification of the need to regulate emotions, selecting an appropriate strategy, and implementing that strategy effectively \cite{matthews_identifying_2021}. The ability to regulate emotions is vital for maintaining goal-directed behavior, as difficulties in this area can impair performance in cognitive tasks \cite{rekar_effect_2023}. For instance, a study suggests that effective emotion regulation strategies support better decision-making by promoting goal-oriented actions \cite{martin_influence_2011}.

In the context of investing, emotional regulation becomes even more crucial due to the complexity and uncertainty involved in financial decisions. Overconfidence, often driven by strong positive emotions like pride, can lead to detrimental outcomes such as increased trading volumes and a stronger disposition effect \cite{im_effect_2016}. When emotions are not adequately regulated, they can overwhelm cognitive processes, leading to poor decision-making and increased risk-taking behaviors \cite{cohen_nature_2008}. As investment decisions are inherently complex, the lack of professional expertise often forces individuals to rely on their emotions rather than rational analysis, further underscoring the importance of emotional regulation in financial contexts \cite{chu_overconfidence_2012}.

\section{Proposed Study Design}

\subsection{Participants}

The study will involve participants aged 18 and above, representing a diverse range of financial literacy, investment experience, emotional regulation skills, and familiarity with AI. Participants will be randomly assigned to one of two groups: positive prediction or negative prediction. This randomization ensures a balanced comparison of the effects of prediction valence on participant behavior and emotions.

\subsection{Experiment protocol}

The study will utilize a simulated investment game, building on the design of previous studies. Each participant will be provided with a virtual currency of \$10,000 and will be asked to invest this amount across three categories: high risk/high return, medium risk/medium return, and low risk/low return. The investment game will consist of two phases:

\begin{itemize}
    \item \textbf{Pre-prophecy phase:} Participants will complete 10 rounds of investment decisions.
    \item \textbf{Post-prophecy phase:} Participants will receive an "AI-generated" prophecy regarding their emotional regulation in investing, after which they will complete additional rounds of investment decisions. We are considering two possible designs for this phase:
    \begin{itemize}
        \item \textbf{Single prophecy:} The prophecy is provided once after the first 10 rounds, followed by 10 more rounds of investment.
        \item \textbf{Multiple prophecies:} The prophecy is provided 2-3 times throughout the second phase, allowing us to observe how repeated feedback might affect behavior and emotions over time.
    \end{itemize}
\end{itemize}

\begin{figure}[h]
\includegraphics[width=16cm]{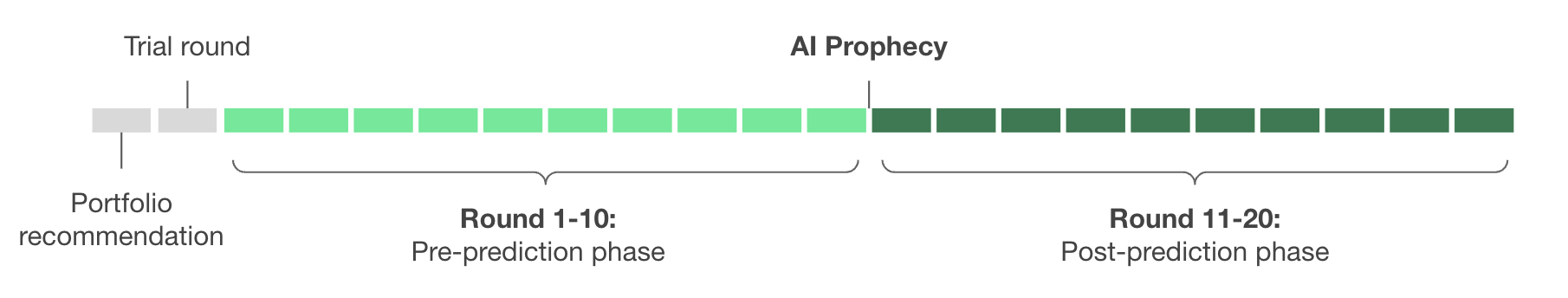}
    \centering
    \caption{Proposed study procedure}
\label{fig:procedure_diagram_2}
\end{figure}

The proposed experimental protocol is provided in Figure \ref{fig:procedure_diagram_2}.

We hypothesize that the valence of the prediction will significantly influence both the participants' investment behavior and emotional responses. Specifically, positive predictions are expected to lead to behavior that reflects higher emotional regulation, while negative predictions may result in behavior consistent with lower emotional regulation, demonstrating a self-fulfilling prophecy effect. Additionally, we anticipate to see significant differences in allocation behavior, emotional state, and decision time post-prediction. While affective state will be measured as a dependent variable, we will also explore its potential role as a mediator or moderator in the relationship between prediction valence and investment behavior, along with the potential influences of other individual differences.

\subsection{Data collection}

The study will collect data on the following variables based on the game interaction:

\begin{itemize}
\setlength\itemsep{0em}
    \item Shift in investment allocation behavior before and after receiving the prediction
    \item Affective state (self-reported integral emotion at each round)
    \item Decision time (time taken to make investment decisions at each round)
\end{itemize}

\noindent Additionally, participants will complete surveys assessing various individual differences, including but not limited to:

\begin{itemize}
\setlength\itemsep{0em}
    \item Self-reported emotional regulation (before and after the game)
    \item Belief in prediction (validity, reliability, usefulness, personalization)
    \item Risk preference
    \item Level of financial literacy/investing experience
    \item Attitude toward AI
    \item Personality traits
    \item Incidental emotion/state affect
\end{itemize}

\subsection{Data analysis}

The data will be analyzed to assess the impact of fictitious AI predictions on investment behavior, emotional responses, and self-perception of emotional regulation. Additionally, we will explore how affective state and changes in self-perception may serve as both mediators and moderators in the relationship between AI predictions and behavioral outcomes.

\subsubsection{Dependent Variables}

We will first examine how AI predictions influence the following key outcomes:

\begin{itemize}
\setlength\itemsep{0em}
    \item \textbf{Behavioral impact:} The shift in investment allocation before and after receiving the prediction, as well as decision time.
    \item \textbf{Emotional impact:} Differences in affective state (integral emotion) and decision time between the positive and negative prediction groups.
    \item \textbf{Self-perception impact:} Changes in self-perception of emotional regulation before and after the study, assessing the reported influence of the prophecy.
\end{itemize}

\subsubsection{Mixed effects model}

We will employ mixed-effects models to account for repeated measures within participants and the hierarchical nature of the data. This approach will allow us to model individual differences (random effects) and the overall effects of the AI prediction, prediction valence, and other relevant factors on our dependent variables: investment behavior, affective state, decision time, and change in self-perception of emotional regulation.

\begin{itemize}
\item \textbf{Fixed effects: } The primary fixed effects will include the presence of a prediction (prediction vs. no prediction), prediction valence (positive vs. negative), and their interaction. Additional factors, such as emotional regulation skills, risk preference, trust, and perceived belief in the AI-generated prophecy (including subscales for validity, personalization, reliability, and usefulness), will also be included to explore their influence on the impact of the AI predictions.

\item \textbf{Random effects: } Random intercepts will be included to account for individual differences in baseline levels of the dependent variables. If variability in responses to the AI predictions is observed, random slopes may be included to model this variability over time.

\end{itemize}

\subsubsection{Mediation analysis}

We will explore whether changes in self-perception of emotional regulation and affective state mediate the relationship between receiving an AI prediction and investment behavior. This analysis will help determine if these internal psychological changes explain how AI predictions influence behavior.

% \begin{itemize}
% \item Mediation pathways:
% \begin{itemize}
% \item Self-perception as a mediator: Investigating whether the change in self-perception of emotional regulation mediates the effect of AI predictions on subsequent investment behavior.
% \item Affective state as a mediator: Assessing whether the impact of AI predictions on investment behavior is mediated by changes in participants' affective state.
% \end{itemize}
% \end{itemize}

\subsubsection{Moderation analysis}

We will investigate whether changes in self-perception, alongside other factors such as emotional regulation skills, risk preference, trust, and perceived belief, moderate the impact of AI predictions on behavior and affective state. This analysis will help identify if the effects of AI predictions on behavior differ depending on how participants’ self-perception changes or how they perceive the AI-generated prophecy.

% \begin{itemize}
% \item Interaction terms:
% \begin{itemize}
% \item Self-perception as a moderator: Testing whether the extent of change in self-perception moderates the relationship between AI predictions and behavioral outcomes.
% \item Other moderators: Evaluating how emotional regulation skills, risk preference, trust, and perceived belief might moderate the effects of AI predictions.
% \end{itemize}
% \end{itemize}

This mixed-effects modeling approach, combined with mediation and moderation analysis, will provide a robust framework for understanding how fictitious AI-generated predictions influence participant behavior and emotional responses, and how these effects are shaped by the valence of the predictions, perceived belief of the predictions, and other individual differences. The analysis plan is flexible and may be adjusted if the data suggest a different approach is more suitable.

%% file: chapter5.tex
% From mitthesis package
% Version: 1.06, 2024/07/09
% Documentation: https://ctan.org/pkg/mitthesis

\chapter{Conclusion}

This thesis explored the complex psychological dynamics underlying human-AI interaction, with a focus on the interplay between belief, perception, and behavior.

Chapter \ref{ch2_paper1} investigated the phenomenon of "rational superstition" in the context of human-AI interaction, revealing that people's perceptions of AI predictions can be irrationally influenced by psychological factors such as paranormal beliefs, cognitive biases, and incorrect mental models. The findings demonstrated that, much like astrology, AI predictions can be seen as trustworthy and reliable even without validation, driven by a combination of idealized portrayals of AI and mental heuristics. This underscores the necessity of critically assessing how AI systems are presented and understood, to avoid unwarranted trust and over-reliance.

Chapter \ref{ch3_paper2} delved into the personal validation effect within AI-generated predictions, particularly focusing on how positive framing can bias user perceptions of validity, personalization, reliability, and usefulness. The study highlighted the substantial influence that positive AI responses have on user perception, which can lead to overconfidence in AI systems and poor decision-making. These findings stress the importance of designing AI systems that are not only accurate but also transparent and balanced, mitigating the risks associated with cognitive biases.

Chapter \ref{ch4_future_work} proposed future work to explore the concept of the self-fulfilling prophecy in human-AI interaction. The proposed study aims to investigate how AI predictions might shape user behavior, particularly in contexts where the predictions are about themselves. This exploration is crucial as it addresses the potential for AI systems to inadvertently influence behavior in ways that reinforce existing biases or expectations, further highlighting the importance of responsible AI design.

Together, this thesis underscores the critical role that psychological factors play in shaping human-AI interaction and highlights the importance of accounting for human irrationality in AI system design. As AI systems become increasingly integrated into daily life, it is essential to design them and educate users in ways that foster informed and critical engagement, rather than blind trust or over-reliance. By understanding and mitigating the effects of cognitive biases and psychological influences, we can ensure that AI serves as a tool for enhancing human decision-making, rather than undermining it.

%% file: appendixa.tex
% From mitthesis package
% Version: 1.01, 2023/07/04
% Documentation: https://ctan.org/pkg/mitthesis

\chapter{Survey Instruments}
\label{appendix_survey}

\section{Consent Form}
\textit{This consent form was administered on Qualtrics.}

\begin{tcolorbox}[
    width=\textwidth,         % box width set to text width
    colframe=black,           % frame color (black)
    colback=white,            % background color (white)
    boxrule=0.75pt,              % line width of the box
    left=5pt,                 % left padding
    right=5pt,                % right padding
    top=5pt,                  % top padding
    bottom=5pt,               % bottom padding
    fontupper=\normalfont,    % use system font, no boldface
    breakable,
    enhanced jigsaw
    ]

Welcome! This study aims to understand what makes people believe in predictions about themselves.

\vspace{10pt}

The study should take around 30 min to complete. You are eligible to participate in this study if you are 18 or above, and you have normal or corrected-to-normal vision and hearing.

\vspace{10pt}

Your responses will be kept confidential. Results from this study may be used for future academic research publications.

\vspace{10pt}

Please note that:

\begin{enumerate}
\setlength\itemsep{0em}
    \item ⁠Your participation is voluntary.
    \item Some questions are mandatory. However, if you do not want to answer them, you can exit the study.
    \item At any time, you may decline further participation without adverse consequences. To do this, simply close out of the study without submitting your answers.
    \item You will remain anonymous in any publications or presentations based on the results of this study. If you feel you have been treated unfairly, or you have questions regarding your rights as a research subject, you may contact the Chairman of the Committee on the Use of Humans as Experimental Subjects, M.I.T., Room E25-143B, 77 Massachusetts Ave, Cambridge, MA 02139, phone 1-617-253 6787.
\end{enumerate}

\vspace{10pt}

By clicking on "Next" below, you acknowledge: Your participation in the study is voluntary. You are 18 years of age or above. You are aware that you may choose to terminate your participation at any time for any reason. Please select if you consent to begin the study.

\end{tcolorbox}

\section{Experiment instructions}
\textit{These instructions were provided on a web app embedded on Qualtrics.}

\begin{tcolorbox}[
    width=\textwidth,         % box width set to text width
    colframe=black,           % frame color (black)
    colback=white,            % background color (white)
    boxrule=0.75pt,              % line width of the box
    left=5pt,                 % left padding
    right=5pt,                % right padding
    top=5pt,                  % top padding
    bottom=5pt,               % bottom padding
    fontupper=\normalfont,    % use system font, no boldface
    breakable,
    enhanced jigsaw
    ]

You will complete a short questionnaire about your birthdate and your personality, then participate in a simulated investing game

\vspace{10pt}

At the end, you will be provided with three different versions of predictions about your future investment behavior, based on analysis based on: 1) your zodiac sign, 2) your personality, and 3) your interaction data monitored by our AI model.

\vspace{10pt}

Please answer honestly and to the best of your ability for accurate predictions. The information you provide in the questionnaire will only be used for the purpose of creating personalized answers for you and will not be shared with anyone else.

\end{tcolorbox}

\section{Astrology and Personality Questionnaires}

\begin{tcolorbox}[
    width=\textwidth,         % box width set to text width
    colframe=black,           % frame color (black)
    colback=white,            % background color (white)
    boxrule=0.75pt,              % line width of the box
    left=5pt,                 % left padding
    right=5pt,                % right padding
    top=5pt,                  % top padding
    bottom=5pt,               % bottom padding
    fontupper=\normalfont,    % use system font, no boldface
    breakable,
    enhanced jigsaw
    ]

\textbf{Zodiac sign}

What is your birthdate? MM/DD/YYYY

(Optional) What is the local time you were born? hh/mm

(Optional) Which city were you born in?

\textbf{MBTI Personality test (shortened version)}

Likert scale of 1 (strongly disagree) to 7 (strongly agree)

You regularly make new friends.

You spend a lot of your free time exploring various random topics that pique your interest.

Seeing other people cry can easily make you feel like you want to cry too.

You often make a backup plan for a backup plan.

You usually stay calm, even under a lot of pressure.

At social events, you rarely try to introduce yourself to new people and mostly talk to the ones you already know.

You prefer to completely finish one project before starting another.

You are very sentimental.

You like to use organizing tools like schedules and lists.

Even a small mistake can cause you to doubt your overall abilities and knowledge.

You feel comfortable just walking up to someone you find interesting and striking up a conversation.

You are not too interested in discussing various interpretations and analyses of creative works.

\end{tcolorbox}

\section{Investment game instructions}
\textit{These instructions were provided on a web app embedded on Qualtrics.}

\begin{tcolorbox}[
    width=\textwidth,         % box width set to text width
    colframe=black,           % frame color (black)
    colback=white,            % background color (white)
    boxrule=0.75pt,              % line width of the box
    left=5pt,                 % left padding
    right=5pt,                % right padding
    top=5pt,                  % top padding
    bottom=5pt,               % bottom padding
    fontupper=\normalfont,    % use system font, no boldface
    breakable,
    enhanced jigsaw
    ]

You will be provided with virtual currency of \$10,000. You may invest across three investment categories (high risk/return, medium risk/return, low risk/return). You will invest over 10 rounds (representing years).

\vspace{10px}

In the next page, you will answer a short risk preference questionnaire and receive a recommended portfolio allocation (\%). You may revise your allocation at any round. After each round, we will ask you how confident you feel about your choices. Good luck!

\vspace{10px}

Important note: Your investment performance will increase your total compensation for participating in the study (10\% increase for every \$100,000).

\end{tcolorbox}

\section{Investment preference survey}
\textit{The survey was administered through a web app embedded in Qualtrics.}

\begin{tcolorbox}[
    width=\textwidth,         % box width set to text width
    colframe=black,           % frame color (black)
    colback=white,            % background color (white)
    boxrule=0.75pt,              % line width of the box
    left=5pt,                 % left padding
    right=5pt,                % right padding
    top=5pt,                  % top padding
    bottom=5pt,               % bottom padding
    fontupper=\normalfont,    % use system font, no boldface
    breakable,
    enhanced jigsaw
    ]

1. When making a long-term investment, I plan to keep the money invested for
\begin{itemize}
\setlength\itemsep{0em}
    \item 1 to 2 years
    \item 3 to 5 years
    \item 6 to 8 years
    \item More than 8 years
\end{itemize}

2. What's your main priority when investing?
\begin{itemize}
\setlength\itemsep{0em}
    \item Avoiding losing money
    \item Preferring safety over returns
    \item Low risk over more money
    \item More money over low risk
    \item Maximizing profits, no matter what
\end{itemize}

3. How do you feel about taking risks, in general?
\begin{itemize}
\setlength\itemsep{0em}
    \item I avoid risks. Better safe than sorry.
    \item Little risk is okay. I'm a bit cautious.
    \item I like to take risks sometimes.
    \item I love risk. I'm a bit of a gambler.
\end{itemize}

4. During market declines, I would sell portions of my riskier assets and invest the money in safe assets. (5-point Likert scale, 1=Strongly disagree to 5=Strongly agree)

\end{tcolorbox}

\section{Simulated investment game}
\textit{The game was administered through a web app embedded in Qualtrics.}

\begin{tcolorbox}[
    width=\textwidth,         % box width set to text width
    colframe=black,           % frame color (black)
    colback=white,            % background color (white)
    boxrule=0.75pt,              % line width of the box
    left=5pt,                 % left padding
    right=5pt,                % right padding
    top=5pt,                  % top padding
    bottom=5pt,               % bottom padding
    fontupper=\normalfont,    % use system font, no boldface
    breakable,
    enhanced jigsaw
    ]

Emotion survey (after each round)

How confident you feel about your choices? (7-point Likert scale, 1=Not at all confident to 7=Very confident)

\end{tcolorbox}

\section{Evaluation of prophecies}
\textit{Administered through a web app embedded in Qualtrics.}

\begin{tcolorbox}[
    width=\textwidth,         % box width set to text width
    colframe=black,           % frame color (black)
    colback=white,            % background color (white)
    boxrule=0.75pt,              % line width of the box
    left=5pt,                 % left padding
    right=5pt,                % right padding
    top=5pt,                  % top padding
    bottom=5pt,               % bottom padding
    fontupper=\normalfont,    % use system font, no boldface
    breakable,
    enhanced jigsaw
    ]

Please tell us what you think about each of the following statements (1=Strongly disagree, 7=Strongly agree).

\begin{itemize}
\setlength\itemsep{0em}
    \item I find the prediction convincing.
    \item I can identify with the prediction.
    \item The source of the prediction is reliable.
    \item I find the prediction helpful.
    \item The prediction is accurate.
    \item The prediction describes me very well.
    \item I trust the source of the prediction.
    \item The prediction is useful for making future decisions.
\end{itemize}

\end{tcolorbox}

\section{Post-study questionnaire}
\textit{Administered on Qualtrics.}

\begin{tcolorbox}[
    width=\textwidth,         % box width set to text width
    colframe=black,           % frame color (black)
    colback=white,            % background color (white)
    boxrule=0.75pt,              % line width of the box
    left=5pt,                 % left padding
    right=5pt,                % right padding
    top=5pt,                  % top padding
    bottom=5pt,               % bottom padding
    fontupper=\normalfont,    % use system font, no boldface
    breakable,
    enhanced jigsaw
    ]

\textbf{Cognitive Reflection Test (CRT-2) \cite{thomson_investigating_2016}}:

\begin{itemize}
\setlength\itemsep{0em}
    \item If you’re running a race and you pass the person in second place, what place are you in? 
    \item A farmer had 15 sheep and all but 8 died. How many are left?
    \item Emily’s father has three daughters. The first two are named April and May. What is the third daughter’s name?
    \item How many cubic feet of dirt are there in a hole that is 3’ deep x 3’ wide x 3’ long?
\end{itemize}

\textbf{Need for Cognition (NCS-6) \cite{lins_de_holanda_coelho_very_2020}}:

\begin{itemize}
\setlength\itemsep{0em}
    \item I would prefer complex to simple problems. 
    \item I like to have the responsibility of handling a situation that requires a lot of thinking. 
    \item Thinking is not my idea of fun. (R) 
    \item I would rather do something that requires little thought than something that is sure to challenge my thinking abilities. (R) 
    \item I really enjoy a task that involves coming up with new solutions to problems. 
    \item I would prefer a task that is intellectual, difficult, and important to one that is somewhat important but does not require much thought.
\end{itemize}

\textbf{Revised Paranormal Belief Scale (R-PBS) \cite{tobacyk_revised_2004}}:

To what extent do you agree with the following (7-point Likert scale, 1=Strongly disagree, 7=Strongly agree):
\begin{itemize}
\setlength\itemsep{0em}
    \item The soul continues to exist though the body may die. [religious beliefs]
    \item Black cats can bring bad luck. [superstition]
    \item Your mind or soul can leave your body and travel (astral projection). [spiritualism]
    \item Astrology is a way to accurately predict the future. [precognition]
    \item There is a devil. [religious beliefs]
    \item If you break a mirror, you will have bad luck. [superstition]
    \item During altered states, such as sleep or trances, the spirit can leave the body. [spiritualism]
    \item The horoscope accurately tells a person’s future. [precognition]
    \item I believe in God. [religious beliefs]
    \item The number “13” is unlucky. [superstition]
    \item Reincarnation does occur. [spiritualism]
    \item Some psychics can accurately predict the future. [precognition]
    \item There is a heaven and a hell. [religious beliefs]
    \item It is possible to communicate with the dead. [spiritualism]
    \item Some people have an unexplained ability to predict the future. [precognition]
\end{itemize}

\textbf{AI attitude scale (AIAS-4) \cite{grassini_development_2023}}:

To what extent do you agree or disagree with the following statements? (1=Not at all, 10=Completely agree)

\begin{itemize}
\setlength\itemsep{0em}
    \item I believe that AI will improve my life.
    \item I believe that AI will improve my work.
    \item I think I will use AI technology in the future.
    \item I think AI technology is positive for humanity.
\end{itemize}

\textbf{Gullibility Scale \cite{teunisse_i_2020}}:

To what extent do you agree or disagree with the following statements? (7-point Likert scale, 1=Strongly disagree, 7=Strongly agree)

\begin{itemize}
\setlength\itemsep{0em}
    \item (G4) I’m not that good at reading the signs that someone is trying to manipulate me. [insensitivity]
    \item (G5) I’m pretty poor at working out if someone is tricking me. [insensitivity]
    \item (G6) It usually takes me a while to “catch on” when someone is deceiving me. [insensitivity]
    \item (G1) I guess I am more gullible than the average person. [persuadability]
    \item (G9) My friends think I’m easily fooled. [persuadability]
    \item (G12) Overall, I’m pretty easily manipulated. [persuadability]
\end{itemize}

\textbf{Big Five Personality Inventory (TIPI) \cite{gosling_very_2003}}:

(7-point Likert scale, 1=Strongly disagree, 7=Strongly agree)

I see myself as:

\begin{itemize}
\setlength\itemsep{0em}
    \item Extraverted, enthusiastic.
    \item Critical, quarrelsome.
    \item Dependable, self-disciplined.
    \item Anxious, easily upset.
    \item Open to new experiences, complex.
    \item Reserved, quiet.
    \item Sympathetic, warm.
    \item Disorganized, careless.
    \item Calm, emotionally stable.
    \item Conventional, uncreative.
\end{itemize}

\end{tcolorbox}

\section{Study debrief}
\textit{This debrief and continued participation agreement form was administered on Qualtrics.}

\begin{tcolorbox}[
    width=\textwidth,         % box width set to text width
    colframe=black,           % frame color (black)
    colback=white,            % background color (white)
    boxrule=0.75pt,              % line width of the box
    left=5pt,                 % left padding
    right=5pt,                % right padding
    top=5pt,                  % top padding
    bottom=5pt,               % bottom padding
    fontupper=\normalfont,    % use system font, no boldface
    breakable,
    enhanced jigsaw
    ]

You are at the end of the study! We appreciate your participation.

\vspace{10px}

Full disclosure: The predictions shown in the experiment were randomly generated and are unrelated to your actual behavior. The data collected for the predictions (birthdate, personality, etc.) are discarded and not used for the study.

\vspace{10px}

If you have any concerns or questions, you may contact the Chairman of the Committee on the Use of Humans as Experimental Subjects, M.I.T., Room E25-143B, 77 Massachusetts Ave, Cambridge, MA 02139, phone 1-617-253 6787

\vspace{10px}

Given this information, if you wish to withdraw your participation and data from this study, please select the option below. Otherwise, please click "Next" to continue.

\vspace{10px}

[Checkbox] I would like to withdraw from the study. I understand that my responses will be deleted and I may not be eligible for any bonus payments based on my game performance.

\end{tcolorbox}

% \section{}

% \begin{tcolorbox}[
%     width=\textwidth,         % box width set to text width
%     colframe=black,           % frame color (black)
%     colback=white,            % background color (white)
%     boxrule=0.75pt,              % line width of the box
%     left=5pt,                 % left padding
%     right=5pt,                % right padding
%     top=5pt,                  % top padding
%     bottom=5pt,               % bottom padding
%     fontupper=\normalfont,    % use system font, no boldface
%     breakable,
%     enhanced jigsaw
%     ]

% \end{tcolorbox}

%% file: appendixb.tex
% From mitthesis package
% Version: 1.01, 2023/07/04
% Documentation: https://ctan.org/pkg/mitthesis

\chapter{Detailed Results and Model Diagnostics for Chapter \ref{ch2_paper1}}
\label{appendix:results1}

\section{Multiple linear regression}

\subsection*{Regression Equation}
\begin{equation}
\begin{split}
\text{ai\_overall\_score} = \beta_0 + \beta_1(\text{zodiac\_overall\_score}) + \\
\beta_2(\text{personality\_overall\_score}) + \cdots + \\
\beta_n(\text{prophecy\_group}) + \epsilon
\end{split}
\end{equation}

\subsection*{Model results}

\begin{table}[h]
\centering
\caption{Linear regression model summary}
\label{table:linear_reg_model_summary}
\begin{tabular}{@{}lr@{}}
\toprule
\textbf{Statistic}      & \textbf{Value}                   \\ \midrule
Residual standard error & 0.7763 on 213 degrees of freedom \\
Multiple R-squared      & 0.7606                           \\
Adjusted R-squared      & 0.7337                           \\
F-statistic             & 28.2 on 24 and 213 DF            \\
p-value                 & \textless 2.2e-16                \\ \bottomrule
\end{tabular}
\end{table}

\begin{table}[h]
\centering
\caption{Linear regression residuals}
\label{table:linear_regression_residuals}
\begin{tabular}{@{}ccccc@{}}
\toprule
\textbf{Min} & \textbf{1Q} & \textbf{Median} & \textbf{3Q} & \textbf{Max} \\ \midrule
-2.52323     & -0.40173    & -0.03516        & 0.34271     & 3.05145      \\ \bottomrule
\end{tabular}
\end{table}

\begin{longtable}[c]{@{}p{7cm}p{1.2cm}p{1.2cm}p{1.2cm}p{1.2cm}l@{}}
\caption{Detailed results of multiple linear regression}
\label{table:linear_reg_results}\\
\toprule
\textbf{Variable}           & \textbf{Estimate} & \textbf{Std. Error} & \textbf{t value} & \textbf{Pr(\textgreater{}|t|)} & \textbf{Signif.} \\* \midrule
\endfirsthead
\endhead
(Intercept)                    & 0.15  & 0.55 & 0.273  & 0.785 &     \\
zodiac\_overall\_score         & 0.31  & 0.05 & 6.690  & 0.000 & *** \\
personality\_overall\_score & 0.46              & 0.05                & 9.585            & \textless 0.000                & ***                   \\
composite\_score               & 0.05  & 0.04 & 1.187  & 0.237 &     \\
paranormal\_score              & 0.00  & 0.00 & 0.970  & 0.333 &     \\
gullibility\_score             & 0.00  & 0.01 & 0.455  & 0.650 &     \\
aias\_score                    & 0.02  & 0.01 & 2.302  & 0.022 & *   \\
interest\_behavior             & 0.10  & 0.05 & 1.889  & 0.060 & .   \\
familiarity\_ai                & 0.06  & 0.07 & 0.854  & 0.394 &     \\
Age                            & -0.00 & 0.00 & -0.395 & 0.693 &     \\
big5\_extraversion             & -0.02 & 0.03 & -0.486 & 0.627 &     \\
big5\_openness                 & -0.03 & 0.05 & -0.536 & 0.593 &     \\
big5\_agreeableness            & 0.06  & 0.05 & 1.286  & 0.200 &     \\
big5\_conscientiousness        & 0.01  & 0.05 & 0.114  & 0.909 &     \\
big5\_emotional\_stability     & -0.01 & 0.04 & -0.169 & 0.866 &     \\
educationBachelor              & -0.29 & 0.23 & -1.241 & 0.216 &     \\
educationDoctorate             & 0.26  & 0.61 & 0.430  & 0.668 &     \\
educationHigh school           & -0.36 & 0.27 & -1.325 & 0.187 &     \\
educationLess than high school & 0.63  & 0.52 & 1.212  & 0.227 &     \\
educationMaster                & -0.59 & 0.27 & -2.206 & 0.029 & *   \\
educationProfessional degree   & -0.65 & 0.52 & -1.262 & 0.208 &     \\
educationSome college          & -0.22 & 0.25 & -0.915 & 0.361 &     \\
genderMale                     & 0.06  & 0.12 & 0.478  & 0.633 &     \\
genderOther                    & 0.03  & 0.31 & 0.086  & 0.932 &     \\
prophecy\_grouprational        & 0.24  & 0.11 & 2.154  & 0.032 & *   \\* \midrule
\multicolumn{6}{l}{Signif. codes: 0 ‘***’ 0.001 ‘**’ 0.01 ‘*’ 0.05 ‘.’ 0.1 ‘ ’ 1}                                    
\end{longtable}

\section{Multiple linear regression: Model diagnostics}
\label{appendix:diagnostics_lm}

\begin{figure*}[h]
    \centering
    \caption{Residuals vs Fitted Plot}
        \label{fig:residual_fitted_lm}
        \includegraphics[width=10cm]{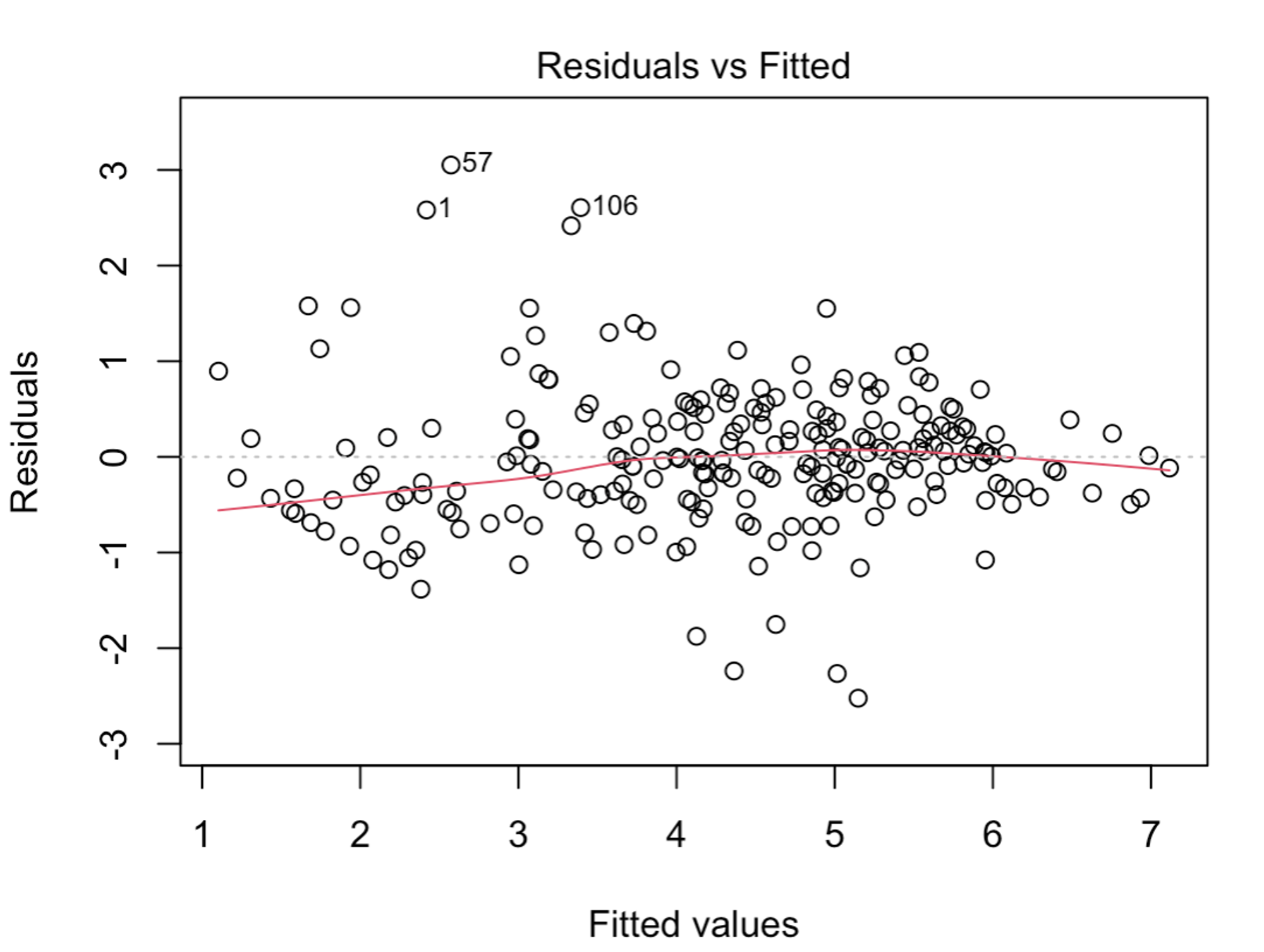}
        \begin{minipage}[b]{\linewidth}
            \vspace{10pt}
            \subcaption*{The residuals scatter randomly around the horizontal line (y=0), indicating that the linearity assumption holds. While there are a few outliers, the plot overall shows no strong patterns, suggesting the assumptions are reasonably met.}
        \end{minipage}
\end{figure*}

\begin{figure*}[h]
    \centering
    \label{fig:qqplot_lm}
    \caption{Q-Q plot of the fitted model}
        \includegraphics[width=10cm]{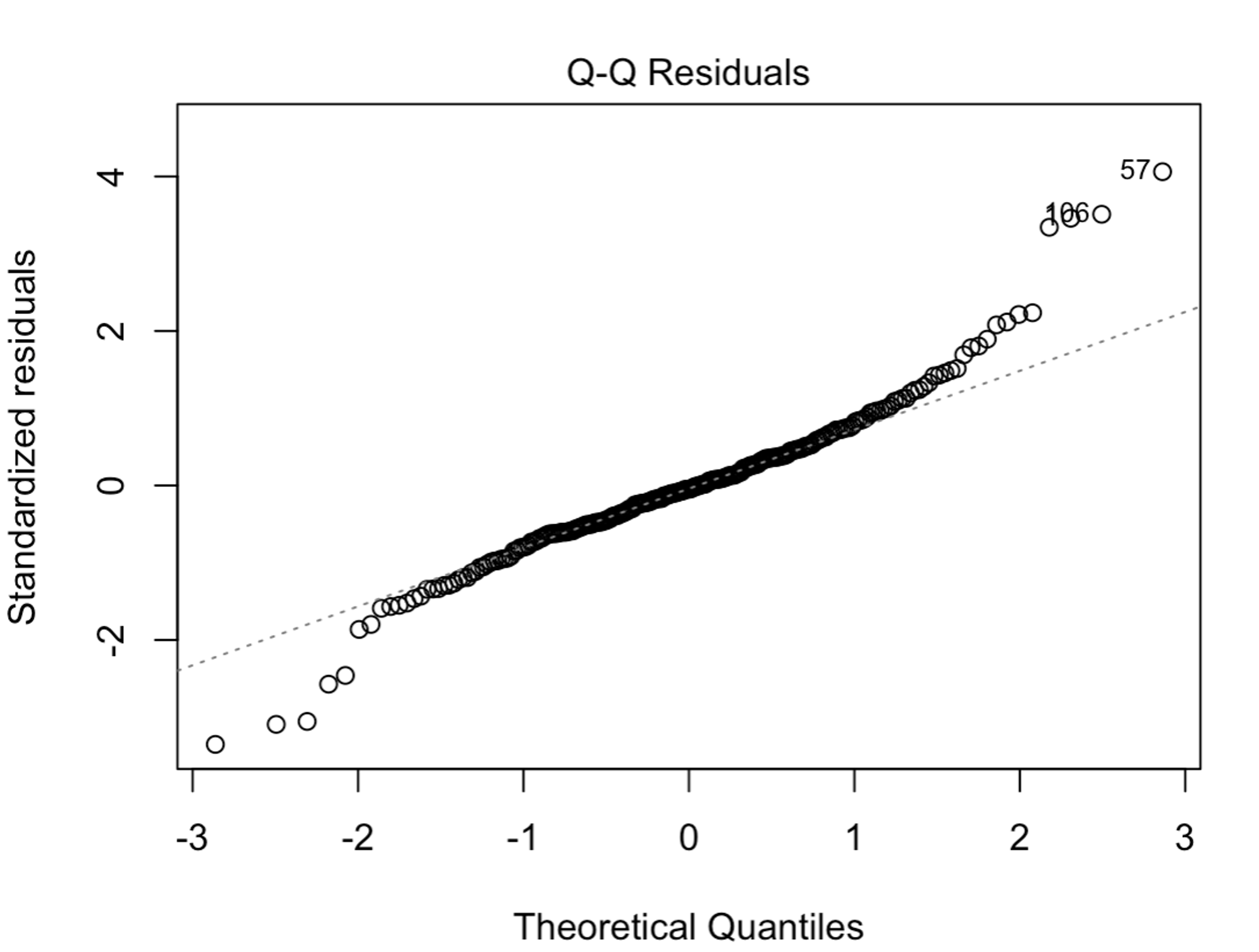}
        \begin{minipage}[b]{\linewidth}
            \vspace{10pt}
            \subcaption*{The points mostly follow the diagonal line, which suggests that the residuals are approximately normally distributed. There are some deviations at the tails, which may indicate some mild non-normality in the residuals.}
        \end{minipage}
\end{figure*}

\begin{table}[H]
\centering
\begin{minipage}{0.48\textwidth}
    \centering
    \caption{VIF values for LM}
    \label{table:vif_lm}
    \begin{tabular}{@{}ll@{}}
    \toprule
    Predictor                   & VIF      \\ \midrule
    zodiac\_overall\_score      & 1.553862 \\
    personality\_overall\_score & 1.570028 \\
    composite\_score            & 1.181011 \\
    paranormal\_score           & 1.234361 \\
    gullibility\_score          & 1.213760 \\
    aias\_score                 & 1.257553 \\
    interest\_behavior          & 1.143645 \\
    familiarity\_ai             & 1.275727 \\
    Age                         & 1.121698 \\
    big5\_extraversion          & 1.173883 \\
    big5\_openness              & 1.205639 \\
    big5\_agreeableness         & 1.195784 \\
    big5\_conscientiousness     & 1.226641 \\
    big5\_emotional\_stability  & 1.372926 \\
    education                   & 1.062415 \\
    gender                      & 1.111215 \\ 
    prophecy\_group             & 1.109082 \\ \bottomrule
    \end{tabular}
    \end{minipage}%
    \hfill
\begin{minipage}{0.48\textwidth}
    \centering
    \caption{Durbin-Watson test results for LM}
    \label{table:dw_test_lm}
    \begin{tabular}{@{}cccc@{}}
    \toprule
    \textbf{Lag} & \textbf{Autocorrelation} & \textbf{D-W Statistic} & \textbf{p-value} \\ \midrule
    1            & 0.0676                   & 1.8128                 & 0.118            \\ \bottomrule
    \end{tabular}
    \begin{minipage}[b]{\linewidth}
        \vspace{10pt}
            \subcaption*{The residuals show no significant autocorrelation, as indicated by a low lag 1 autocorrelation (0.0676), a near-ideal Durbin-Watson statistic (1.8128), and a p-value (0.118) that suggests the independence of errors in the regression model is likely satisfied.}
    \end{minipage}
\end{minipage}
\end{table}

\section{Mixed effects model: Results}
\label{appendix:results_lme1}

This is the detailed results table from the model used in Section \ref{ch2-methods}. All reported coefficients are based on the reference categories: Prophecy Source = AI, Prophecy Group = Positive, Subscale = Validity, Gender = Female, and Education = Bachelor.

\begin{longtable}[c]{@{}p{10cm}p{1cm}p{1cm}p{1cm}p{1cm}@{}}
\caption{Detailed mixed effects model results for Chapter 2}
\label{table:model_results_12f_c}\\
\toprule
Predictor                                                               & Estimate & PValue & Lower & Upper \\* \midrule
\endfirsthead
\multicolumn{5}{c}%
{{\bfseries Table \thetable\ continued from previous page}} \\
\toprule
Predictor                                                               & Estimate & PValue & Lower & Upper \\* \midrule
\endhead
\bottomrule
\endfoot
\endlastfoot
(Intercept)                                                             & 5.10     & 0.000  & 4.75  & 5.44  \\
subscalePersonalization                                                 & 0.23     & 0.000  & 0.08  & 0.37  \\
subscaleReliability                                                     & -0.91    & 0.000  & -1.13 & -0.68 \\
subscaleUsefulness                                                      & -0.63    & 0.000  & -0.83 & -0.44 \\
prophecy\_sourceAstrology                                               & -0.66    & 0.000  & -0.93 & -0.39 \\
prophecy\_sourcePersonality                                             & 0.06     & 0.620  & -0.19 & 0.31  \\
prophecy\_groupNegative                                                 & -1.19    & 0.000  & -1.53 & -0.85 \\
composite\_score                                                        & 0.13     & 0.060  & -0.01 & 0.26  \\
paranormal\_score                                                       & 0.02     & 0.000  & 0.01  & 0.03  \\
aias\_score                                                             & 0.04     & 0.000  & 0.01  & 0.06  \\
gullibility\_score                                                      & -0.02    & 0.130  & -0.04 & 0.01  \\
big5\_extraversion                                                      & -0.02    & 0.650  & -0.12 & 0.08  \\
big5\_openness                                                          & -0.10    & 0.210  & -0.25 & 0.05  \\
big5\_agreeableness                                                     & 0.09     & 0.240  & -0.06 & 0.24  \\
big5\_conscientiousness                                                 & -0.15    & 0.030  & -0.30 & -0.01 \\
big5\_emotional\_stability                                              & 0.00     & 0.980  & -0.13 & 0.13  \\
interest\_behavior                                                      & 0.27     & 0.000  & 0.11  & 0.43  \\
familiarity                                                             & -0.02    & 0.650  & -0.11 & 0.07  \\
Age                                                                     & -0.02    & 0.010  & -0.03 & -0.00 \\
educationAssociate                                                      & 0.33     & 0.360  & -0.38 & 1.03  \\
educationDoctorate                                                      & 0.91     & 0.310  & -0.84 & 2.65  \\
educationHigh school                                                    & 0.39     & 0.170  & -0.18 & 0.96  \\
educationLess than high school                                          & 0.66     & 0.370  & -0.80 & 2.12  \\
educationMaster                                                         & 0.19     & 0.460  & -0.31 & 0.69  \\
educationProfessional degree                                            & -0.27    & 0.720  & -1.71 & 1.18  \\
educationSome college                                                   & 0.02     & 0.910  & -0.40 & 0.44  \\
genderMale                                                              & -0.31    & 0.100  & -0.69 & 0.06  \\
genderOther                                                             & -0.06    & 0.910  & -1.06 & 0.95  \\
prophecy\_sourceAstrology:prophecy\_groupNegative                       & 0.14     & 0.360  & -0.17 & 0.45  \\
prophecy\_sourcePersonality:prophecy\_groupNegative                     & -0.00    & 0.980  & -0.29 & 0.28  \\
prophecy\_sourceAstrology:composite\_score                              & -0.10    & 0.080  & -0.21 & 0.01  \\
prophecy\_sourcePersonality:composite\_score                            & -0.07    & 0.170  & -0.18 & 0.03  \\
prophecy\_sourceAstrology:paranormal\_score                             & 0.01     & 0.000  & 0.01  & 0.02  \\
prophecy\_sourcePersonality:paranormal\_score                           & -0.00    & 0.270  & -0.01 & 0.00  \\
prophecy\_sourceAstrology:aias\_score                                   & -0.03    & 0.000  & -0.05 & -0.01 \\
prophecy\_sourcePersonality:aias\_score                                 & 0.01     & 0.570  & -0.01 & 0.02  \\
prophecy\_sourceAstrology:gullibility\_score                            & -0.00    & 0.980  & -0.02 & 0.02  \\
prophecy\_sourcePersonality:gullibility\_score                          & -0.02    & 0.100  & -0.04 & 0.00  \\
prophecy\_sourceAstrology:Age                                           & -0.00    & 0.530  & -0.02 & 0.01  \\
prophecy\_sourcePersonality:Age                                         & 0.01     & 0.380  & -0.01 & 0.02  \\
prophecy\_sourceAstrology:genderMale                                    & -0.06    & 0.740  & -0.39 & 0.27  \\
prophecy\_sourcePersonality:genderMale                                  & 0.01     & 0.930  & -0.30 & 0.32  \\
prophecy\_sourceAstrology:genderOther                                   & -0.20    & 0.650  & -1.08 & 0.68  \\
prophecy\_sourcePersonality:genderOther                                 & 0.20     & 0.630  & -0.62 & 1.01  \\
subscalePersonalization:prophecy\_sourceAstrology                       & 0.23     & 0.030  & 0.02  & 0.44  \\
subscaleReliability:prophecy\_sourceAstrology                           & -0.31    & 0.050  & -0.62 & -0.00 \\
subscaleUsefulness:prophecy\_sourceAstrology                            & -0.20    & 0.150  & -0.47 & 0.07  \\
subscalePersonalization:prophecy\_sourcePersonality                     & 0.08     & 0.440  & -0.12 & 0.27  \\
subscaleReliability:prophecy\_sourcePersonality                         & 0.18     & 0.240  & -0.12 & 0.47  \\
subscaleUsefulness:prophecy\_sourcePersonality                          & 0.10     & 0.440  & -0.16 & 0.36  \\
subscalePersonalization:prophecy\_groupNegative                         & -0.24    & 0.000  & -0.40 & -0.08 \\
subscaleReliability:prophecy\_groupNegative                             & 0.44     & 0.000  & 0.19  & 0.69  \\
subscaleUsefulness:prophecy\_groupNegative                              & 0.53     & 0.000  & 0.31  & 0.74  \\
subscalePersonalization:composite\_score                                & -0.05    & 0.090  & -0.11 & 0.01  \\
subscaleReliability:composite\_score                                    & -0.11    & 0.020  & -0.20 & -0.02 \\
subscaleUsefulness:composite\_score                                     & -0.12    & 0.000  & -0.20 & -0.04 \\
subscalePersonalization:paranormal\_score                               & 0.00     & 0.430  & -0.00 & 0.01  \\
subscaleReliability:paranormal\_score                                   & 0.01     & 0.020  & 0.00  & 0.01  \\
subscaleUsefulness:paranormal\_score                                    & 0.01     & 0.000  & 0.01  & 0.02  \\
subscalePersonalization:aias\_score                                     & -0.00    & 0.940  & -0.01 & 0.01  \\
subscaleReliability:aias\_score                                         & 0.03     & 0.000  & 0.02  & 0.05  \\
subscaleUsefulness:aias\_score                                          & 0.01     & 0.070  & -0.00 & 0.03  \\
subscalePersonalization:gullibility\_score                              & 0.01     & 0.020  & 0.00  & 0.02  \\
subscaleReliability:gullibility\_score                                  & -0.01    & 0.480  & -0.02 & 0.01  \\
subscaleUsefulness:gullibility\_score                                   & -0.01    & 0.070  & -0.03 & 0.00  \\
subscalePersonalization:big5\_extraversion                              & 0.00     & 0.890  & -0.03 & 0.03  \\
subscaleReliability:big5\_extraversion                                  & 0.04     & 0.090  & -0.01 & 0.09  \\
subscaleUsefulness:big5\_extraversion                                   & 0.06     & 0.000  & 0.02  & 0.10  \\
subscalePersonalization:big5\_openness                                  & -0.06    & 0.010  & -0.10 & -0.01 \\
subscaleReliability:big5\_openness                                      & -0.02    & 0.640  & -0.09 & 0.05  \\
subscaleUsefulness:big5\_openness                                       & -0.02    & 0.550  & -0.08 & 0.04  \\
subscalePersonalization:big5\_agreeableness                             & 0.02     & 0.370  & -0.02 & 0.07  \\
subscaleReliability:big5\_agreeableness                                 & -0.01    & 0.690  & -0.08 & 0.06  \\
subscaleUsefulness:big5\_agreeableness                                  & 0.02     & 0.470  & -0.04 & 0.08  \\
subscalePersonalization:big5\_conscientiousness                         & 0.01     & 0.770  & -0.04 & 0.05  \\
subscaleReliability:big5\_conscientiousness                             & -0.00    & 0.890  & -0.07 & 0.06  \\
subscaleUsefulness:big5\_conscientiousness                              & -0.00    & 0.920  & -0.06 & 0.05  \\
subscalePersonalization:big5\_emotional\_stability                      & 0.02     & 0.390  & -0.02 & 0.05  \\
subscaleReliability:big5\_emotional\_stability                          & 0.01     & 0.700  & -0.05 & 0.07  \\
subscaleUsefulness:big5\_emotional\_stability                           & 0.00     & 0.870  & -0.05 & 0.06  \\
subscalePersonalization:interest\_behavior                              & -0.06    & 0.020  & -0.11 & -0.01 \\
subscaleReliability:interest\_behavior                                  & -0.07    & 0.070  & -0.14 & 0.01  \\
subscaleUsefulness:interest\_behavior                                   & 0.02     & 0.480  & -0.04 & 0.09  \\
subscalePersonalization:familiarity                                     & 0.05     & 0.040  & 0.00  & 0.11  \\
subscaleReliability:familiarity                                         & 0.02     & 0.690  & -0.06 & 0.10  \\
subscaleUsefulness:familiarity                                          & 0.01     & 0.780  & -0.06 & 0.08  \\
subscalePersonalization:Age                                             & 0.00     & 0.720  & -0.01 & 0.01  \\
subscaleReliability:Age                                                 & 0.01     & 0.200  & -0.00 & 0.02  \\
subscaleUsefulness:Age                                                  & 0.00     & 0.420  & -0.01 & 0.01  \\
subscalePersonalization:educationAssociate                              & -0.03    & 0.780  & -0.25 & 0.18  \\
subscaleReliability:educationAssociate                                  & 0.11     & 0.500  & -0.22 & 0.44  \\
subscaleUsefulness:educationAssociate                                   & 0.21     & 0.140  & -0.07 & 0.49  \\
subscalePersonalization:educationDoctorate                              & 0.39     & 0.140  & -0.13 & 0.91  \\
subscaleReliability:educationDoctorate                                  & -0.36    & 0.390  & -1.17 & 0.46  \\
subscaleUsefulness:educationDoctorate                                   & -0.30    & 0.400  & -0.99 & 0.40  \\
subscalePersonalization:educationHigh school                            & -0.23    & 0.010  & -0.40 & -0.06 \\
subscaleReliability:educationHigh school                                & -0.06    & 0.680  & -0.32 & 0.21  \\
subscaleUsefulness:educationHigh school                                 & -0.01    & 0.960  & -0.23 & 0.22  \\
subscalePersonalization:educationLess than high school                  & -0.03    & 0.880  & -0.49 & 0.43  \\
subscaleReliability:educationLess than high school                      & -0.89    & 0.010  & -1.56 & -0.22 \\
subscaleUsefulness:educationLess than high school                       & 0.08     & 0.790  & -0.51 & 0.67  \\
subscalePersonalization:educationMaster                                 & -0.03    & 0.680  & -0.18 & 0.12  \\
subscaleReliability:educationMaster                                     & 0.15     & 0.210  & -0.08 & 0.39  \\
subscaleUsefulness:educationMaster                                      & 0.05     & 0.630  & -0.15 & 0.25  \\
subscalePersonalization:educationProfessional degree                    & 0.07     & 0.760  & -0.37 & 0.51  \\
subscaleReliability:educationProfessional degree                        & -0.35    & 0.310  & -1.02 & 0.32  \\
subscaleUsefulness:educationProfessional degree                         & -0.17    & 0.570  & -0.74 & 0.41  \\
subscalePersonalization:educationSome college                           & -0.04    & 0.510  & -0.17 & 0.08  \\
subscaleReliability:educationSome college                               & -0.05    & 0.650  & -0.24 & 0.15  \\
subscaleUsefulness:educationSome college                                & 0.03     & 0.730  & -0.14 & 0.20  \\
subscalePersonalization:genderMale                                      & 0.11     & 0.210  & -0.06 & 0.28  \\
subscaleReliability:genderMale                                          & 0.41     & 0.000  & 0.14  & 0.68  \\
subscaleUsefulness:genderMale                                           & 0.18     & 0.130  & -0.05 & 0.42  \\
subscalePersonalization:genderOther                                     & 0.38     & 0.100  & -0.07 & 0.84  \\
subscaleReliability:genderOther                                         & -0.15    & 0.680  & -0.87 & 0.57  \\
subscaleUsefulness:genderOther                                          & -0.14    & 0.670  & -0.76 & 0.49  \\
subscalePersonalization:prophecy\_sourceAstrology: prophecy\_groupNegative   & -0.10 & 0.420 & -0.33 & 0.14 \\
subscaleReliability:prophecy\_sourceAstrology: prophecy\_groupNegative   & 0.11     & 0.530  & -0.24 & 0.47  \\
subscaleUsefulness:prophecy\_sourceAstrology: prophecy\_groupNegative    & 0.07     & 0.630  & -0.23 & 0.38  \\
subscalePersonalization:prophecy\_sourcePersonality: prophecy\_groupNegative & -0.07 & 0.550 & -0.29 & 0.16 \\
subscaleReliability:prophecy\_sourcePersonality: prophecy\_groupNegative & -0.09    & 0.630  & -0.44 & 0.27  \\
subscaleUsefulness:prophecy\_sourcePersonality: prophecy\_groupNegative  & -0.19    & 0.230  & -0.49 & 0.12  \\
subscalePersonalization:prophecy\_sourceAstrology:composite\_score      & 0.04     & 0.400  & -0.05 & 0.12  \\
subscaleReliability:prophecy\_sourceAstrology:composite\_score          & 0.01     & 0.870  & -0.12 & 0.14  \\
subscaleUsefulness:prophecy\_sourceAstrology:composite\_score           & 0.06     & 0.320  & -0.05 & 0.16  \\
subscalePersonalization:prophecy\_sourcePersonality:composite\_score    & 0.04     & 0.360  & -0.04 & 0.12  \\
subscaleReliability:prophecy\_sourcePersonality:composite\_score        & -0.03    & 0.640  & -0.16 & 0.10  \\
subscaleUsefulness:prophecy\_sourcePersonality:composite\_score         & 0.05     & 0.410  & -0.06 & 0.16  \\
subscalePersonalization:prophecy\_sourceAstrology:paranormal\_score     & -0.01    & 0.060  & -0.01 & 0.00  \\
subscaleReliability:prophecy\_sourceAstrology:paranormal\_score         & 0.00     & 0.840  & -0.01 & 0.01  \\
subscaleUsefulness:prophecy\_sourceAstrology:paranormal\_score          & -0.00    & 0.530  & -0.01 & 0.01  \\
subscalePersonalization:prophecy\_sourcePersonality:paranormal\_score   & -0.00    & 0.130  & -0.01 & 0.00  \\
subscaleReliability:prophecy\_sourcePersonality:paranormal\_score       & -0.00    & 0.830  & -0.01 & 0.01  \\
subscaleUsefulness:prophecy\_sourcePersonality:paranormal\_score        & -0.00    & 0.600  & -0.01 & 0.01  \\
subscalePersonalization:prophecy\_sourceAstrology:aias\_score           & 0.01     & 0.110  & -0.00 & 0.03  \\
subscaleReliability:prophecy\_sourceAstrology:aias\_score               & -0.01    & 0.190  & -0.04 & 0.01  \\
subscaleUsefulness:prophecy\_sourceAstrology:aias\_score                & 0.00     & 0.810  & -0.02 & 0.02  \\
subscalePersonalization:prophecy\_sourcePersonality:aias\_score         & 0.00     & 0.690  & -0.01 & 0.02  \\
subscaleReliability:prophecy\_sourcePersonality:aias\_score             & -0.02    & 0.110  & -0.04 & 0.00  \\
subscaleUsefulness:prophecy\_sourcePersonality:aias\_score              & -0.01    & 0.330  & -0.03 & 0.01  \\
subscalePersonalization:prophecy\_sourceAstrology:gullibility\_score    & 0.00     & 0.950  & -0.01 & 0.02  \\
subscaleReliability:prophecy\_sourceAstrology:gullibility\_score        & -0.01    & 0.560  & -0.03 & 0.02  \\
subscaleUsefulness:prophecy\_sourceAstrology:gullibility\_score         & 0.01     & 0.370  & -0.01 & 0.03  \\
subscalePersonalization:prophecy\_sourcePersonality:gullibility\_score  & -0.02    & 0.030  & -0.03 & -0.00 \\
subscaleReliability:prophecy\_sourcePersonality:gullibility\_score      & -0.01    & 0.520  & -0.03 & 0.02  \\
subscaleUsefulness:prophecy\_sourcePersonality:gullibility\_score       & 0.01     & 0.610  & -0.02 & 0.03  \\
subscalePersonalization:prophecy\_sourceAstrology:Age                   & -0.00    & 0.570  & -0.01 & 0.01  \\
subscaleReliability:prophecy\_sourceAstrology:Age                       & 0.01     & 0.420  & -0.01 & 0.02  \\
subscaleUsefulness:prophecy\_sourceAstrology:Age                        & -0.00    & 0.830  & -0.01 & 0.01  \\
subscalePersonalization:prophecy\_sourcePersonality:Age                 & -0.00    & 0.940  & -0.01 & 0.01  \\
subscaleReliability:prophecy\_sourcePersonality:Age                     & -0.00    & 0.750  & -0.02 & 0.01  \\
subscaleUsefulness:prophecy\_sourcePersonality:Age                      & -0.01    & 0.330  & -0.02 & 0.01  \\
subscalePersonalization:prophecy\_sourceAstrology:genderMale            & -0.21    & 0.100  & -0.45 & 0.04  \\
subscaleReliability:prophecy\_sourceAstrology:genderMale                & 0.00     & 1.000  & -0.38 & 0.38  \\
subscaleUsefulness:prophecy\_sourceAstrology:genderMale                 & 0.08     & 0.630  & -0.24 & 0.40  \\
subscalePersonalization:prophecy\_sourcePersonality:genderMale          & -0.17    & 0.150  & -0.41 & 0.07  \\
subscaleReliability:prophecy\_sourcePersonality:genderMale              & -0.29    & 0.140  & -0.66 & 0.09  \\
subscaleUsefulness:prophecy\_sourcePersonality:genderMale               & -0.08    & 0.650  & -0.40 & 0.25  \\
subscalePersonalization:prophecy\_sourceAstrology:genderOther           & -0.06    & 0.870  & -0.71 & 0.59  \\
subscaleReliability:prophecy\_sourceAstrology:genderOther               & 0.36     & 0.490  & -0.65 & 1.37  \\
subscaleUsefulness:prophecy\_sourceAstrology:genderOther                & 0.48     & 0.270  & -0.37 & 1.33  \\
subscalePersonalization:prophecy\_sourcePersonality:genderOther         & -0.12    & 0.720  & -0.75 & 0.52  \\
subscaleReliability:prophecy\_sourcePersonality:genderOther             & -0.17    & 0.730  & -1.18 & 0.83  \\
subscaleUsefulness:prophecy\_sourcePersonality:genderOther              & 0.03     & 0.940  & -0.84 & 0.91  \\* \bottomrule
\end{longtable}

% \section{Model fit statistics}

\begin{table}[h]
\centering
\begin{minipage}{0.48\textwidth}
    \centering
    \caption{Model fit statistics for mixed effects model}
    \label{table:model_fit_stats_12f_c}
    \begin{tabular}{@{}lc@{}}
    \toprule
    \textbf{Statistic}     & \textbf{Value} \\ \midrule
    AIC                    & 7949.6         \\
    BIC                    & 9129.2         \\
    Log-Likelihood         & -3774.8        \\
    Number of Observations & 2856           \\ 
    Number of Groups       & 238            \\ \bottomrule
    \end{tabular}
\end{minipage}%
\hfill
\begin{minipage}{0.48\textwidth}
    \centering
    \caption{Standardized residuals}
    \label{table:standardized_res_12f_c}
    \begin{tabular}{@{}lcccc@{}}
    \toprule
    \textbf{Min} & \textbf{Q1} & \textbf{Med} & \textbf{Q3} & \textbf{Max} \\ \midrule
    -4.5316      & -0.5581     & -0.0065      & 0.5654      & 3.8666       \\ \bottomrule
    \end{tabular}
\end{minipage}
\end{table}

% \begin{table}[h]
% \centering
% \caption{Model fit statistics for mixed effects model}
% \label{table:model_fit_stats_12f_c}
% \begin{tabular}{@{}lc@{}}
% \toprule
% \textbf{Statistic}     & \textbf{Value} \\ \midrule
% AIC                    & 7949.6         \\
% BIC                    & 9129.2         \\
% Log-Likelihood         & -3774.8        \\
% Number of Observations & 2856           \\ 
% Number of Groups       & 238            \\ \bottomrule
% \end{tabular}
% \end{table}

\begin{table}[h]
\centering
\caption{Correlation structure specified at the lowest level}
\label{table:corr_structure_12f_c}
\begin{tabular}{@{}c|ccc@{}}
\toprule
           & \textbf{1} & \textbf{2} & \textbf{3} \\ \midrule
\textbf{2} & -0.474     &            &            \\
\textbf{3} & -0.737     & 0.263      &            \\
\textbf{4} & 0.131      & -0.365     & -0.596     \\ \bottomrule
\end{tabular}
\begin{minipage}[b]{\linewidth}
        \vspace{10pt}
        \subcaption*{This models the correlations within \texttt{prophecy\_source} nested within \texttt{qualtrics\_code}. This structure accounts for the dependency of observations grouped by \texttt{prophecy\_source} within each \texttt{qualtrics\_code}, allowing for different intercepts and slopes across these nested groups.}
    \end{minipage}
\end{table}

% \subsection{Standardized residuals}

% \begin{table}[h]
% \centering
% \caption{Standardized residuals}
% \label{}
% \begin{tabular}{@{}lcccc@{}}
% \toprule
% \textbf{Min} & \textbf{Q1} & \textbf{Med} & \textbf{Q3} & \textbf{Max} \\ \midrule
% -4.5316      & -0.5581     & -0.0065      & 0.5654      & 3.8666       \\ \bottomrule
% \end{tabular}
% \end{table}

\section{Mixed effects model: Model diagnostics for Section \ref{ch2-methods}}
\label{appendix:diagnostics_1}

% \subsection{Normality of residuals}

\begin{figure}[h]
\centering
    \caption{Q-Q plot of the fitted model}
    \includegraphics[width=10cm]{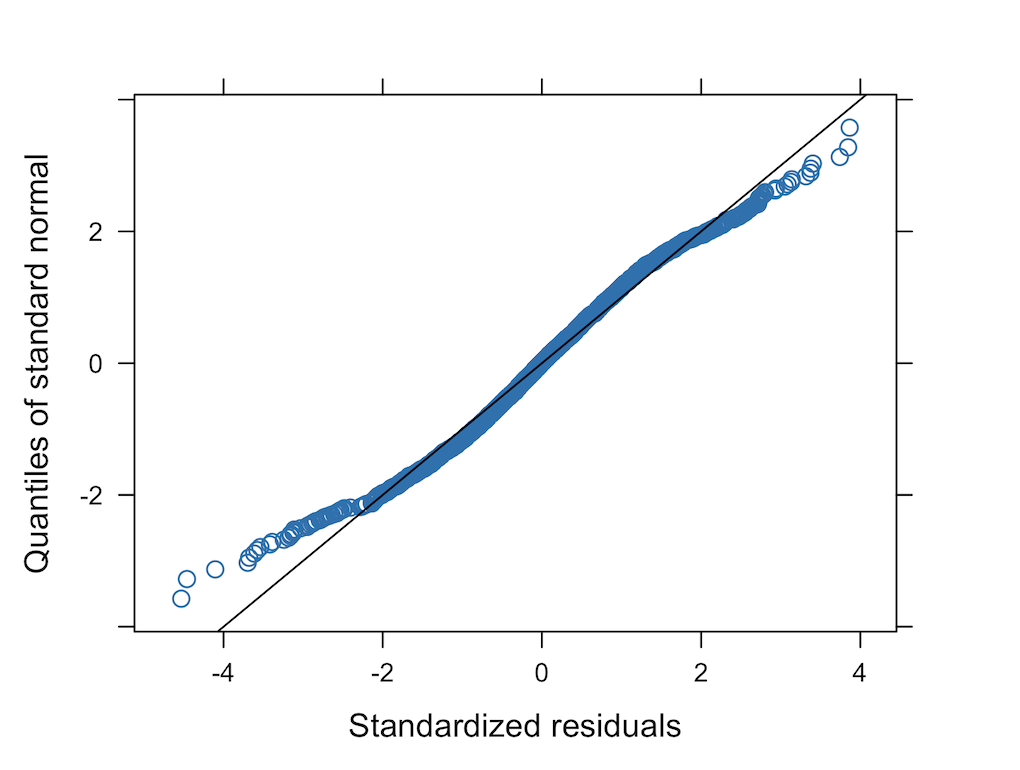}
    \begin{minipage}[b]{\linewidth}
        \vspace{10pt}
        \subcaption*{Normality of residuals was assessed using a Q-Q plot, using standardized residuals to account for both fixed and random effects. While there were slight deviations at the extremes, they were not severe and thus considered as normal deviations from real-world data.}
    \end{minipage}
    \label{fig:qqplot_12f_c}
\end{figure}

% \subsection{Homoscedasticity and linearity}

\begin{figure}[h]
    \centering
    \caption{A residuals plot comparing standardized residuals and fitted values}
    
    \label{fig:residual_fitted_12f_c}
    \includegraphics[width=16cm]{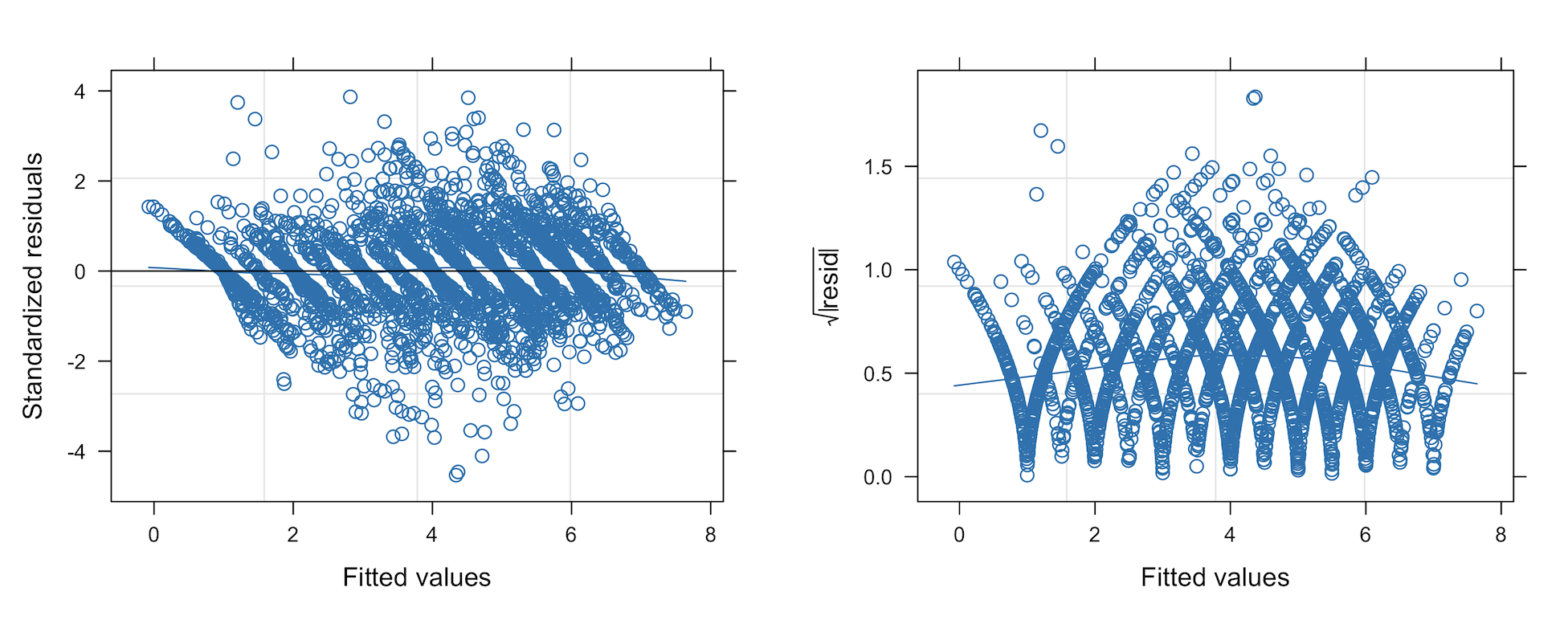}
    \begin{minipage}[b]{\linewidth}
        \vspace{10pt}
        \subcaption*{The plot showed a random spread around the horizontal line at zero, confirming that heteroscedasticity was successfully addressed using the variance function (varIdent) and that the linearity assumption is satisfied. One limitation to this diagnostics was the less continuous nature of the outcome variable (subscale score), which was multilevel with 13 levels (value of 1 to 7 with an increment of 0.5), which caused the residual plot to exhibit a slight diagonal pattern.}
    \end{minipage}
\end{figure}

% \subsection{Multicollinearity}

\begin{longtable}{@{}ll@{}}
\caption{Variance Inflation Factors (VIFs) values}
\label{table:vif_12f_c}\\
\toprule
\textbf{Predictor}                     & \textbf{VIF}      \\* \midrule
\endfirsthead
\multicolumn{2}{c}%
{{\bfseries Table \thetable\ continued from previous page}} \\
\toprule
\textbf{Predictors}                     & \textbf{VIF}      \\* \midrule
\endhead
\bottomrule
\endfoot
\endlastfoot
subscale                                     & 3.285782 \\
prophecy\_source                             & 3.509332 \\
prophecy\_group                              & 3.511723 \\
composite\_score                             & 3.776002 \\
paranormal\_score                            & 3.771773 \\
aias\_score                                  & 3.762766 \\
gullibility\_score                           & 3.734846 \\
big5\_extraversion                           & 2.268579 \\
big5\_openness                               & 2.383225 \\
big5\_agreeableness                          & 2.361957 \\
big5\_conscientiousness                      & 2.420150 \\
big5\_emotional\_stability                   & 2.723458 \\
interest\_behavior                           & 2.253065 \\
familiarity                                  & 2.264958 \\
Age                                          & 3.618152 \\
education                                    & 2.107854 \\
gender                                       & 3.674199 \\
prophecy\_source:prophecy\_group             & 3.386351 \\
prophecy\_source:composite\_score            & 2.745626 \\
prophecy\_source:paranormal\_score           & 2.846607 \\
prophecy\_source:aias\_score                 & 2.758390 \\
prophecy\_source:gullibility\_score          & 2.691879 \\
prophecy\_source:Age                         & 2.697924 \\
prophecy\_source:gender                      & 3.048713 \\
subscale:prophecy\_source                    & 2.746200 \\
subscale:prophecy\_group                     & 2.892868 \\
subscale:composite\_score                    & 2.378733 \\
subscale:paranormal\_score                   & 2.376068 \\
subscale:aias\_score                         & 2.370394 \\
subscale:gullibility\_score                  & 2.352806 \\
subscale:big5\_extraversion                  & 1.429115 \\
subscale:big5\_openness                      & 1.501338 \\
subscale:big5\_agreeableness                 & 1.487939 \\
subscale:big5\_conscientiousness             & 1.524599 \\
subscale:big5\_emotional\_stability          & 1.715671 \\
subscale:interest\_behavior                  & 1.419342 \\
subscale:familiarity                         & 1.426834 \\
subscale:Age                                 & 2.279293 \\
subscale:education                           & 1.391251 \\
subscale:gender                              & 2.612706 \\
subscale:prophecy\_source:prophecy\_group    & 2.304924 \\
subscale:prophecy\_source:composite\_score   & 1.729636 \\
subscale:prophecy\_source:paranormal\_score  & 1.793250 \\
subscale:prophecy\_source:aias\_score        & 1.737677 \\
subscale:prophecy\_source:gullibility\_score & 1.695778 \\
subscale:prophecy\_source:Age                & 1.699586 \\
subscale:prophecy\_source:gender             & 1.989948 \\* \bottomrule
\end{longtable}

\noindent All VIF values after centering the continuous variables were under 5, suggesting that multicollinearity was not a concern. This method adjusts the GVIF for predictors with multiple degrees of freedom, making multicollinearity measures more comparable across variables.

% \subsection{Random effects}

\begin{figure}[h]
    \centering
    \caption{Q-Q plot of the random effects}
    \label{fig:random_qq_12f_c}
    \includegraphics[width=10cm]{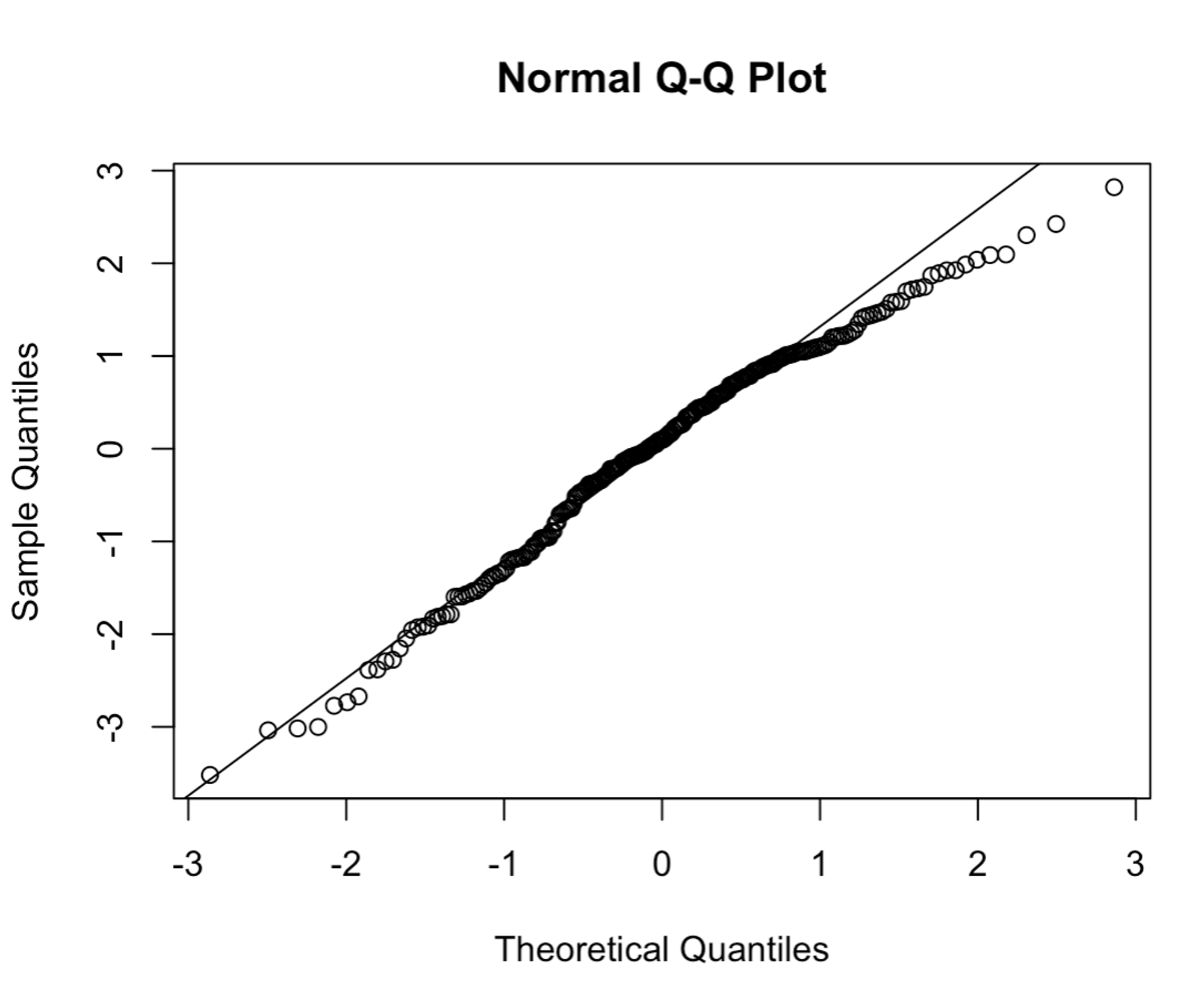}
    \begin{minipage}[b]{\linewidth}
        \vspace{10pt}
        \subcaption*{This indicates that the random effects are approximately normally distributed, as the points closely follow the theoretical normal line. However, there are slight deviations at the tails, suggesting minor departures from normality. These deviations were not deemed significant enough to invalidate the model assumptions.}
    \end{minipage}
\end{figure}

% \section{Independence of residuals / Autocorrelation}

\begin{table}[h]
\centering
\caption{Results of the Durbin-Watson test}
\label{table:dw_test_12f_c}
\begin{tabular}{@{}lcc@{}}
\toprule
\textbf{Group}  & \textbf{DW Statistic} & \textbf{p-value} \\ \midrule
Validity        & 1.6175                & 1.547e-07        \\ 
Personalization & 1.4457                & 6.027e-14        \\ 
Reliability     & 1.3706                & 2.2e-16          \\ 
Usefulness      & 1.4651                & 4.158e-13        \\ \bottomrule
\end{tabular} 
    \begin{minipage}[b]{\linewidth}
    \vspace{10pt}
        \subcaption*{We conducted the Durbin-Watson test within each subscale group to check for autocorrelation, finding significant positive autocorrelation in all groups. We initially used an autoregressive correlation structure (corAR1) to mitigate this, but model fit statistics indicated that an unstructured symmetric correlation structure (corSymm) provided a better fit, leading us to adopt corSymm as the final model. Although corSymm improved the model fit, it did not fully resolve the autocorrelation, which could limit the validity of our results and suggests a need for alternative approaches in future research.}
    \end{minipage}
\end{table}

%% file: appendixc.tex
% From mitthesis package
% Version: 1.01, 2023/07/04
% Documentation: https://ctan.org/pkg/mitthesis

\chapter{Detailed Results and Model Diagnostics for Chapter \ref{ch3_paper2}}
\label{appendix:results2}

\section{Model results}

This is the detailed results table from the model used in Section \ref{ch3-methods}. All reported coefficients are based on the reference categories: Prophecy Source = AI, Prophecy Group = Positive, Subscale = Validity, Gender = Female, and Education = Bachelor.

\begin{longtable}[c]{@{}p{10cm}p{1cm}p{1cm}p{1cm}p{1cm}@{}}
\caption{Detailed mixed effects model results for Chapter 3}
\label{table:model_results_14aa}\\
\toprule
Predictor                                                               & Estimate & PValue & Lower & Upper \\* \midrule
\endfirsthead
\multicolumn{5}{c}%
{{\bfseries Table \thetable\ continued from previous page}} \\
\toprule
Predictor                                                               & Estimate & PValue & Lower & Upper \\* \midrule
\endhead
\bottomrule
\endfoot
\endlastfoot
(Intercept)                                                             & 5.10     & 0.000  & 4.75  & 5.44  \\
subscalePersonalization                                                 & 0.23     & 0.000  & 0.08  & 0.37  \\
subscaleReliability                                                     & -0.91    & 0.000  & -1.13 & -0.68 \\
subscaleUsefulness                                                      & -0.63    & 0.000  & -0.83 & -0.44 \\
prophecy\_sourceAstrology                                               & -0.66    & 0.000  & -0.93 & -0.39 \\
prophecy\_sourcePersonality                                             & 0.06     & 0.620  & -0.19 & 0.31  \\
prophecy\_groupNegative                                                 & -1.19    & 0.000  & -1.53 & -0.85 \\
composite\_score                                                        & 0.13     & 0.060  & -0.01 & 0.26  \\
paranormal\_score                                                       & 0.02     & 0.000  & 0.01  & 0.03  \\
aias\_score                                                             & 0.04     & 0.000  & 0.01  & 0.06  \\
gullibility\_score                                                      & -0.02    & 0.130  & -0.04 & 0.01  \\
big5\_extraversion                                                      & -0.02    & 0.650  & -0.12 & 0.08  \\
big5\_openness                                                          & -0.10    & 0.210  & -0.25 & 0.05  \\
big5\_agreeableness                                                     & 0.09     & 0.240  & -0.06 & 0.24  \\
big5\_conscientiousness                                                 & -0.15    & 0.030  & -0.30 & -0.01 \\
big5\_emotional\_stability                                              & 0.00     & 0.980  & -0.13 & 0.13  \\
interest\_behavior                                                      & 0.27     & 0.000  & 0.11  & 0.43  \\
familiarity                                                             & -0.02    & 0.650  & -0.11 & 0.07  \\
Age                                                                     & -0.02    & 0.010  & -0.03 & -0.00 \\
educationAssociate                                                      & 0.33     & 0.360  & -0.38 & 1.03  \\
educationDoctorate                                                      & 0.91     & 0.310  & -0.84 & 2.65  \\
educationHigh school                                                    & 0.39     & 0.170  & -0.18 & 0.96  \\
educationLess than high school                                          & 0.66     & 0.370  & -0.80 & 2.12  \\
educationMaster                                                         & 0.19     & 0.460  & -0.31 & 0.69  \\
educationProfessional degree                                            & -0.27    & 0.720  & -1.71 & 1.18  \\
educationSome college                                                   & 0.02     & 0.910  & -0.40 & 0.44  \\
genderMale                                                              & -0.31    & 0.100  & -0.69 & 0.06  \\
genderOther                                                             & -0.06    & 0.910  & -1.06 & 0.95  \\
prophecy\_sourceAstrology:prophecy\_groupNegative                       & 0.14     & 0.360  & -0.17 & 0.45  \\
prophecy\_sourcePersonality:prophecy\_groupNegative                     & -0.00    & 0.980  & -0.29 & 0.28  \\
prophecy\_sourceAstrology:composite\_score                              & -0.10    & 0.080  & -0.21 & 0.01  \\
prophecy\_sourcePersonality:composite\_score                            & -0.07    & 0.170  & -0.18 & 0.03  \\
prophecy\_sourceAstrology:paranormal\_score                             & 0.01     & 0.000  & 0.01  & 0.02  \\
prophecy\_sourcePersonality:paranormal\_score                           & -0.00    & 0.270  & -0.01 & 0.00  \\
prophecy\_sourceAstrology:aias\_score                                   & -0.03    & 0.000  & -0.05 & -0.01 \\
prophecy\_sourcePersonality:aias\_score                                 & 0.01     & 0.570  & -0.01 & 0.02  \\
prophecy\_sourceAstrology:gullibility\_score                            & -0.00    & 0.980  & -0.02 & 0.02  \\
prophecy\_sourcePersonality:gullibility\_score                          & -0.02    & 0.100  & -0.04 & 0.00  \\
prophecy\_sourceAstrology:Age                                           & -0.00    & 0.530  & -0.02 & 0.01  \\
prophecy\_sourcePersonality:Age                                         & 0.01     & 0.380  & -0.01 & 0.02  \\
prophecy\_sourceAstrology:genderMale                                    & -0.06    & 0.740  & -0.39 & 0.27  \\
prophecy\_sourcePersonality:genderMale                                  & 0.01     & 0.930  & -0.30 & 0.32  \\
prophecy\_sourceAstrology:genderOther                                   & -0.20    & 0.650  & -1.08 & 0.68  \\
prophecy\_sourcePersonality:genderOther                                 & 0.20     & 0.630  & -0.62 & 1.01  \\
subscalePersonalization:prophecy\_sourceAstrology                       & 0.23     & 0.030  & 0.02  & 0.44  \\
subscaleReliability:prophecy\_sourceAstrology                           & -0.31    & 0.050  & -0.62 & -0.00 \\
subscaleUsefulness:prophecy\_sourceAstrology                            & -0.20    & 0.150  & -0.47 & 0.07  \\
subscalePersonalization:prophecy\_sourcePersonality                     & 0.08     & 0.440  & -0.12 & 0.27  \\
subscaleReliability:prophecy\_sourcePersonality                         & 0.18     & 0.240  & -0.12 & 0.47  \\
subscaleUsefulness:prophecy\_sourcePersonality                          & 0.10     & 0.440  & -0.16 & 0.36  \\
subscalePersonalization:prophecy\_groupNegative                         & -0.24    & 0.000  & -0.40 & -0.08 \\
subscaleReliability:prophecy\_groupNegative                             & 0.44     & 0.000  & 0.19  & 0.69  \\
subscaleUsefulness:prophecy\_groupNegative                              & 0.53     & 0.000  & 0.31  & 0.74  \\
subscalePersonalization:composite\_score                                & -0.05    & 0.090  & -0.11 & 0.01  \\
subscaleReliability:composite\_score                                    & -0.11    & 0.020  & -0.20 & -0.02 \\
subscaleUsefulness:composite\_score                                     & -0.12    & 0.000  & -0.20 & -0.04 \\
subscalePersonalization:paranormal\_score                               & 0.00     & 0.430  & -0.00 & 0.01  \\
subscaleReliability:paranormal\_score                                   & 0.01     & 0.020  & 0.00  & 0.01  \\
subscaleUsefulness:paranormal\_score                                    & 0.01     & 0.000  & 0.01  & 0.02  \\
subscalePersonalization:aias\_score                                     & -0.00    & 0.940  & -0.01 & 0.01  \\
subscaleReliability:aias\_score                                         & 0.03     & 0.000  & 0.02  & 0.05  \\
subscaleUsefulness:aias\_score                                          & 0.01     & 0.070  & -0.00 & 0.03  \\
subscalePersonalization:gullibility\_score                              & 0.01     & 0.020  & 0.00  & 0.02  \\
subscaleReliability:gullibility\_score                                  & -0.01    & 0.480  & -0.02 & 0.01  \\
subscaleUsefulness:gullibility\_score                                   & -0.01    & 0.070  & -0.03 & 0.00  \\
subscalePersonalization:big5\_extraversion                              & 0.00     & 0.890  & -0.03 & 0.03  \\
subscaleReliability:big5\_extraversion                                  & 0.04     & 0.090  & -0.01 & 0.09  \\
subscaleUsefulness:big5\_extraversion                                   & 0.06     & 0.000  & 0.02  & 0.10  \\
subscalePersonalization:big5\_openness                                  & -0.06    & 0.010  & -0.10 & -0.01 \\
subscaleReliability:big5\_openness                                      & -0.02    & 0.640  & -0.09 & 0.05  \\
subscaleUsefulness:big5\_openness                                       & -0.02    & 0.550  & -0.08 & 0.04  \\
subscalePersonalization:big5\_agreeableness                             & 0.02     & 0.370  & -0.02 & 0.07  \\
subscaleReliability:big5\_agreeableness                                 & -0.01    & 0.690  & -0.08 & 0.06  \\
subscaleUsefulness:big5\_agreeableness                                  & 0.02     & 0.470  & -0.04 & 0.08  \\
subscalePersonalization:big5\_conscientiousness                         & 0.01     & 0.770  & -0.04 & 0.05  \\
subscaleReliability:big5\_conscientiousness                             & -0.00    & 0.890  & -0.07 & 0.06  \\
subscaleUsefulness:big5\_conscientiousness                              & -0.00    & 0.920  & -0.06 & 0.05  \\
subscalePersonalization:big5\_emotional\_stability                      & 0.02     & 0.390  & -0.02 & 0.05  \\
subscaleReliability:big5\_emotional\_stability                          & 0.01     & 0.700  & -0.05 & 0.07  \\
subscaleUsefulness:big5\_emotional\_stability                           & 0.00     & 0.870  & -0.05 & 0.06  \\
subscalePersonalization:interest\_behavior                              & -0.06    & 0.020  & -0.11 & -0.01 \\
subscaleReliability:interest\_behavior                                  & -0.07    & 0.070  & -0.14 & 0.01  \\
subscaleUsefulness:interest\_behavior                                   & 0.02     & 0.480  & -0.04 & 0.09  \\
subscalePersonalization:familiarity                                     & 0.05     & 0.040  & 0.00  & 0.11  \\
subscaleReliability:familiarity                                         & 0.02     & 0.690  & -0.06 & 0.10  \\
subscaleUsefulness:familiarity                                          & 0.01     & 0.780  & -0.06 & 0.08  \\
subscalePersonalization:Age                                             & 0.00     & 0.720  & -0.01 & 0.01  \\
subscaleReliability:Age                                                 & 0.01     & 0.200  & -0.00 & 0.02  \\
subscaleUsefulness:Age                                                  & 0.00     & 0.420  & -0.01 & 0.01  \\
subscalePersonalization:educationAssociate                              & -0.03    & 0.780  & -0.25 & 0.18  \\
subscaleReliability:educationAssociate                                  & 0.11     & 0.500  & -0.22 & 0.44  \\
subscaleUsefulness:educationAssociate                                   & 0.21     & 0.140  & -0.07 & 0.49  \\
subscalePersonalization:educationDoctorate                              & 0.39     & 0.140  & -0.13 & 0.91  \\
subscaleReliability:educationDoctorate                                  & -0.36    & 0.390  & -1.17 & 0.46  \\
subscaleUsefulness:educationDoctorate                                   & -0.30    & 0.400  & -0.99 & 0.40  \\
subscalePersonalization:educationHigh school                            & -0.23    & 0.010  & -0.40 & -0.06 \\
subscaleReliability:educationHigh school                                & -0.06    & 0.680  & -0.32 & 0.21  \\
subscaleUsefulness:educationHigh school                                 & -0.01    & 0.960  & -0.23 & 0.22  \\
subscalePersonalization:educationLess than high school                  & -0.03    & 0.880  & -0.49 & 0.43  \\
subscaleReliability:educationLess than high school                      & -0.89    & 0.010  & -1.56 & -0.22 \\
subscaleUsefulness:educationLess than high school                       & 0.08     & 0.790  & -0.51 & 0.67  \\
subscalePersonalization:educationMaster                                 & -0.03    & 0.680  & -0.18 & 0.12  \\
subscaleReliability:educationMaster                                     & 0.15     & 0.210  & -0.08 & 0.39  \\
subscaleUsefulness:educationMaster                                      & 0.05     & 0.630  & -0.15 & 0.25  \\
subscalePersonalization:educationProfessional degree                    & 0.07     & 0.760  & -0.37 & 0.51  \\
subscaleReliability:educationProfessional degree                        & -0.35    & 0.310  & -1.02 & 0.32  \\
subscaleUsefulness:educationProfessional degree                         & -0.17    & 0.570  & -0.74 & 0.41  \\
subscalePersonalization:educationSome college                           & -0.04    & 0.510  & -0.17 & 0.08  \\
subscaleReliability:educationSome college                               & -0.05    & 0.650  & -0.24 & 0.15  \\
subscaleUsefulness:educationSome college                                & 0.03     & 0.730  & -0.14 & 0.20  \\
subscalePersonalization:genderMale                                      & 0.11     & 0.210  & -0.06 & 0.28  \\
subscaleReliability:genderMale                                          & 0.41     & 0.000  & 0.14  & 0.68  \\
subscaleUsefulness:genderMale                                           & 0.18     & 0.130  & -0.05 & 0.42  \\
subscalePersonalization:genderOther                                     & 0.38     & 0.100  & -0.07 & 0.84  \\
subscaleReliability:genderOther                                         & -0.15    & 0.680  & -0.87 & 0.57  \\
subscaleUsefulness:genderOther                                          & -0.14    & 0.670  & -0.76 & 0.49  \\
subscalePersonalization:prophecy\_sourceAstrology:prophecy\_groupNegative   & -0.10 & 0.420 & -0.33 & 0.14 \\
subscaleReliability:prophecy\_sourceAstrology:prophecy\_groupNegative   & 0.11     & 0.530  & -0.24 & 0.47  \\
subscaleUsefulness:prophecy\_sourceAstrology:prophecy\_groupNegative    & 0.07     & 0.630  & -0.23 & 0.38  \\
subscalePersonalization:prophecy\_sourcePersonality:prophecy\_groupNegative & -0.07 & 0.550 & -0.29 & 0.16 \\
subscaleReliability:prophecy\_sourcePersonality:prophecy\_groupNegative & -0.09    & 0.630  & -0.44 & 0.27  \\
subscaleUsefulness:prophecy\_sourcePersonality:prophecy\_groupNegative  & -0.19    & 0.230  & -0.49 & 0.12  \\
subscalePersonalization:prophecy\_sourceAstrology:composite\_score      & 0.04     & 0.400  & -0.05 & 0.12  \\
subscaleReliability:prophecy\_sourceAstrology:composite\_score          & 0.01     & 0.870  & -0.12 & 0.14  \\
subscaleUsefulness:prophecy\_sourceAstrology:composite\_score           & 0.06     & 0.320  & -0.05 & 0.16  \\
subscalePersonalization:prophecy\_sourcePersonality:composite\_score    & 0.04     & 0.360  & -0.04 & 0.12  \\
subscaleReliability:prophecy\_sourcePersonality:composite\_score        & -0.03    & 0.640  & -0.16 & 0.10  \\
subscaleUsefulness:prophecy\_sourcePersonality:composite\_score         & 0.05     & 0.410  & -0.06 & 0.16  \\
subscalePersonalization:prophecy\_sourceAstrology:paranormal\_score     & -0.01    & 0.060  & -0.01 & 0.00  \\
subscaleReliability:prophecy\_sourceAstrology:paranormal\_score         & 0.00     & 0.840  & -0.01 & 0.01  \\
subscaleUsefulness:prophecy\_sourceAstrology:paranormal\_score          & -0.00    & 0.530  & -0.01 & 0.01  \\
subscalePersonalization:prophecy\_sourcePersonality:paranormal\_score   & -0.00    & 0.130  & -0.01 & 0.00  \\
subscaleReliability:prophecy\_sourcePersonality:paranormal\_score       & -0.00    & 0.830  & -0.01 & 0.01  \\
subscaleUsefulness:prophecy\_sourcePersonality:paranormal\_score        & -0.00    & 0.600  & -0.01 & 0.01  \\
subscalePersonalization:prophecy\_sourceAstrology:aias\_score           & 0.01     & 0.110  & -0.00 & 0.03  \\
subscaleReliability:prophecy\_sourceAstrology:aias\_score               & -0.01    & 0.190  & -0.04 & 0.01  \\
subscaleUsefulness:prophecy\_sourceAstrology:aias\_score                & 0.00     & 0.810  & -0.02 & 0.02  \\
subscalePersonalization:prophecy\_sourcePersonality:aias\_score         & 0.00     & 0.690  & -0.01 & 0.02  \\
subscaleReliability:prophecy\_sourcePersonality:aias\_score             & -0.02    & 0.110  & -0.04 & 0.00  \\
subscaleUsefulness:prophecy\_sourcePersonality:aias\_score              & -0.01    & 0.330  & -0.03 & 0.01  \\
subscalePersonalization:prophecy\_sourceAstrology:gullibility\_score    & 0.00     & 0.950  & -0.01 & 0.02  \\
subscaleReliability:prophecy\_sourceAstrology:gullibility\_score        & -0.01    & 0.560  & -0.03 & 0.02  \\
subscaleUsefulness:prophecy\_sourceAstrology:gullibility\_score         & 0.01     & 0.370  & -0.01 & 0.03  \\
subscalePersonalization:prophecy\_sourcePersonality:gullibility\_score  & -0.02    & 0.030  & -0.03 & -0.00 \\
subscaleReliability:prophecy\_sourcePersonality:gullibility\_score      & -0.01    & 0.520  & -0.03 & 0.02  \\
subscaleUsefulness:prophecy\_sourcePersonality:gullibility\_score       & 0.01     & 0.610  & -0.02 & 0.03  \\
subscalePersonalization:prophecy\_sourceAstrology:Age                   & -0.00    & 0.570  & -0.01 & 0.01  \\
subscaleReliability:prophecy\_sourceAstrology:Age                       & 0.01     & 0.420  & -0.01 & 0.02  \\
subscaleUsefulness:prophecy\_sourceAstrology:Age                        & -0.00    & 0.830  & -0.01 & 0.01  \\
subscalePersonalization:prophecy\_sourcePersonality:Age                 & -0.00    & 0.940  & -0.01 & 0.01  \\
subscaleReliability:prophecy\_sourcePersonality:Age                     & -0.00    & 0.750  & -0.02 & 0.01  \\
subscaleUsefulness:prophecy\_sourcePersonality:Age                      & -0.01    & 0.330  & -0.02 & 0.01  \\
subscalePersonalization:prophecy\_sourceAstrology:genderMale            & -0.21    & 0.100  & -0.45 & 0.04  \\
subscaleReliability:prophecy\_sourceAstrology:genderMale                & 0.00     & 1.000  & -0.38 & 0.38  \\
subscaleUsefulness:prophecy\_sourceAstrology:genderMale                 & 0.08     & 0.630  & -0.24 & 0.40  \\
subscalePersonalization:prophecy\_sourcePersonality:genderMale          & -0.17    & 0.150  & -0.41 & 0.07  \\
subscaleReliability:prophecy\_sourcePersonality:genderMale              & -0.29    & 0.140  & -0.66 & 0.09  \\
subscaleUsefulness:prophecy\_sourcePersonality:genderMale               & -0.08    & 0.650  & -0.40 & 0.25  \\
subscalePersonalization:prophecy\_sourceAstrology:genderOther           & -0.06    & 0.870  & -0.71 & 0.59  \\
subscaleReliability:prophecy\_sourceAstrology:genderOther               & 0.36     & 0.490  & -0.65 & 1.37  \\
subscaleUsefulness:prophecy\_sourceAstrology:genderOther                & 0.48     & 0.270  & -0.37 & 1.33  \\
subscalePersonalization:prophecy\_sourcePersonality:genderOther         & -0.12    & 0.720  & -0.75 & 0.52  \\
subscaleReliability:prophecy\_sourcePersonality:genderOther             & -0.17    & 0.730  & -1.18 & 0.83  \\
subscaleUsefulness:prophecy\_sourcePersonality:genderOther              & 0.03     & 0.940  & -0.84 & 0.91  \\* \bottomrule
\end{longtable}

\section{Model diagnostics}
\label{appendix:diagnostics_2}

% \subsection{Normality of residuals}

\begin{figure}[h]
\centering
    \caption{Q-Q plot of the fitted model}
    \includegraphics[width=10cm]{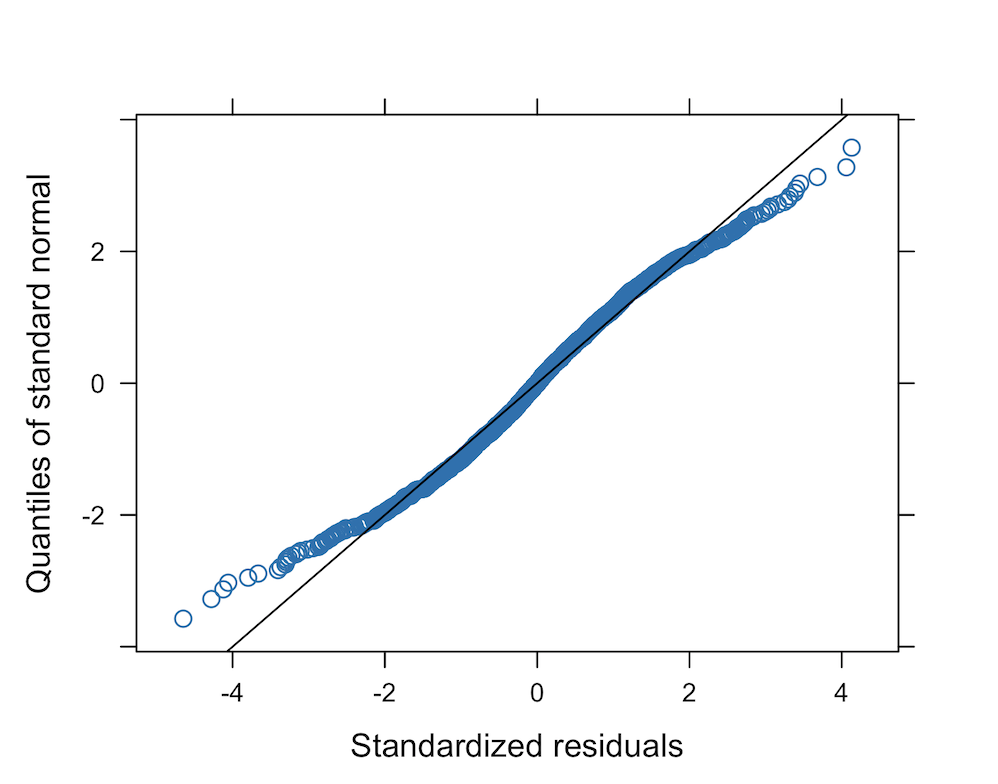}
    \begin{minipage}[b]{\linewidth}
        \vspace{10pt}
        \subcaption*{Normality of residuals was assessed using a Q-Q plot, using standardized residuals to account for both fixed and random effects. While there were slight deviations at the extremes, they were not severe and thus considered as normal deviations from real-world data.}
    \end{minipage}
    \label{fig:qqplot_14aa}
\end{figure}

% \subsection{Homoscedasticity and linearity}

\begin{figure}[h]
    \centering
    \caption{A residuals plot comparing standardized residuals and fitted values}
    
    \label{fig:residual_fitted_14aa}
    \includegraphics[width=15cm]{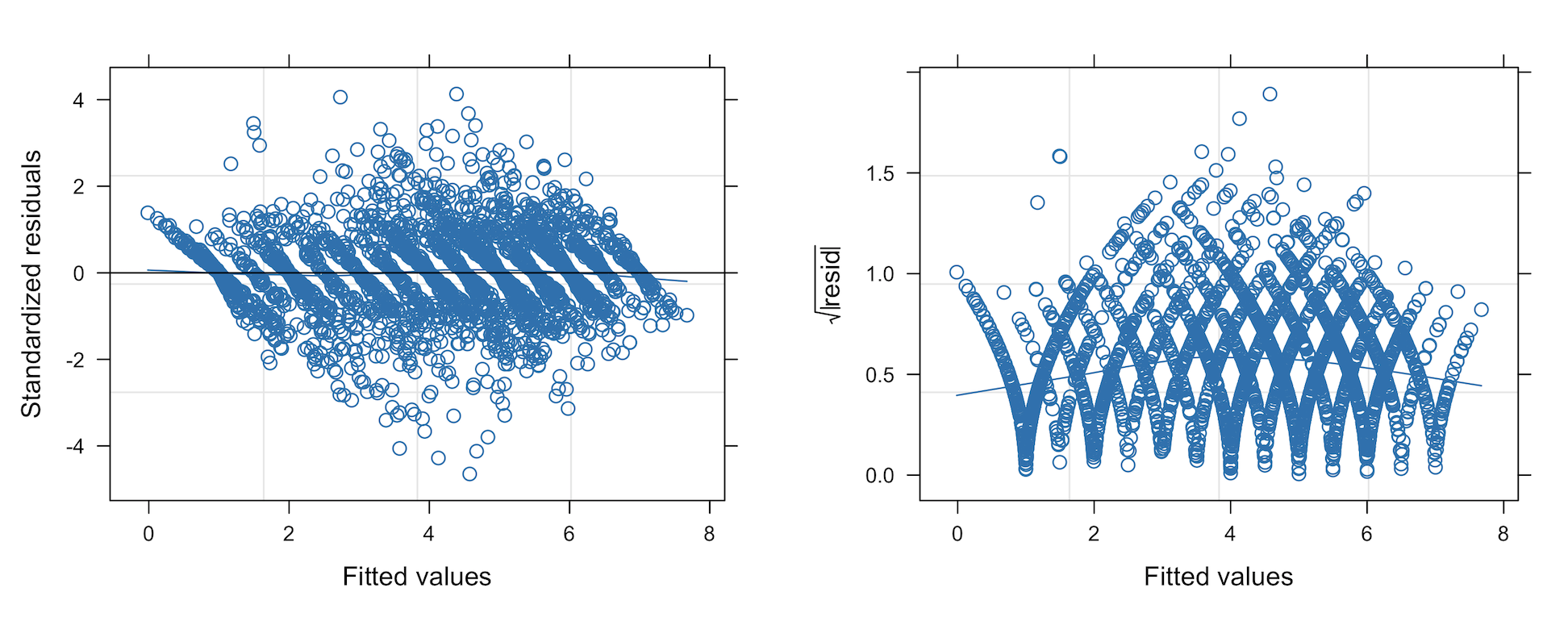}
    \begin{minipage}[b]{\linewidth}
        \vspace{10pt}
        \subcaption*{The plot showed a random spread around the horizontal line at zero, confirming that heteroscedasticity was successfully addressed using the variance function (varIdent) and that the linearity assumption is satisfied. One limitation to this diagnostics was the less continuous nature of the outcome variable (subscale score), which was multilevel with 13 levels (value of 1 to 7 with an increment of 0.5), which caused the residual plot to exhibit a slight diagonal pattern.}
    \end{minipage}
\end{figure}

% \subsection{Multicollinearity}

\begin{longtable}[c]{@{}ll@{}}
\caption{Variance Inflation Factors (VIFs) values}
\label{tab:my-table}\\
\toprule
\textbf{Predictor}                     & \textbf{VIF}      \\* \midrule
\endfirsthead
\multicolumn{2}{c}%
{{\bfseries Table \thetable\ continued from previous page}} \\
\toprule
\textbf{Predictors}                     & \textbf{VIF}      \\* \midrule
\endhead
\bottomrule
\endfoot
\endlastfoot
subscale                                    & 3.209298 \\
prophecy\_group                             & 4.055251 \\
prophecy\_source                            & 2.829113 \\
composite\_score                            & 3.265803 \\
paranormal\_score                           & 3.141498 \\
aias\_score                                 & 3.319988 \\
gullibility\_score                          & 3.214735 \\
big5\_extraversion                          & 2.363846 \\
big5\_openness                              & 2.398541 \\
big5\_agreeableness                         & 2.420504 \\
big5\_conscientiousness                     & 2.465333 \\
big5\_emotional\_stability                  & 2.754594 \\
interest\_behavior                          & 2.283370 \\
familiarity                                 & 2.120407 \\
Age                                         & 2.989669 \\
education                                   & 2.163890 \\
gender                                      & 3.146641 \\
prophecy\_group:prophecy\_source            & 3.365217 \\
prophecy\_group:composite\_score            & 3.169956 \\
prophecy\_group:paranormal\_score           & 3.159193 \\
prophecy\_group:aias\_score                 & 3.188012 \\
prophecy\_group:gullibility\_score          & 3.077286 \\
prophecy\_group:Age                         & 3.053249 \\
prophecy\_group:gender                      & 3.342625 \\
subscale:prophecy\_group                    & 3.340613 \\
subscale:prophecy\_source                   & 2.213901 \\
subscale:composite\_score                   & 2.057327 \\
subscale:paranormal\_score                  & 1.979020 \\
subscale:aias\_score                        & 2.091462 \\
subscale:gullibility\_score                 & 2.025156 \\
subscale:big5\_extraversion                 & 1.489130 \\
subscale:big5\_openness                     & 1.510986 \\
subscale:big5\_agreeableness                & 1.524822 \\
subscale:big5\_conscientiousness            & 1.553062 \\
subscale:big5\_emotional\_stability         & 1.735285 \\
subscale:interest\_behavior                 & 1.438433 \\
subscale:familiarity                        & 1.335773 \\
subscale:Age                                & 1.883374 \\
subscale:education                          & 1.428237 \\
subscale:gender                             & 2.237562 \\
subscale:prophecy\_group:prophecy\_source   & 2.290539 \\
subscale:prophecy\_group:composite\_score   & 1.997536 \\
subscale:prophecy\_group:paranormal\_score  & 1.992019 \\
subscale:prophecy\_group:aias\_score        & 2.008683 \\
subscale:prophecy\_group:gullibility\_score & 1.940316 \\
subscale:prophecy\_group:Age                & 1.923432 \\
subscale:prophecy\_group:gender             & 2.211006 \\* \bottomrule
\end{longtable}

\noindent All VIF values after centering the continuous variables were under 5, suggesting that multicollinearity was not a concern. This method adjusts the GVIF for predictors with multiple degrees of freedom, making multicollinearity measures more comparable across variables. 

% \subsection{Random effects}

\begin{figure}[h]
    \centering
    \caption{Q-Q plots of the random effects}
    \label{fig:random_qq_14aa}
    \includegraphics[width=10cm]{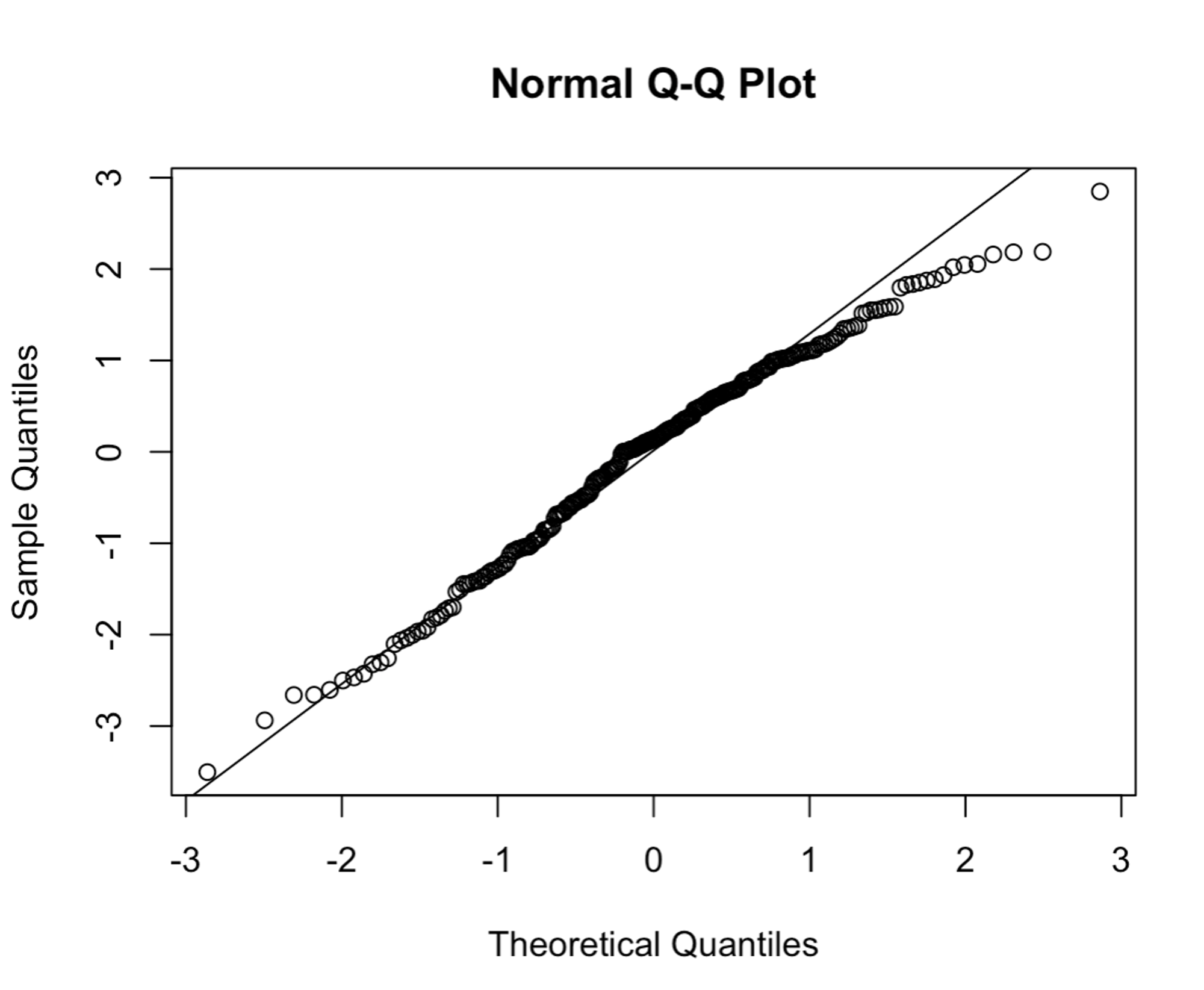}
    \begin{minipage}[b]{\linewidth}
        \vspace{10pt}
        \subcaption*{This indicates that the random effects are approximately normally distributed, as the points closely follow the theoretical normal line. However, there are slight deviations at the tails, suggesting minor departures from normality. These deviations were not deemed significant enough to invalidate the model assumptions.}
    \end{minipage}
\end{figure}

% \section{Independence of residuals / Autocorrelation}

\begin{table}[h]
\centering
\caption{Results of the Durbin-Watson test}
\label{table:dw_test_14aa}
\begin{tabular}{@{}lcc@{}}
\toprule
\textbf{Group}  & \textbf{DW Statistic} & \textbf{p-value} \\ \midrule
Validity        & 1.6917                & 1.86e-05        \\ 
Personalization & 1.5129                & 3.593e-11        \\ 
Reliability     & 1.4308                & 1.314e-14        \\ 
Usefulness      & 1.5136                & 3.813e-11        \\ \bottomrule
\end{tabular} 
    \begin{minipage}[b]{\linewidth}
    \vspace{10pt}
        \subcaption*{We conducted the Durbin-Watson test within each subscale group to check for autocorrelation, finding significant positive autocorrelation in all groups. We initially used an autoregressive correlation structure (corAR1) to mitigate this, but model fit statistics indicated that an unstructured symmetric correlation structure (corSymm) provided a better fit, leading us to adopt corSymm as the final model. Although corSymm improved the model fit, it did not fully resolve the autocorrelation, which could limit the validity of our results and suggests a need for alternative approaches in future research.}
    \end{minipage}
\end{table}

%% file: appendixd.tex
% From mitthesis package
% Version: 1.01, 2023/07/04
% Documentation: https://ctan.org/pkg/mitthesis

\chapter{Simulated investment game design: Additional details}
\label{appendix:game_details}

\section{Porfolio allocation scenario}

A dynamic allocation strategy was chosen as the default user action scenario for the simulated investment game. In contrast to a fixed allocation approach, a dynamic allocation strategy allows users to adjust their portfolio based on outcomes over time. In this scenario, investors react to market changes, new information, or shifts in their investment thesis by modifying their asset allocation. Dynamic allocation is particularly suitable for simulating tactical asset allocation, sector rotation, or active management strategies. It mirrors real-world practices employed by both funds and individual investors, especially in volatile markets.

\section{Market scenario}
The investment game scenario was set to be consistent for everyone to control the condition.

\noindent The sequence of user actions for each round included: 
\begin{enumerate}
\setlength\itemsep{0em}
    \item User sees the investment news.
    \item User sees the market change.
    \item User edits allocation based for that round.
\end{enumerate}

\begin{table}[h]
\centering
\caption{Overview of the news sentiment, expected user behavior, and market scenario in the simulated investment game}
\label{table:investment_news}
\begin{tabular}{@{}llll@{}}
\toprule
\textbf{Round} & \textbf{News sentiment} & \textbf{Expected behavior }                    & \textbf{Market Scenario} \\ \midrule
1     & n/a            & n/a                                   & Bull            \\ \midrule
2     & Positive       & Increase high risk; Decrease low risk & Bull            \\ \midrule
3     & Positive       & Increase high risk; Decrease low risk & Bull            \\ \midrule
4     & Positive       & Increase high risk; Decrease low risk & Bear            \\ \midrule
5     & Negative       & Decrease high risk; Increase low risk & Bear            \\ \midrule
6     & Negative       & Decrease high risk; Increase low risk & Bull            \\ \midrule
7     & Positive       & Increase high risk; Decrease low risk & Bull            \\ \midrule
8     & Positive       & Increase high risk; Decrease low risk & Bear            \\ \midrule
9     & Negative       & Decrease high risk; Increase low risk & Bear            \\ \midrule
10    & Negative       & Decrease high risk; Increase low risk & Bear            \\ \bottomrule
\end{tabular}
\end{table}

\section{Investment News}

\begin{table}[h]
\centering
\caption{Example market news that was presented to users at each round}
\label{table:market_news}
\begin{tabular}{@{}p{7.5cm}p{7.5cm}@{}}
\toprule
\textbf{Positive sentiment} &
  \textbf{Negative sentiment} \\ \midrule
The high risk high return category is expected to generate significant growth in the next few months. &
  Due to an unexpected market downturn, some projects in the high risk high return category appear to be much riskier; some investors are selling. \\
Financial experts project a strong bull market for the high risk high return category for the next few months. &
  Due to continuing volatile market conditions, the performance of high risk high return investment category looks likely to plummet. \\ 
Warren Buffett reveals that he has just invested \$1 million in a company in the high risk high return category. &
  A leading company that has attracted huge amounts of high risk high return investment looks shaky as their stock price takes a dive after the recession announcement. \\ \bottomrule
\end{tabular}
\end{table}

%% file: appendixe.tex
% From mitthesis package
% Version: 1.01, 2023/07/04
% Documentation: https://ctan.org/pkg/mitthesis

\chapter{Code listing}
\label{appendix:code}

The full code is available on GitHub (\url{https://github.com/mitmedialab/ai-superstition}).

\section{Multiple Linear Regression (Chapter \ref{ch2_paper1})}

\lstdefinestyle{mystyle}{
    backgroundcolor=\color{CadetBlue!15!white},   
    commentstyle=\color{Red3},
    numberstyle=\tiny\color{gray},
    stringstyle=\color{Blue3},
    keywordstyle=\color{Purple3}, % Adjust for R keywords
    basicstyle=\small\ttfamily,
    breakatwhitespace=false,         
    breaklines=true,                 
    numbers=left,                    
    numbersep=5pt,                  
    showspaces=false,                
    showstringspaces=false,
    showtabs=false,                  
    tabsize=2
}%
\lstset{language=R,style={mystyle}}%

\begin{lstlisting}
model <- lm(ai_overall_score ~ zodiac_overall_score + personality_overall_score + composite_score + paranormal_score + gullibility_score + interest_behavior  + Age + education + gender + prophecy_group, data = wide_df)

\end{lstlisting}

\section{Mixed Effects Model (Chapter \ref{ch2_paper1})}

\lstdefinestyle{mystyle}{
    backgroundcolor=\color{CadetBlue!15!white},   
    commentstyle=\color{Red3},
    numberstyle=\tiny\color{gray},
    stringstyle=\color{Blue3},
    keywordstyle=\color{Purple3}, % Adjust for R keywords
    basicstyle=\small\ttfamily,
    breakatwhitespace=false,         
    breaklines=true,                 
    numbers=left,                    
    numbersep=5pt,                  
    showspaces=false,                
    showstringspaces=false,
    showtabs=false,                  
    tabsize=2
}%
\lstset{language=R,style={mystyle}}%

\begin{lstlisting}
model <- lme(fixed = subscale_score ~ subscale * (prophecy_source * prophecy_group +
                prophecy_source * composite_score +
                prophecy_source * paranormal_score +
                prophecy_source * aias_score + 
                prophecy_source * gullibility_score + 
                big5_extraversion + 
                big5_openness + 
                big5_agreeableness + 
                big5_conscientiousness + 
                big5_emotional_stability + 
                interest_behavior + 
                familiarity +
                prophecy_source * Age +
                education + 
                prophecy_source * gender),
    random = ~ prophecy_source | qualtrics_code,
    correlation = corSymm(form = ~ 1 | qualtrics_code/prophecy_source),
    weights = varIdent(form = ~1 | subscale * prophecy_source * prophecy_group),
    data = long_df_centered,
    na.action = na.exclude,
    control = list(msMaxIter = 1000, msMaxEval = 1000, opt="optim"))

\end{lstlisting}

\section{Mixed Effects Model (Chapter \ref{ch3_paper2})}

\lstdefinestyle{mystyle}{
    backgroundcolor=\color{CadetBlue!15!white},   
    commentstyle=\color{Red3},
    numberstyle=\tiny\color{gray},
    stringstyle=\color{Blue3},
    keywordstyle=\color{Purple3}, % Adjust for R keywords
    basicstyle=\small\ttfamily,
    breakatwhitespace=false,         
    breaklines=true,                 
    numbers=left,                    
    numbersep=5pt,                  
    showspaces=false,                
    showstringspaces=false,
    showtabs=false,                  
    tabsize=2
}%
\lstset{language=R,style={mystyle}}%

\begin{lstlisting}
model <- lme(fixed = subscale_score ~ subscale * (prophecy_group * prophecy_source + 
                prophecy_group * composite_score + 
                prophecy_group * paranormal_score + 
                prophecy_group * aias_score + 
                prophecy_group * gullibility_score + 
                big5_extraversion + 
                big5_openness + 
                big5_agreeableness + 
                big5_conscientiousness + 
                big5_emotional_stability + 
                interest_behavior + 
                familiarity +
                prophecy_group * Age + 
                education + 
                prophecy_group * gender), 
    random = ~ prophecy_source | qualtrics_code, 
    correlation = corSymm(form = ~1 | qualtrics_code/prophecy_source),
    weights = varIdent(form = ~1 | subscale * prophecy_source * prophecy_group),
    data = long_df_centered,
    na.action = na.exclude,
    control = list(msMaxIter = 1000, msMaxEval = 1000, opt="optim"))

\end{lstlisting}